\documentclass[preprint2]{aastex}

\usepackage{graphicx,natbib}
\usepackage{xspace}

\newcommand{\mic}{\ensuremath{{\rm \kern 0.2em \mu m}}\xspace}
\newcommand{\microns}{\ensuremath{{\rm \kern 0.2em \mu m}}\xspace}
\newcommand{\mj}{\hbox{\ensuremath{{\rm \kern 0.2em MJy \kern 0.1em sr^{-1}}}}}
\newcommand{\wms}{\hbox{\ensuremath{{\rm \kern 0.2em W \kern 0.1em m^{-2} \kern 0.1em sr^{-1}}}}}
\newcommand{\wm}{\hbox{\ensuremath{{\rm \kern 0.2em W \kern 0.1em m^{-2}}}}}
\newcommand{\wmm}{\hbox{\ensuremath{{\rm \kern 0.2em W \kern 0.1em m^{-2} \kern 0.1em \mu m^{-1}}}}}
\newcommand{\km}{\hbox{\ensuremath{{\rm \kern 0.2em km \kern 0.1em s^{-1}}}}}
\newcommand{\ccm}{\ensuremath{{\rm \kern 0.2em cm^{-3}}}}
\newcommand{\scm}{\ensuremath{{\rm \kern 0.2em cm^{-2}}}}
\newcommand{\sscm}{\ensuremath{{\rm \kern 0.2em cm^{-2}}}}
\newcommand{\icm}{\ensuremath{{\rm \kern 0.2em cm^{-1}}}}
\newcommand{\is}{\ensuremath{{\rm \kern 0.2em s^{-1}}}}
\newcommand{\htwo}{\hbox{\ensuremath{\rm H_2}}}
\newcommand{\water}{\hbox{\ensuremath{\rm H_2 \kern -0.1em O}}}
\newcommand{\co}{\hbox{\ensuremath{\rm C \kern -0.1em O}}}
\newcommand{\cotwo}{\hbox{\ensuremath{\rm C \kern -0.1em O_2}}}

\newcommand{\sif}{\hbox{\ion{[S}{1}]}}

\newcommand{\siliif}{\hbox{\ion{[Si}{2}]}}
\newcommand{\hi}{\hbox{\ion{H}{1}}}
\newcommand{\feiif}{\hbox{\ion{[Fe}{2}]}}

\newcommand{\neiif}{\hbox{\ion{[Ne}{2}]}}
\newcommand{\neiiif}{\hbox{\ion{[Ne}{3}]}}

\newcommand{\msun}{\ensuremath{{\kern 0.2em M_\odot}}}
\newcommand{\ngc}{\ensuremath{\rm NGC\,2264}}
\newcommand{\ic}{\ensuremath{\rm IC\,348}}
\newcommand{\ha}{\hbox{\ensuremath{{\rm H \kern 0.1em \alpha}}}}
\newcommand{\lya}{\hbox{\ensuremath{{\rm Ly \kern 0.1em \alpha}}}}
\newcommand{\spi}{\ensuremath{\it Spitzer}}
\newcommand{\bg}{\hbox{\ensuremath{{\rm Br \kern 0.1em \gamma}}}}
\newcommand{\ammpl}{\hbox{\ensuremath{\rm NH_4^+}}}

\shorttitle{Disentangling protostellar evolutionary stages}
\begin{document}


\title{Disentangling protostellar evolutionary stages in clustered environments using \textsl{Spitzer}-IRS spectra and comprehensive SED modeling}


\author{Jan Forbrich\altaffilmark{1}, Achim Tappe\altaffilmark{1}, Thomas Robitaille\altaffilmark{1}, August~A. Muench\altaffilmark{1}, Paula S. Teixeira\altaffilmark{2}, Elizabeth A. Lada\altaffilmark{3}, Andrea Stolte\altaffilmark{4}, \& Charles J. Lada\altaffilmark{1}}

\altaffiltext{1}{Harvard-Smithsonian Center for Astrophysics, 60 Garden Street, Cambridge, MA 02138, USA}
\altaffiltext{2}{European Southern Observatory, Karl-Schwarzschild-Stra{\ss}e 2, D-85748 Garching bei M\"unchen, Germany}
\altaffiltext{3}{Department of Astronomy, University of Florida, Gainesville, FL 32611, USA}
\altaffiltext{4}{I. Physikalisches Institut, Universit\"at zu K\"oln, Z\"ulpicher Str. 77, 50937 K\"oln, Germany}


%% \altaffiltext{1}{Visiting Astronomer, Cerro Tololo Inter-American Observatory.
%% CTIO is operated by AURA, Inc.\ under contract to the National Science
%% Foundation.}
%% \altaffiltext{2}{Society of Fellows, Harvard University.}
%% \altaffiltext{3}{present address: Center for Astrophysics,
%%     60 Garden Street, Cambridge, MA 02138}
%% \altaffiltext{4}{Visiting Programmer, Space Telescope Science Institute}
%% \altaffiltext{5}{Patron, Alonso's Bar and Grill}


\begin{abstract}
When studying the evolutionary stages of protostars that form in clusters,
the role of any intracluster medium cannot be neglected. High foreground
extinction can lead to situations where young stellar objects (YSOs) appear 
to be in earlier evolutionary stages than they actually are, particularly
when using simple criteria like spectral indices. To address this
issue, we have assembled detailed SED characterizations of a sample of 56
\textsl{Spitzer}-identified candidate
YSOs in the clusters NGC~2264 and IC~348. For these, we use spectra obtained
with the Infrared Spectrograph onboard the \textsl{Spitzer} Space Telescope and
ancillary multi-wavelength photometry.
The primary aim is twofold: 1) to discuss the role of spectral features,
particularly those due to ices and silicates, in determining a YSO's
evolutionary stage, and 2) to perform comprehensive modeling of spectral
energy distributions (SEDs) enhanced by the IRS data. The SEDs consist of 
ancillary optical-to-submillimeter multi-wavelength data as well as an 
accurate description of the 9.7~$\mu$m silicate feature and of the mid-infrared
continuum derived from line-free parts of the IRS spectra. We find that using
this approach, we can distinguish genuine protostars in the cluster from
T Tauri stars masquerading as protostars due to external foreground extinction. 
Our results underline the importance of photometric data in
the far-infrared/submillimeter wavelength range, at sufficiently high angular resolution 
to more accurately classify cluster members. Such observations are becoming possible now
with the advent of the \textsl{Herschel} Space Observatory.
\end{abstract}





\section{Introduction}

Finding and characterizing individual Young Stellar Objects (YSOs) in embedded clustered environments is challenging because the observed YSO spectral energy distributions (SEDs) can easily be influenced by dense material that is distributed throughout the cluster, but physically unrelated to an individual YSO or its envelope. In that situation, the extra extinction can mislead the observer since an affected source may appear to be in an earlier evolutionary stage than it actually is. For example, the SED of a T Tauri star behind foreground extinction can in some circumstances look protostellar. In a more isolated environment, this situation would only occur in case of edge-on disks, aggravated by the presence of disk flaring. To avoid confusion, \citet{rob06} proposed that sources be assigned designations that refer to the actual evolutionary stages, rather than spectral classes defined by observations, since the latter are affected by extinction. In most instances, the evolutionary stages (I, II, III) correspond to the SED classes. However, in heavily extincted regions the SED classes may not always correspond to the appropriate evolutionary stages. In embedded clustered environments, the difference between YSO classes and evolutionary stages becomes particularly important.

Infrared spectroscopy allows us to go beyond protostellar classifications based on photometry. At near-infrared wavelengths, varying degrees of veiling can be used to distinguish protostars from PMS stars \citep{cas92,gre96,whi07}. At mid-infrared wavelengths, silicate dust and ices cause telltale spectral features that can be used to refine spectral classes. Many of these features occur in the wavelength range that was accessible with the Infrared Spectrograph (IRS) onboard the \textsl{Spitzer} Space Telescope \citep{hou04}. In the context of star formation, the power of this instrument has been shown most notably by studies of T Tauri stars and protostars in Taurus, Chamaeleon, and Ophiuchus \citep{fur06,fur08,fur09}. Most recently, \citet{mcc10} study an extensive sample of YSOs in Ophiuchus, using IRS data to determine their evolutionary stages.

To find out whether the effect of intracluster or external foreground extinction can be disentangled from the actual evolutionary stage of a YSO, we have obtained IRS spectra and assembled detailed multi-wavelength photometry for a sample of 56 candidate YSOs in NGC~2264 and IC~348. We then use these data for comprehensive SED modeling with the primary aim of ascertaining the evolutionary stages for the sources in our sample. These can then finally be compared to previous classification attempts that are mainly based on less broad mid-infrared wavelength coverage. For each of the 56 sources, we have obtained mid-infrared spectra using \textsl{Spitzer}-IRS. These data not only allow us to study spectral features that trace the evolutionary stage of a YSO (silicates and ices), but they also allow us to derive photometry from line-free parts of the spectrum to accurately reflect the underlying continuum. This technique also allows us to accurately describe the main silicate spectral feature by photometric data at selected wavelengths. Combined with ancillary photometric data spanning optical to submillimeter wavelengths, the IRS data thereby allow us to construct SEDs that are tailored toward the requirements of detailed modeling. The accuracy of the modeling depends on the SEDs not containing spectral lines that are not part of the underlying model (e.g., PAH features); these features can fall into the band passes of broadband filters. Our IRS-derived photometry attempts to depict the continuum and the silicate features. Compared to the usual approach of using photometry from \textsl{Spitzer}-IRAC in SEDs for subsequent modeling, we discard bands 3 and 4 of the IRAC data that we have for every source in favor of corresponding line-free photometry derived from the IRS data.

\subsection{Source selection}

We selected 56 candidate protostars in NGC 2264 and IC\,348 -- 40 in the former and 16 in the latter region -- from previous \textsl{Spitzer} photometric studies of these regions by \citet{mue07} and \citet{tei08}. Our target sources are listed in Table~\ref{tab_srclist}. 

\begin{figure*}
\centering
 \includegraphics[width=0.8\linewidth]{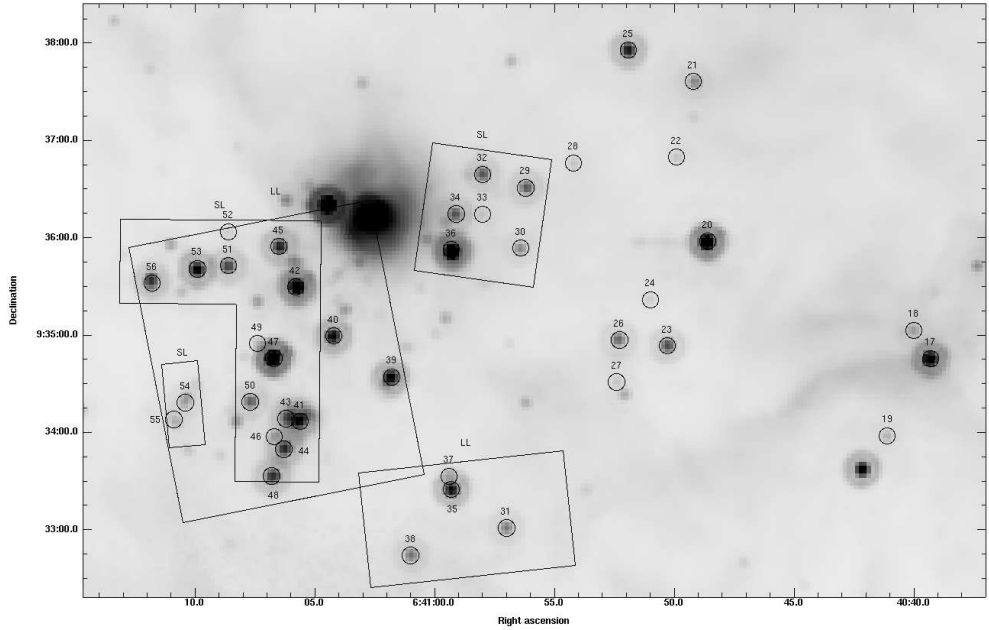}
 \caption{Objects selected in NGC\,2264 for IRS survey, overlaid on a MIPS 24\,$\mu$m image. The boxes mark the location of spectral maps in the indicated modules. The large spectral map covers the \textsl{Spokes} cluster. The brightest source is NGC\,2264~IRS\,2, and it is not included in our sample.}
 \label{mips_n2264}
\end{figure*}

\begin{figure*}
\centering
 \includegraphics[width=0.8\linewidth]{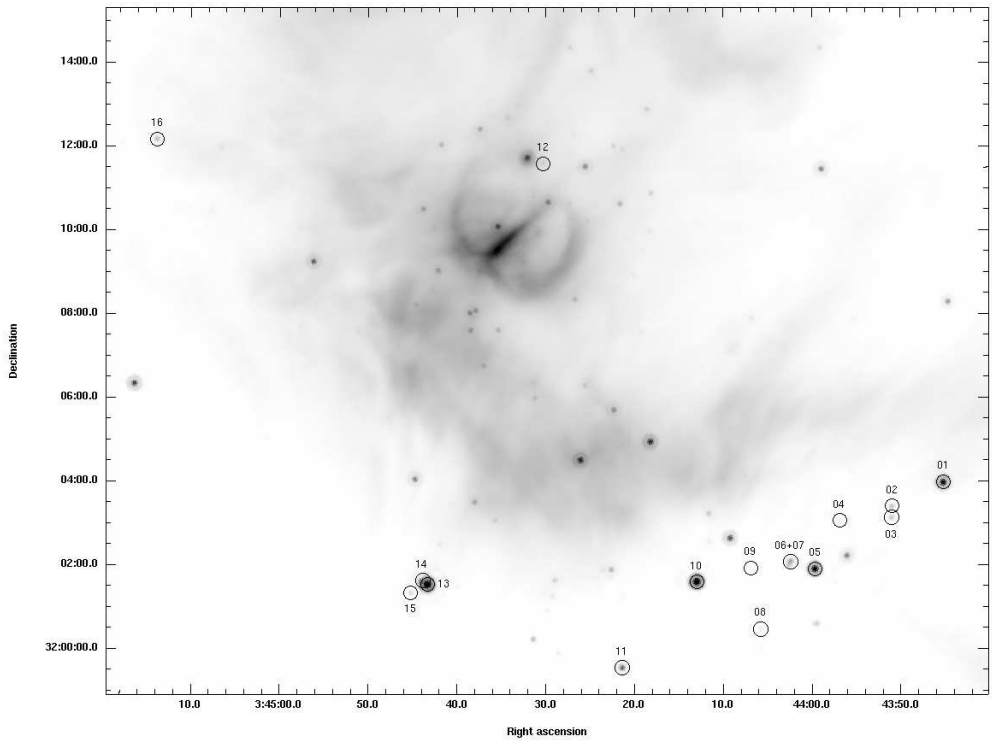}
 \caption{Objects selected in IC\,348 for IRS survey, overlaid on a MIPS 24\,$\mu$m image.}
 \label{mips_ic348}
\end{figure*}

NGC 2264 is a prominent, clustered star-forming region located at a distance of $\sim$913~pc \citep{bax09}; for a recent review, see \citet{dah08}. Here, we focus on the population of embedded protostars in the vicinity of IRS~2 \citep{you06} where we selected 40 sources for \textsl{Spitzer}-IRS mid-infrared spectroscopic observations. This region is one of the two most prominent regions of ongoing star formation in NGC~2264, next to IRS~1. All selected sources are located in the \textsl{Spokes} cluster \citep{tei06}. Figure~\ref{mips_n2264} shows the selected sources in a mid-infrared view of the region. IC\,348 is located in Perseus at a distance of $\sim$260~pc \citep[e.g.,][]{lom10}; for a recent review, see \citet{her08}. \citet{mue07} conducted a \textsl{Spitzer} census of this star-forming region. The IC\,348 sources discussed in this paper are among the most embedded in the cluster and are predominantly protostars in a narrow filamentary ridge located about 1~pc southwest of the central cluster. In Figure~\ref{mips_ic348}, we mark the selected sources in a mid-infrared view of the region.

A first comparison within our sample can be accomplished by using the IRAC spectral index $\alpha_{\rm IRAC}$, as determined from a least-squares fit to the photometry of all four IRAC bands \citep{lad06}; see section~\ref{sec_anc} for a discussion of the photometry. The sources in our NGC 2264 subsample show diverse IRAC spectral indices, ranging from $\alpha=-1.41$ (source 27) to $\alpha=4.21$ (source 45). Eighteen sources in this subsample have positive spectral indices. \citet{sun09} identify 37 out of the 40 sources as candidate cluster members in various stages, 20 of them are listed as class~I sources; their classification is included in Table~\ref{tab_srclist}. Note that while \citet{sun09} list no membership status for the three remaining sources, our results suggest that they are cluster members, too. 
In the IC\,348 sample, the IRAC spectral indices range from $\alpha=-0.93$ (source 14) to $\alpha=1.92$ (source 6). The classification from \citet{lad06} and \citet{mue07} is listed in Table~\ref{tab_srclist}. Most of the sources in this subsample are class I objects. Since we do not find significant differences between these two cluster subsamples -- both (unevenly) sample the same YSO classes --,  we will discuss them as one sample in the following, focusing more on a discussion of evolutionary stages in general. Note, however, that the source table allows the reader to identify individual YSOs in both clusters.

In this paper, we first discuss the \textsl{Spitzer}-IRS observations and their data reduction. We then describe the ancillary multi-wavelength data, ranging from optical to submillimeter wavelengths, and how we assembled composite SEDs from both IRS spectra and ancillary data. Subsequently, we perform comprehensive modeling of these SEDs before concluding with a discussion and summary of the results.

\section{Observations and Data Reduction}
\label{sec_obs}
\subsection{Spectroscopic observations with \textsl{Spitzer}-IRS}
\label{sec:observations}

\ic\ and \ngc\ were observed with the \spi\ Infrared Spectrograph (IRS) between 2006 September~16 and 2007 October~30 (\spi\ program 30033). We mapped the majority of YSOs using the IRS short-low (SL2/SL1, 5.2--8.7/7.4--14.5\mic, $R=64$--128), short-high (SH, 9.9--19.6\mic, $R=600$), and long-high (LH, 18.7--37.2\mic, $R=600$ ) or long-low (LL, 14.0--38.0\mic, $R=64$--128) modules.

For \ic, the sources are covered by individual small maps of typically 2 by 3 dithered pointings, i.e., three steps perpendicular and two steps along the slit, for both low and high resolution modules. We included separate high-resolution background observations for each source. For \ngc, which shows a much higher degree of source clustering, we designed extended SL/LL maps that cover a larger number of sources simultaneously, with individual high-resolution 2 by 3 dithered maps for the brightest sources. For the observations in NGC~2264, only a single high-resolution background observation was obtained; it is used for all NGC~2264 high-resolution maps.

All maps in this program are optimized for data reduction with CUBISM\footnote{\url{http://ssc.spitzer.caltech.edu/archanaly/contributed/cubism/}} v1.60 \citep{smi07}, i.e., they were obtained with half slit-width stepping perpendicular to the slit and additional dithering along the slit direction in most cases, resulting in quadruple data redundancy for each pixel in the map center. The individual steps of our data reduction are as follows:

\begin{enumerate}
  \item Acquisition of the BCDs (Basic Calibrated Data) as provided by the \spi\ IRS pipeline S17.2 to S18.2
  \item Importation of all sequential observations taken at the same date into CUBISM projects, separately for each module
  \item Selection and subtraction of an average background (dedicated for SH/LH; selected from outrigger slit positions of the same and/or other sources judged by variability of the background level in IRAC/MIPS (Infrared Array Camera, Multiband Imaging Photometer for Spitzer) maps
  \item Identification and excision of remaining bad pixels using CUBISM's statistical algorithms
  \item Construction of the data cube without using CUBISM's SLCF (Slit Loss Correction Function), i.e., keeping the original IRS pipeline calibration (see $\S$\ref{sec:cubismvsspice})
  \item Extraction of spectra for each source using a square aperture of typically 4 by 4 module pixels (nominal pixel sizes for the IRS SL/LL/SH/LH modules are 1.8, 5.1, 2.3, and 4.5\arcsec, respectively)
  \item Conversion of spectra from the default \mj\ unit to Jy by multiplying by the aperture area in steradian
  \item Importation of spectra into SMART \citep{hig04} for joining of the orders and scaling to match the MIPS 24\,$\mu$m photometry (used as reference, see below)
\end{enumerate}

\subsubsection{CUBISM}
\label{sec:cubismvsspice}
The \spi\ IRS pipeline is optimized for isolated stellar point sources. It is correcting for light losses due to the limited size of the slit and the extraction aperture compared to the point-spread-function. The flux calibration is accomplished by observing standard stars and matching their spectra to stellar models. Hence, the pipeline flux calibration is invalid if: a) the observed source is not a stellar point source, b) the slit is not centered on the observed point source, or c), the default pipeline extraction apertures are changed. It is possible to correct for items b) and c) by deriving a new calibration using the pipeline approach.

However, if the observed source is not a stellar or `true' point source, e.g., a YSO with an accretion disk and/or an extended envelope, we cannot expect the flux calibration to be correct. To make matters worse, the effects change with wavelength and also among the different IRS modules because of their different sizes and the resulting uneven coverage on an asymmetric, extended source. Since all YSOs have an individual and a priori unknown extended structure and, in addition, slit coverage will change as a function of distance to the source, no reliable flux calibration can be derived from standards. Our approach to this dilemma is to use CUBISM (see $\S$\ref{sec:observations}), which is optimized for handling \spi\ IRS mapping data of homogenous extended sources, and scale the resulting spectra to the measured MIPS-24\,$\mu$m photometry of each source.
As a consistency check of the photometric calibration, we have extracted\footnote{using IDL code provided by the \textsl{Spitzer} Science Center at \url{http://ssc.spitzer.caltech.edu/dataanalysistools/cookbook/10/}} color-correct magnitudes for IRAC bands 3 and 4 from the IRS spectra in order to compare the results with the IRAC aperture photometry. For spectra that are not too noisy, the IRAC photometry yields fluxes that are slightly higher than those from the synthetic IRS photometry, with mean offsets of 12\% in band~3 and 18\% in band~4, although offsets of up to a factor of two are occurring. While the two sets of photometry are equivalent in their spectral coverage, this comparison does not account for slit losses. Additionally, any source variability would cause further differences since the IRAC and IRS data were not obtained simultaneously. Given these limitations, the two sets of photometry thus appear to be consistent.

We compared the extracted spectra of CUBISM and SPICE for a representative, mapped YSO in our sample. This example, source 56, is shown in Figure~\ref{fig:cubismspice}.  We selected this source since it is fairly bright and shows complex spectral structure due to ices, silicate grains, \htwo\ emission lines, and a rising continuum due to the presence of an envelope. SPICE is the IRS data reduction tool supported by the Spitzer Science Center, and it is optimized for single pointing (staring) observations of stellar point sources. The SPICE spectra were extracted by using only the map slits centered on the YSO, whereas the CUBISM spectra were extracted using the approach described in $\S$\ref{sec:observations}. Reassuringly, both methods yield very similar spectral features, continuum shapes, and signal-to noise ratios. In both cases additional scaling is needed to join the spectral orders and match the MIPS photometry. The main advantages of CUBISM for reducing IRS mapping data of a large sample of diverse
YSO's are:

\begin{itemize}
  \item The data reduction is greatly facilitated due to integrated BCD bookkeeping, accurate visualization and accounting of all slit pointings, background subtraction, and statistical bad pixel handling.
  \item The spatial maps cover all or a large fraction of an extended YSO and therefore minimize the effects of uneven slit coverage or varying slit orientation.
  \item There is no restriction on the size and shape of the extraction aperture for the source and background spectra.
  \item The spatial maps provide increased data redundancy compared to single slit staring observations.
\end{itemize}

The resulting spectra, concatenated from the different modules, are shown in Figures~\ref{fig_irsplots1} to \ref{fig_irsplots4}, where they are sorted by evolutionary stage, as informed by subsequent analysis\footnote{Note that sources 14 and 49 are not shown because nearby bright sources blend with their data at longer wavelengths. The SEDs of these sources therefore are not modeled at all.}. Overall, they represent a wide variety both in evolutionary stage as well an in the spectral features present.

\subsubsection{Continuum and ice features}
\label{sec:ice}
\begin{figure}
\centering
\includegraphics[width=\linewidth]{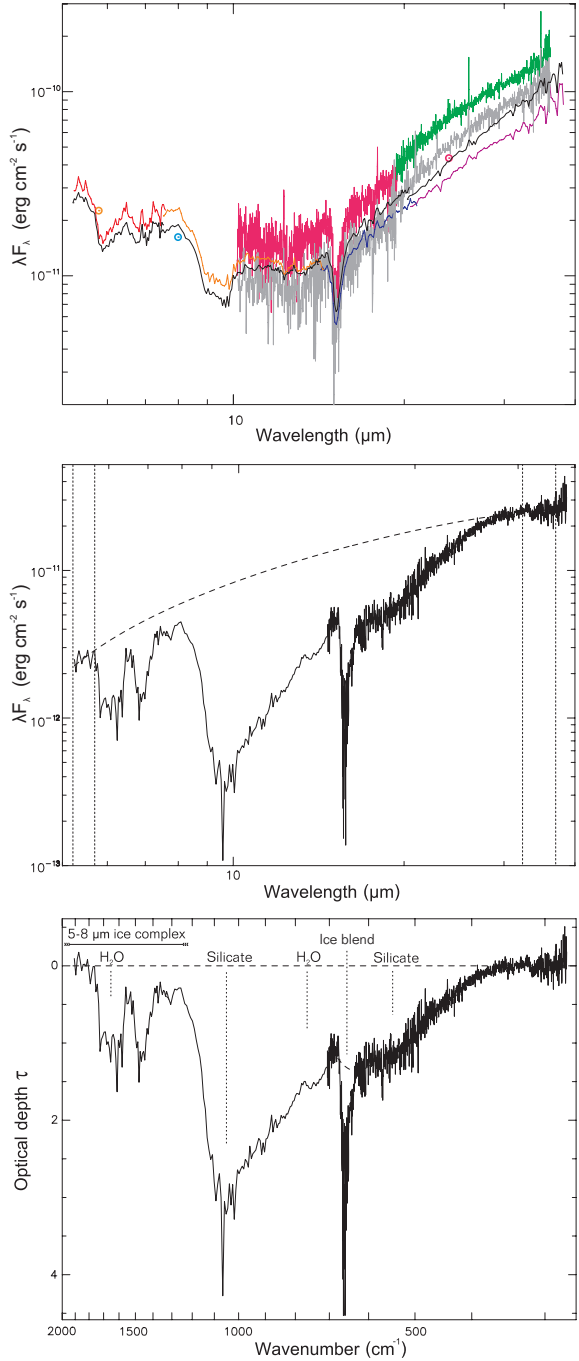}
  \caption{Top: \spi\ IRS low and high resolution spectrum of source 56: CUBISM (color)
 compared to SPICE (grey/black) and IRAC/MIPS photometry (colored circles). No scalings have been applied.\label{fig:cubismspice} Center: Continuum fit for \spi\ spectrum of source 26. The IRS low- and high-resolution modules have been scaled and merged. Vertical dashed lines mark the regions used for the continuum fit.\label{fig:contfit} Bottom: Optical depth spectrum for source~26.\label{fig:odepth}}
\end{figure}

Many of the YSOs in our sample show prominent ice absorption features at 5--8 and 15\mic\ as well as silicate absorption at 9.7 and 18\mic\ (see, e.g.,~Figure~\ref{fig:cubismspice}). In order to quantify and compare the ice and silicate features between different sources, we measured the center optical depths $\tau_\lambda$ of the 6.0\mic\ and 9.7\mic\ \water\ ice and silicate absorption features, respectively, and the integrated optical depth $\int\tau\,d \tilde{\nu}$ of the 15\mic\ ice feature which is dominated by CO$_2$ \citep{ger99} and has been described as a \cotwo:\co:\water\ blend \citep{pon08}. The latter can be converted to a column density, $N_{\rm ice}= A_{\rm ice}^{-1}\int\tau\,d \tilde{\nu}$, where $A_{\rm ice}$ is the ice band strength of $1.1\times10^{-17}\,{\rm cm\,molecule^{-1}}$ assuming pure CO$_2$ ice \citep{ger95,pon08}. In order to convert the spectra to an optical depth scale, $\tau_\lambda=\ln(I_{\rm cont,\lambda}I_{\lambda}^{-1})$, we determined the continuum by fitting a low-order polynomial on a log-log scale to the spectral regions below wavelengths of about 5.8\mic\ and above 30\mic\ (see Figure~\ref{fig:contfit}). These regions are least affected by absorption features (see also \citealp{boo08}). In addition, we determined a local continuum for the 15\mic\ ice feature by fitting a low-order polynomial to the adjacent spectral continuum of that feature. A detailed analysis of the individual ice feature components is beyond the scope of this paper. We note that the distribution of spectral feature strengths is similar to previously published surveys of other regions, even without taking into account the different sets of criteria that were used to define the various samples.

\subsection{Ancillary optical, infrared, and submillimeter photometry}
\label{sec_anc}

Our IRS spectra complement a wealth of ancillary photometric data that we use for subsequent comprehensive SED modeling. For IC\,348, we use data compiled by \citet{mue07}. The only exceptions are MIPS mid-infrared and SCUBA submillimeter data where we have converted several flux densities for individual sources into upper limits if there is any indication that several sources contribute to the flux density. 

Our main source of optical \textsl{UBVRI} photometry for NGC 2264 is the ancillary data catalog compiled by \citet{reb06}. The photometry for the sources presented here is from various sources cited therein \citep{sun97,fla99,par00,reb02,lam04}. To resolve inconsistencies (for a given source, the \textsl{UBVRI} photometry typically is from at least two different sources), the USNO-B1 catalog has been used additionally.

In the near-infrared, we use FLAMINGOS \citep{els98} imaging data in \textsl{JHK} to surpass the 2MASS catalog in sensitivity. The observations were taken on 01 January 2005 when the instrument was mounted to the 2.1m telescope on Kitt Peak. In this data, all but seven of our sources in NGC 2264 are detected at least in \textsl{K} band. For the undetected sources, the estimated sensitivity limits of $J=19$~mag, $H=18$~mag, and $K=17.5$~mag were assumed as upper limits. Aperture photometry was performed on these images with an aperture of five pixels or 1.5$''$ with a surrounding annulus of five pixels width to subtract the background. To avoid saturation problems, we have used 2MASS photometry for three bright sources (sources 17, 36, and 47).

In the mid-infrared range covered by \textsl{Spitzer}, we have obtained aperture photometry from IRAC and PRF-fitting photometry from MIPS imaging data. The IRAC data were obtained in March 2006 (program ID 37) when both 0.6\,sec and 12\,sec exposures were taken. Aperture photometry was performed with an aperture radius of five pixels and a background annulus with inner and outer radii of 5 and 10 pixels. For all but two sources, we used the 12\,sec exposures. Sources 20 and 47 required correcting for saturation in the first two IRAC channels. We used a comparison of IRAC and longer-wavelength data to find obvious blends of multiple objects and exclude the corresponding SEDs from modeling; this concerns sources 6+7 (a blend), 41, and 43. From the IRAC photometry, a spectral index $\alpha_{\rm IRAC}$ was determined from a least-squares fit to all four IRAC channels in $\lambda F_\lambda$. To ensure comparability within our full sample, the same method was used to also determine spectral indices for the sources IC\,348, using the photometric data from \citet{mue07}.

In the submillimeter range, we use aperture photometry to estimate upper limits in SCUBA data (450~$\mu$m and 850~$\mu$m) obtained by \citet{wol03} for NGC 2264 in 1999 and 2000. For NGC\,2264 and most of IC\,348, we decided to only use upper limits due the crowding of the cluster relative to the angular resolution. In determining the photometry, we read out the pixel value at the source position (calibrated in Jy/beam) and subsequently added five times the rms of the surrounding 25 pixels.
For the use in SED modeling, in order to use strict upper limits, we have doubled the upper limits that were determined in the mid-infrared (MIPS) and submillimeter bands.

\subsection{Constructing photometric SEDs}

As a preparation for the SED model fits, we have constructed photometric SEDs for all observed sources using the ancillary photometric data as well as the IRS observations. The SEDs cover the optical (\textsl{UBVRI}), the near infrared (\textsl{JHK}), the mid-infrared (3\,$\mu$m up to 35\,$\mu$m from IRAC and IRS) and the submillimeter ranges (450\,$\mu$m and 850\,$\mu$m). We have extracted photometry from the IRS data to complement the broadband data from the literature, using bands that characterize the continuum as well as the 9.7~$\mu$m and 18~$\mu$m silicate features -- these are taken into account in the SED models that we use -- while being least affected by molecular or atomic lines and ice features that are not taken into account in the model. The wavelengths for which photometry was derived using the IRS data are 5.580, 7.650, 9.95, 12.93, 17.72, 24.28, 29.95, and 35.06 $\mu$m, where the first two bands replace IRAC bands 3 and 4 since they are less affected by spectral features than these two IRAC bands. The widths of the IRS spectral windows that are converted into photometry are set to 2\% of the center wavelengths. In this process, a few photometry points were flagged and not taken into account in the modeling if the IRS data were too noisy. Overall, fluxes from 21 separate wavelength bands were used to construct the SEDs for individual sources.

\section{Analysis}
\subsection{IRS spectra}
\subsubsection{Dust and ice features}
\label{sec:iceresults}

In the following, we will focus on the silicate and ice features and their correlation with the sources' infrared spectral indices while also noting the detection of PAH emission in only three sources in NGC\,2264, perhaps indicating the presence of relatively massive stars. The spectral analysis for the sources in our sample is summarized in Table~\ref{tab_icefit}, where the sources are already sorted by evolutionary stage, as informed by subsequent SED modeling. The plots of the spectra in Figures~\ref{fig_irsplots1} to \ref{fig_irsplots4} are sorted in the same fashion. The detailed SED modeling that is required to sort the sources by evolutionary stage will be described subsequently.

For the purpose of comparing spectral features to an infrared spectral index as a more easily observable source parameter, we use the IRAC spectral index simply because it is available for more sources in our sample than another frequently used index, spanning from the near-infrared $K$ band to 24~$\mu$m. Several, presumably deeply embedded sources in our sample remain undetected in the near-infrared wavelength range. We note, however, that the trends discussed here are also apparent when using the \textsl{K}--24\,$\mu$m spectral index (not shown). We remind the reader that the IRAC spectral index was calculated from all four IRAC bands; to ensure the comparability with other datasets, it does not use any IRS data in replacement for IRAC channels. The scatter of the four IRAC channels is reflected in the uncertainty of the spectral index, also derived from the fit.

We now discuss the relationship of the 9.7\,$\mu$m silicate feature and the total ice column density with the IRAC spectral index. We also directly compare silicate emission and ice absorption and compare this relationship to the results of the c2d survey. In Fig.~\ref{fig_speccorr}, we show the corresponding plots in six panels that are described in more detail in the following. As shown in Fig.~\ref{fig_speccorr}a/b, a source with a positive spectral index tends to have stronger silicate absorption and ice features than a source with a negative spectral index. Note that the silicate feature can appear in emission (indicative of disks) or in absorption, while the ice features always are absorption features. 
Our discussion of silicate features focuses on the prominent 9.7\,$\mu$m silicate feature, but we note that the optical depth of the 18\,$\mu$m silicate features is correlated to the one at 9.7\,$\mu$m, even if any more detailed comparison is limited by the uncertainties. The measured full-width half-maximum (FWHM) line widths of the  9.7\,$\mu$m silicate feature range from 1.8\,$\mu$m to 3.6\,$\mu$m. Fig.~\ref{fig_speccorr}c shows that the broader line widths correspond to negative silicate optical depths, i.e., silicate emission, while the line width does not change much for the silicate features in absorption. There is a tentative anti-correlation of the width of the silicate feature with the IRAC spectral index; the broadest lines are found in emission and in the sources with the lowest spectral indices (see Fig.~\ref{fig_speccorr}d), possibly indicative of grain growth in the later evolutionary stages. However, \citet{sar09} and \citet{wat09} point out that the effects of grain growth on the silicate feature are difficult to observe unambiguously even for class~II sources where one might expect the effect to be more discernible than in protostars. 
Finally, there also appears to be a correlation between the optical depth of the 9.7\,$\mu$m silicate absorption features and the total ice column density (shown in Fig.~\ref{fig_speccorr}e), similar to the results of \citet[][Fig.~29]{fur08}. As pointed out by \citet{fur08}, the silicate optical depth and the ice column density are expected to be correlated since a higher silicate optical depth should also provide shielding for ice mantles to grow.  While generally, sources with larger optical depths of the silicate feature tend to have stronger ice features, the class~I source 53 is the only source not following this trend; it has a much lower ice column density than what could be expected from this correlation, indicating that the central star might be able to evaporate the ice mantles. 

None of the sources with silicate in emission show ice features. This probably indicates that the envelopes of these sources have dissipated and ices are no longer observable. In Fig.~\ref{fig_speccorr}f, we compare the silicate optical depth and the total ice column density from this study with results from the \textsl{Spitzer} c2d survey \citep{pon08,boo08}. Again there is a correlation of large silicate optical depths and high ice column densities even though there are two exceptions similar to source 53. These two sources (W33A and W3\,IRS\,5) have high silicate optical depths, but comparably low total ice column densities. W33A is an H\,{\sc II} region and W3\,IRS\,5 is a massive-star--forming region. Possibly, both sources as well as our source 53 produce strong enough radiation to evaporate the ice mantles. 

A quantitative comparison of the spectral features with the literature is difficult since particularly the methods used to determine the continuum vary considerably. However, we can directly compare our results with the c2d
survey \citep{pon08,boo08}. The range and distribution of the optical depths of the silicate features ($\tau_{6\mu{\rm m}}$ and $\tau_{9.7\mu{\rm m}}$) and of the total ice optical depth are very similar.

\subsubsection{PAH and Gaseous Spectral Lines}

As noted above, a detailed discussion of all spectral features that were observed in our sample is beyond the scope of this paper. However, we would like to point out the most important IRS results beyond silicate and ice features.
We detected strong PAH emission in sources 25, 36, and 42, all in NGC~2264. The emission is clearly associated with the YSOs, since the surrounding background is devoid of extended PAH emission in all three cases. However, the \textsl{Spitzer} resolution does not permit to resolve the origin of the PAH emission within the YSO envelope/disk system. We can exclude extended PAH emission beyond the \textsl{Spitzer} resolution limit for sources 25 and 42. In these cases, the PAH emission is hence located within a radius of about 1000 AU from the YSO sources. In the case of source 36, the high luminosity as well as blending of the PAH 6.2~$\mu$m feature with ice features make the spatial PAH analysis uncertain. These three sources might represent the upper end of the mass function in the \textsl{Spokes} cluster. We will later see that among these sources only the SED of source 42 can be modeled reliably. The fit indeed suggests that source 42 is in the upper luminosity range among the objects whose SEDs we have studied in more detail. Note that even though NGC\,2264~IRS\,2 is not included in our spectral map, its immediate vicinity also shows strong PAH emission.

Within the \textsl{Spokes} cluster, \citet{you06} and \citet{tei07} discovered a dense cluster of protostars called the \textsl{microcluster}. This region is included in our sample as two separate objects, 41 and 43, each being a blend of several sources that can be resolved in the near-infrared wavelength range. For these regions, we can take advantage of the spectral mapping capabilities of \textsl{Spitzer} to characterize the region. Even though the
\textsl{microcluster} remains spatially unresolved at the longer wavelengths, we can still study a variety of spectral features in the region; this is shown in several panels in Fig.~\ref{fig_microclmaps}. The interpretation of the spatial features is complicated by the highly variable extinction across the region, most strongly affecting the near-IR wavelengths. In addition, the absorption maps are constrained by the presence of a background continuum source at the respective wavelength, i.e., the absence of absorption does not necessarily indicate the absence of the absorbing species.

Fig.~\ref{fig_microclmaps} (panel b) shows spatially extended \htwo\ emission, which together with the detection of \neiif\ indicates outflow activity in the vicinity of sources 41, 43, and 44. The \htwo\ detection is consistent with the narrowband \htwo\ (1--0) 2.12\mic\ image presented by \citet{you06}. The 30\mic\ continuum emission and the 15.2\mic\ ice absorption (Fig.~\ref{fig_microclmaps}, panel c) and f), respectively) are consistent with the positions of the protostars SMA--2, 3, 5, and 6 reported by \citet{tei07}, when taking into account the larger \textsl{Spitzer} beam. Note that SMA--5 and 6, the class 0 protostars with the highest observed 230\,GHz flux in the microcluster, have no near-IR counterpart due to the high extinction, but are detected in our longer wavelength \textsl{Spitzer} maps.

The aim of a quantitative analysis of spectral lines is usually to determine the local physical conditions, primarily density and temperature of the gas, and to infer chemical abundances. Infrared emission lines are a useful gas diagnostic for optically obscured regions such as embedded YSOs. However, a detailed, quantitative analysis of infrared emission lines in YSOs is hampered for the following reasons:

\begin{itemize}
  \item Typically, only a limited set of infrared lines is observed for a given source (probing certain ionization stages and energy levels).

  \item Inadequate angular resolution results in limited knowledge of the origin, physical geometry, and location of the emissive region (outflow shocks, an accretion disk, an envelope, etc., or a combination of those).

  \item Collisionally excited lines from an inhomogeneous medium will be biased toward regions of high temperature, where collisionally excited emission lines will be strongest (e.g., shocked gas).

  \item The UV radiation field (outflow shocks, disk and outflow cavity irradiation) and infrared continuum are hard to quantify, particularly if the location of the emitting material is unknown.

  \item The line-of-sight extinction is difficult to quantify accurately.
\end{itemize}

In light of these difficulties, a meaningful discussion of the relative line intensities of the sources in our sample is beyond the scope of this paper and would ideally require spatially resolved observations. However, a qualitative assessment, in particular of the presence or absence of infrared emission lines, can still give useful insights into the physical characteristics of a given YSO. Table~\ref{tab;specdata} shows an overview of the infrared emission lines observed in our sample and lists important atomic and molecular line parameters.

The last column in Table~\ref{tab;specdata} gives the critical density, $n_{{\rm crit,}i}= {{\sum\nolimits_{{{j < i}}} {A_{{{ij}}} }} / {\sum\nolimits_{{{j}} \ne {{i}}} {\gamma _{{{ij}}} } }}  $, where $i$ is the index for the initial level with energy $E_i$, $j$ is the final level, $\gamma$ is the collisional rate coefficient out of level $i$, and $A$ is the Einstein coefficient for spontaneous emission (\citealp{ost89}, eq. 3.31). If the gas density is much larger than the critical density of a given energy level $i$, then collisional de-excitation dominates over spontaneous emission, and the infrared lines originating from this level get collisionally suppressed. If the density is much smaller than the critical density, the system exhibits a non-LTE (local thermodynamic equilibrium) behavior, where line emission is the dominant de-excitation mechanism.

The initial energy level of a given infrared line is usually populated by collisions with the most abundant gas species (electrons, H, or \htwo, depending on the conditions) and is depopulated either via spontaneous radiative emission giving rise to the observed spectral line or via collisional de-excitation. The general conditions necessary for an infrared emission line to be strong are: a) a large abundance of the species, b) a large population of the initial energy level (a kinetic gas temperature similar or larger than $E_{ i}/k_B$), and c) a large Einstein $A$ coefficient for spontaneous emission. Note that these are qualitative conditions, i.e., substantial populations and abundances can compensate for small Einstein $A$-values (e.g., \htwo\ pure rotational lines).

The most common infrared lines observed in our sample of 56 sources are pure rotational transitions of \htwo\ (17 sources, of which 10 sources show S(4) and higher) and atomic fine-structure transitions of Ne$^+$ (9 sources) and Fe$^+$ (7 sources). Among these sources, there is no clear correlation between the detection of any one of these features with evolutionary stage, partly certainly due to the low number of sources in each category. For example, out of the 17 sources with \htwo\ transitions, four are in stage I, six are in stage II(ex), and three are in stage II (while the stages of four sources were not conclusively determined). 

The excited \htwo\ lines trace the presence of hot molecular gas, typically from slower, non-dissociating outflow shocks. Good examples are sources 41 and 43, which are part of the microcluster, an association of young sources with clear signs of outflow activity (see above and Fig.~\ref{fig_microclmaps}).
The \neiif\ 12.8\mic\ line is a good example of a fine-structure line with a relatively large $A$-value, a low-lying energy level, and a large critical density. In addition, neon is not depleted in the gas phase since it is not incorporated into dust grains. Therefore, Ne$^+$ is abundant if sufficiently energetic UV photons are present to ionize neutral neon ($\lambda \leq 57 \rm \,nm$). This may be the case in the irradiated surface layers of accretion disks or in fast outflow shocks that are strong enough to ionize atomic hydrogen.

Iron is strongly depleted in the gas phase and therefore it is typically observed in fast, dissociative shocks, which destroy dust grains and liberate the iron into the gas phase (e.g., SNR shocks and fast protostellar outflows). Fe$^+$ has a complex energy level structure and features both low and high-lying energy levels, which makes it a very useful density and temperature diagnostic.

Combining all the information above, we attempt a simplified interpretation scheme:
\begin{description}
  \item[\htwo\ 0--0 S(4) and higher] traces a hot molecular environment (elevated gas temperature of a few 1000\,K or larger, low UV radiation, non- dissociating shock).
  \item[\neiif\ 12.8\mic] traces a high UV radiation environment leading to high ionization and \htwo\ photodissociation.
  \item[\feiif\ lines] trace faster, dissociating shocks that partially destroy dust grains and produce UV radiation/ionization.
\end{description}

Of the eleven sources in our sample that show \neiif\ and/or \feiif\ lines, five show lines of both species, four show only \neiif\, and two show only \feiif. The \neiif\ 12.8\mic\ emission in sources with no observed \feiif\ lines is unlikely to be due to strong outflow shocks. These are probably cases of irradiated disks with little or no outflow shock activity, consistent with the stage II/II(ex) classification of three out of these sources (while we did not determine a stage for the fourth source). The two sources with \feiif\ lines but no \neiif\ 12.8\mic\ both seem to be young stage~I sources, one of them being part of the \textsl{microcluster}, with strong outflow shock activity.

Sources that show both \neiif\ and \feiif\ lines are probably due to irradiated disks and/or strong outflow shocks. For example, source 41, which is part of the microcluster and thus has no assigned stage due to source crowding, shows strong \htwo, \feiif, and \neiif. It is unlikely that \htwo\ and Ne$^+$ coexist, and indeed our spectral mapping indicates a spatial offset between the \neiif\ 12.8\mic\ and the extended \htwo\ emission, consistent with an irradiated disk and an outflow (see Fig.~\ref{fig_microclmaps}, panel b). Note as a caveat that the near-infrared imaging in the figure shows the complex, blended emission in the microcluster and the clear limitations of the spatial resolution of our mid-infrared observations. An advantage is, however, that the longer wavelength observations penetrate the highly obscured region and show the deeply embedded protostellar sources.

\subsection{SED Modeling}

\subsubsection{Procedure}

The SEDs of all the sources were fit using the large set of model SEDs from \citet[][hereafter R06]{rob06} and the SED fitting tool from \citet[][hereafter R07]{rob07}. These models were computed using the radiation transfer code from \citet{whi03a,whi03b}. The basic model consists of a pre-main-sequence star surrounded by a flared accretion disk and a rotationally flattened envelope with cavities carved out by a bipolar outflow. Both scattering and reprocessing of the stellar radiation by dust, as well as the disk accretion luminosity, are included. The set of models covers a large range of stellar masses (from 0.1 to 50\,M$_{\odot}$) and evolutionary stages (from deeply embedded protostars to `anemic' and `transitional' disks). In total, 14 parameters describe the stellar properties and the circumstellar dust geometry, such as the stellar luminosity, temperature, disk mass, radius, inner radius, envelope infall rate, or bipolar cavity opening angle. For each set of parameters, SEDs were computed at 10 different viewing angles. In total, 20,000 sets of parameters, and therefore 200,000 model SEDs were computed.

The R07 SED fitting tool fits each of the models in the R06 model set to the data, allowing both the distance and external foreground A$_{\rm V}$ to be free parameters (within specified ranges) and characterizes each fit by a $\chi^2$ value. By choosing a $\chi^2$ threshold for `good' fits, one can then look at how well constrained different parameters are.

For the fitting, we ignored models with viewing angles $i>80^\circ$ (i.e., edge-on). Model fits with edge-on models tend to require unrealistically high stellar luminosities and seriously affect the resulting parameter ranges for many of the sources. However, if the sources in our sample are randomly oriented, we would expect $56/9\approx6$ sources to have $i>80^\circ$. However, this number is an upper limit because edge-on sources are fainter, and therefore a flux-limited sample such as ours will be biased towards non--edge-on sources. Therefore, while we may be incorrectly fitting a couple of edge-on sources with non--edge-on models, we prevent the majority of the sources from being fit by edge-on models, which seriously skew the resulting parameter ranges.

In this paper, we define the goodness-of-fit criterion to be $\chi^2 - \chi^2_{\rm best} < 3 \times n_{\rm data}$. This criterion is identical to that chosen in R07, and is arbitrary, but models that fit better than this criterion generally fit the data well. While this criterion is fairly `loose' in statistical terms, as described in R07, the coarse coverage of 14-dimensional parameter space, intrinsic uncertainties in the models, and variability all mean that adopting a stricter criterion would lead to over-interpretation. For the purposes of the modeling, we allow the external A$_{\rm V}$ to vary between 0 and 40\,mag and the distance to vary within 5\% ranges of 913\,pc for NGC\,2264 and 260\,pc for IC\,348.

The interstellar extinction from the sun to the outer `edge' of the star forming regions is not likely to exceed a magnitude of visual extinction. As will be shown further on in this paper, the external foreground (intracluster) extinction inside both star forming regions is what dominates the overall extinction toward most of the sources. Therefore, rather than assuming the same dust model as in R07, we choose to use a dust model with larger grains (M. Wolff, private communication, 2009). The dust model contains a mixture of astronomical silicates and graphite in solar abundance, using the optical constants of \citet{lao93}, and a size distribution derived from a maximum entropy method \citep{kim94,cla03} to reproduce an R$_{\rm V}$ value of 5.5. Such high R$_{\rm V}$ values are more suited to dense star forming regions than the typical diffuse ISM value of R$_{\rm V}$=3-3.5 \citep[e.g.,][]{car89}. In order to assess the influence of the extinction law that we assume for the external foreground extinction on the source classification, we have also used the two observational mid-infrared extinction laws derived by \citet{mcc09}. As noted in Table~\ref{tab:sedmodeling}, we find that only six to seven out of 44 sources would be classified differently. There seems to be no systematic effect on the number counts of the three different evolutionary stages since the changes largely cancel out. Conclusions concerning source populations are thus not significantly affected. The choice of extinction law does, however, alter the parameters derived for individual YSOs from the SED fits, but these are not considered in detail in this paper.

Figures \ref{fig:g1seds} to \ref{fig:g3seds} show the SEDs of most of the sources along with the model fits. These sources are very well fit by the models, and the derived parameters are reliable (within the framework of the assumed model geometry). On the other hand, for the sources shown in Figure~\ref{fig:g4seds}, the derived parameters cannot be trusted. In the case of sources 8 and 32, strong ice features in the 10 to 20\microns range and the lack of longer wavelength emission means that the model fits cannot be trusted. Sources 12, 16, 25, and 36 are not well fit by the models. In particular, source 25 is likely to be an edge-on source. This source, which has strong PAH features, has a very similar overall SED and mid-infrared spectrum as Gomez' Hamburger which is a known edge-on disk \citep[e.g., ][]{buj08,woo08}. Finally, in the case of sources 6+7 (a blend), 41, and 43, the broadband fluxes at most wavelengths contain contributions from multiple sources. Therefore, while the SEDs appear well fit, the resulting parameters are not meaningful.

Table \ref{tab:sedmodeling} lists the lower and upper values of each parameter of the good fits for each source. The best-fit value is not meaningful, and is therefore not shown. Only sources with reliable parameters (as discussed above) are listed. The modeling results are discussed in the next section. We note that the fit results are generally compatible with additional constraints from the literature (spectral types, X-ray--derived column densities); deviations are not significant.

\subsubsection{Results}

One particularity about the R06 set of models is that parameter space was not sampled in a uniform way due to computational limitations. Instead, the parameter space coverage was designed to include reasonable regions of parameter space suggested by observations and theory. Therefore, there exist intrinsic correlations between the parameters in the set of models that need to be taken into account when using the models to look for trends \citep{rob08}. For example, if in this sample we were to find an inverse correlation between stellar temperature and envelope mass - which would make sense as one expects the stellar temperature to rise and the envelope mass to decrease as the source evolves - such a trend was \textit{assumed} in the R06 set of models and therefore the result may just reflect this assumption rather than a real trend. We find no correlation between model parameters that is not likely to be due to the assumptions built into the R06 models and therefore in the remainder of this section we concentrate on the determination of the evolutionary stage of the sources.

The sources in the present sample were originally targeted as preferentially being candidate embedded protostars, based in part on the spectral indices derived from \textsl{Spitzer} IRAC and MIPS photometry. As shown in R06, while such a spectral index provides a way of statistically estimating the evolutionary stage of sources in a sample, it is not necessarily reliable on a source-by-source basis especially in regions that can be expected to contain large amounts of intersource extinction such as massive, cluster-forming cloud cores. Therefore, some of the sources in this embedded cluster sample may not actually be embedded protostars. In this section, we use the results of the SED modeling along with information extracted from the IRS spectra and the spatial distribution of the sources to examine in detail the physical nature of each of the sources.

Based on the results of the broadband SED modeling, we separate the sources into sources that require significant envelopes, and are therefore are likely to be genuine protostars, and ones that can be fit by disk-only models, and therefore may be more evolved. We refer to the former as stage I sources and the latter as stage II sources as in R06, but define the separation between the two groups of sources differently to R06. We define the genuine embedded sources (stage I) as those which cannot be fit by models which have envelope masses less than $10^{-3}$\,M$_\odot$, and the potentially more evolved sources (stage II) as those that can be fit with models that have envelope masses below $10^{-3}$M$_\odot$. Stage I sources are expected to display class 0 or class I SEDs. Typically, we would expect stage II sources to display class II SEDs, but in the presence of significant (unrelated) foreground extinction they can display class~I-like SEDs. The same effect can be caused by flared disks that are seen edge-on, but this should only affect a minor fraction of a given sample.

As can be seen in Table \ref{tab:sedmodeling}, some sources require very little external foreground extinction while others require at least 10 or 20 magnitudes of such extinction in addition to the self-extinction by the circumstellar dust.  We therefore separate the stage II sources into ones that require an external extinction of at least 7\,mag (stage II(ex)), and sources that do not require an external A$_{\rm V}$ of more than 7\,mag (stage II). Stage II(ex) sources typically display flat-spectrum/class~I-like SEDs while the stage~II sources typically display more traditional class~II SEDs. The exact values to separate the sources into the three categories sources is fairly arbitrary, but there are only a few borderline cases. Most sources clearly fall into one of these categories. We note that some stage II(ex) sources have a minimum interstellar extinction of less than 7\,mag: these are cases where only the stage I models require external extinctions below 7\,mag, and all stage II models require higher external extinctions.

Figure \ref{fig:scuba_map} shows the spatial distribution of the sources overlaid on a SCUBA 850\microns image for each cluster. One striking result in NGC 2264 is that the stage II sources clearly do not follow the same spatial distribution as the other sources, and are preferentially located to the West. The stage I sources are preferentially associated with brighter SCUBA 850\microns emission, while the stage II sources are preferentially associated with undetected or faint SCUBA 850\microns emission, with the stage II(ex) sources associated with intermediate brightness sub-mm emission. 
The mean 850\microns fluxes associated with the sources and the corresponding standard errors of these means are 0.69$\pm$0.10, 0.34$\pm$0.06, and 0.14$\pm$0.06 Jy/beam for stages I, II(ex), and II respectively.

Taken together, these results suggest that the stage II sources are a population of T-Tauri-like stars that are in their protoplanetary disk phase, and furthermore are not deeply embedded in the molecular cloud, as they show fairly low extinctions.
On the other hand, the sources in stage II(ex) require at least 7\,mag of external foreground extinction and are spatially coincident with the massive dense SCUBA core in the East. This indicates that the sources in the East are more deeply embedded in the cloud than those in the West. Both the fact that the stage II(ex) sources can be fit by disk-dominated models and their lower associated SCUBA 850\microns fluxes suggest that these are in fact a similar population to stage II, but with larger amounts of external foreground extinction. Finally, the stage I sources are genuine embedded protostars.

% --------------   END MODELING SECTION

\section{Discussion}

After the analysis of the IRS spectra and subsequent detailed SED modeling, we can characterize the YSOs in our sample more accurately than what was previously possible from YSO spectral classes alone by assigning evolutionary stages next to their SED classes. As we have seen, the comprehensive SED modeling suggests that the sources fall into three classes: 1) genuine protostars (stage I), 2) disk sources behind high foreground extinction (stage II(ex)), and 3) disk sources with less external foreground extinction (stage II). Of the 44 sources for which we determine evolutionary stages, 17 are stage I protostars. Most of these were already classified as class~I in the literature (one, source 4,  is a class 0 source, but a stage earlier than stage~I has not been defined), with the exceptions of sources 28, 45, and 51 which have not been classified, or have been classified differently (source 35) in \citet{sun09}. On the other hand, as summarized in Table~\ref{tab_srclist}, the sample contains many previous class~I sources that instead appear to be in stage II(ex). Thus, a number of sources appear to be in later evolutionary stages than what was previously suspected. Most previously determined class~II sources turn out to be in stage II, although a few are in stage II(ex).

With regard to the use of spectral indices as an indicator of YSO class, our sample contains a few objects that can serve as instructive examples of the limits of using a single spectral index for the classification of individual sources. Sources 3 and 56 both have negative IRAC spectral indices, but it turns out that they are stage I protostars and were classified as class~I protostars in the literature.

The IRS data are a significant help to constrain the SEDs of these sources both by allowing us to make use of spectral properties in YSO characterization, and by extending the SED wavelength coverage. The latter is further improved since the IRS data allow us to avoid spectral ranges contaminated by unwanted line features for photometry and subsequent SED modeling. In particular, the youngest sources show strong silicate and ice features in their mid-infrared spectra; both features are correlated with the sources' spectral indices, both IRAC and \textsl{K}--24\,$\mu$m. A correlation of the silicate optical depth at 9.7\,$\mu$m and the total ice column density is present, but has more scatter than the correlations of the two features with the IRAC spectral index as a measure of evolutionary stage. The main exception appear to be objects with high silicate optical depths but low total ice column densities. These might be more massive stars with radiation fields able to evaporate the ice mantles. A similar correlation is found in the c2d data and by \citet{fur08}. We detect three sources with PAH emission (all in NGC~2264), a small subset of our sample. Presumably, these are more massive YSOs. For the crowded region of the \textsl{microcluster} of NGC~2264, we can make use of IRS spectral maps, for example to identify regions with jet activity.

Based on the IRS and modeling results, we can check whether the three evolutionary stages that were identified by SED modeling in turn have distinctive features in their IRS spectra. We have therefore marked the sources in Fig.~\ref{fig_speccorr} according to their evolutionary stage. Indeed, the stage I sources tend to have large optical depths of the silicate feature and strong ice features. Stage~II(ex) sources have less strong silicate and ice features and stage II sources tend to show the silicate feature in emission while not showing any ice features. It is interesting to note that the IRS spectral features also appear to be distinctly different when comparing stage II(ex) and stage II sources. Fig.~\ref{fig_speccorr} shows that while virtually all stage II sources have the 9.7~$\mu$m in \textsl{emission}, indicative of disks, that is \textsl{not} the case for the stage II(ex) sources, presumably due to the additional extinction which may `cancel out' the emission. Also, none of the stage II sources show ice features, while stage II(ex) interestingly do show ice absorption, although at smaller total column densities than the stage I sources. In our interpretation, the stage II(ex) ice features would be related to the additional extinction by intracluster foreground material and not the sources themselves. In summary, while IRS data are of crucial importance, they alone are not sufficient to distinguish the evolutionary stages. For this purpose, detailed SED modeling is required. 

We point out that in the context of our comprehensive SED modeling, the archival submillimeter data were very important for the separation of the sample into the different evolutionary stages. Dense dust and gas in an envelope manifest themselves in submillimeter maps as concentrated emission in contrast to more large-scale foreground extinction. Future high-angular resolution observations in the far-infrared/submillimeter wavelength range, as forthcoming from the \textsl{Herschel} Space Observatory, will be even better suited to help determining the evolutionary stages in YSO samples.

Finally, we discuss the observable quantities that correlate best with the different evolutionary stages. We have seen that this is certainly not a spectral index since the correlation of the evolutionary stages with spectral indices is not unambiguous. The genuine protostars (stage I) all have submillimeter detections and the highest silicate optical depths as well as the highest ice column densities. The stage II sources are broadly characterized by a silicate feature in emission and no ice features. The stage II(ex) sources lie between these two extremes, but their SEDs can be easily mistaken for class I sources (in fact, some were previously determined to be class I sources, see above); in particular, they do not show the silicate feature in emission. Once an individual SED shape is well characterized, far-infrared and submillimeter continuum data are the most important photometric data to distinguish the different evolutionary stages.

\section{Summary and Lessons Learned}
We have combined \textsl{Spitzer} IRS spectra, SCUBA submillimeter, \textsl{Spitzer} mid-infrared, FLAMINGOS near-infrared and published UBVRI photometry with detailed radiative transfer models to investigate the nature of the most embedded members in the NGC\,2264 and IC\,348 clusters. We summarize our results as follows: 
\begin{itemize}
\item Empirical SED classes and mid-infrared spectral features do not provide unambiguous identifications of YSO evolutionary stages in clustered environments.
\item In particular, significant amounts of foreground extinction can transform a class II SED into a flat/class I SED and suppress silicate emission characteristic of disks.
\item SED modeling incorporating the silicate features and mm/submm photometry provides a more reliable measure of the evolutionary stage.
\item In our sample of IC\,348 sources, we identify three stage~I, six stage II(ex), and one stage II source. In our NGC\,2264 sample, covering the \textsl{Spokes} cluster, we identify 14 stage~I, 9 stage II(ex), and 11 stage~II sources. Among the previously known 20 class~I sources in our NGC~2264 sample, we find 16 stage~I sources whereas the reminder are classified as stage~II(ex).
\item Only in NGC\,2264, we identify three sources with PAH emission, potentially tracing the upper end of the mass function in this cluster.
\item We find various YSOs with rotational \htwo\ transitions as well as sources with neon and iron fine structure transitions, but for these features, we do not find a clear correlation with evolutionary stage in our sample.
\end{itemize}

\bibliographystyle{aa} % style aa.bst
\bibliography{bib_n2264} % your references Yourfile.bib

\acknowledgments{We would like to thank Grace Wolf-Chase, for providing us with files of her submillimeter data of NGC~2264, as well as the referee, Dan Watson, whose constructive comments led to improvements of this paper. This project is based, in large part, on observations made with the \textsl{Spitzer} Space Telescope, which is operated by the Jet Propulsion Laboratory, California Institute of Technology under a contract with NASA. Support for this work was provided by NASA through award no. 1288815 issued by JPL/Caltech.}

{\it Facilities:} \facility{Spitzer (IRS)}

\clearpage

\begin{deluxetable}{rrrrrrrrrrrrr}
\tabletypesize{\scriptsize}
%\rotate
\tablecaption{Source list\label{tab_srclist}}
\tablewidth{0pt}
\tablecolumns{13}
\tablehead{
\colhead{}   & \colhead{ID} & \colhead{RA} & \colhead{Dec} & \colhead{$\alpha_{\rm IRAC}$} & \colhead{$\alpha_{\rm K-24}$} & \colhead{class\tablenotemark{a}} & \colhead{Stage}\\
\colhead{}   & \colhead{}   & \colhead{}   & \colhead{}   & \colhead{} & \colhead{} & \colhead{(lit.)}  & \colhead{(this paper)}\\
}
\startdata
 1 &LRL 00245& 03:43:45.17 & +32:03:58.7&   0.44  & 0.42   & I       & II(ex)\\
 2 &LRL 54362& 03:43:50.97 & +32:03:24.0&   0.95  & 0.04   & I       & I     \\
 3 &LRL 54361& 03:43:51.03 & +32:03:07.7& --0.83  & n.a.   & I       & I     \\
 4 &LRL 57025& 03:43:56.89 & +32:03:03.4& n.a.    & n.a.   & 0       & I     \\
 5 &LRL 00013& 03:43:59.64 & +32:01:54.2& --0.31  & --0.27 & thick   & II(ex)\\
 6 &LRL 54459& 03:44:02.41 & +32:02:04.5&   1.92  & n.a.   & I       & \tablenotemark{b}\\
 7 &LRL 54460& 03:44:02.62 & +32:01:59.6&   0.97  & n.a.   & I       & \tablenotemark{b}\\
 8 &LRL 01916& 03:44:05.78 & +32:00:28.5& --0.29  & 0.00   & I       &       \\
 9 &LRL 04011& 03:44:06.91 & +32:01:55.4& --0.15  & 0.20   & I       & II(ex)\\
10 &LRL 00051& 03:44:12.97 & +32:01:35.4&   0.37  & 0.42   & thick   & II(ex)\\
11 &LRL 01889& 03:44:21.35 & +31:59:32.7& --0.01  & 0.58   & I       & II(ex)\\
12 &LRL 00435& 03:44:30.28 & +32:11:35.3&   0.47  & 0.51   & flat    &       \\
13 &LRL 01872& 03:44:43.31 & +32:01:31.6&   0.40  & 0.76   & I       & II(ex)\\
14 &LRL 01898& 03:44:43.89 & +32:01:37.4& --0.93  & 0.65   & 0/I     & \tablenotemark{c}\\
15 &LRL 00234& 03:44:45.22 & +32:01:20.0& --0.42  & --0.32 & I       & II    \\
16 &LRL 00904& 03:45:13.81 & +32:12:10.1&   1.20  & 0.53   & I       &       \\
17 &SSB 08807& 06:40:39.34 & +09:34:45.6&   0.03  & 0.01   & II,+    & II    \\
18 &SSB 08910& 06:40:40.06 & +09:35:02.9& --0.66  & --0.58 & II,+    & II    \\
19 &SSB 09077& 06:40:41.15 & +09:33:57.9& --1.02  & --0.63 & II,+    & II    \\
20 &SSB 10207& 06:40:48.62 & +09:35:57.7& --0.04  & 0.13   & I,+     & II(ex)\\
21 &SSB 10281& 06:40:49.15 & +09:37:36.4& --0.57  & --0.24 & pre-TD  & II    \\
22 &SSB 10387& 06:40:49.89 & +09:36:49.5& --1.31  & --0.89 & II,+    & II    \\
23 &SSB 10455& 06:40:50.29 & +09:34:53.7&  2.63   & 2.45   & I	     & I     \\
24 &SSB 10587& 06:40:51.06 & +09:35:22.2&  1.73   & 1.30   & I	     & I     \\
25 &SSB 10710& 06:40:51.91 & +09:37:55.9&  1.89   & 0.92   & I	     &       \\
26 &SSB 10769& 06:40:52.34 & +09:34:57.7&  1.75   & 1.60   & I	     & I     \\
27 &SSB 10775& 06:40:52.38 & +09:34:31.4& --1.41  & 0.41   & I	     & II(ex)\\
28 &SSB 11017& 06:40:54.19 & +09:36:46.2&  1.45   & 1.44   & member? & I     \\
29 &SSB 11314& 06:40:56.17 & +09:36:31.0& --0.48  & --0.42 & II,+    & II    \\
30 &SSB 11347& 06:40:56.40 & +09:35:53.3& --1.16  & --0.42 & II,+    & II    \\
31 &SSB 11445& 06:40:57.00 & +09:33:01.4&  0.52   & --0.04 & II,P    & II    \\
32 &SSB 11592& 06:40:57.98 & +09:36:39.5&  1.20   & 1.62   & I,X     &       \\
33 &SSB 11598& 06:40:58.01 & +09:36:14.6& --0.45  & 0.44   & I,x     & II(ex)\\
34 &SSB 11802& 06:40:59.14 & +09:36:14.9&  0.83   & 1.27   & I	     & II(ex)\\
35 &SSB 11825& 06:40:59.27 & +09:33:25.1& --0.48  & 0.07   & pre-TD  & I     \\
36 &SSB 11829& 06:40:59.30 & +09:35:52.4& --0.94  & --0.61 & II,X    &       \\
37 &SSB 11847& 06:40:59.37 & +09:33:33.5& --1.22  & --0.02 & II,+    & II    \\
38 &SSB 12070& 06:41:00.98 & +09:32:44.5& --1.21  & --0.89 & II,+    & II    \\
39 &SSB 12208& 06:41:01.83 & +09:34:34.3& --0.35  & 0.53   & I,x     & II(ex)\\
40 &SSB 12583& 06:41:04.25 & +09:34:59.7& --0.06  & 0.77   & I	     & I     \\
41 &SSB 12780& 06:41:05.56 & +09:34:08.0&  0.21   & 1.64   & I,X     &       \\
42 &SSB 12820& 06:41:05.77 & +09:35:29.6&  0.62   & 0.36   & I,X     & I     \\
43 &SSB 12875& 06:41:06.19 & +09:34:08.8&  0.33   & 1.53   & I,x     &       \\
44 &SSB 12893& 06:41:06.29 & +09:33:50.0&  1.09   & 1.38   & I	     & I     \\
45 &SSB 12913& 06:41:06.42 & +09:35:54.5&  4.21   & 2.58   & member? & I     \\
46 &SSB 12951& 06:41:06.66 & +09:33:57.8& --0.08  & 0.42   & I,x     & II(ex)\\
47 &SSB 12966& 06:41:06.74 & +09:34:45.9& --0.38  & 0.05   & II,+    & II    \\
48 &SSB 12968& 06:41:06.77 & +09:33:34.9&  2.29   & 2.56   & I	     & I     \\
49 &SSB 13069& 06:41:07.40 & +09:34:54.9& --0.98  & 0.20   & II,x    & \tablenotemark{c}\\
50 &SSB 13111& 06:41:07.67 & +09:34:19.1&  1.63   & 2.24   & I,X     & I     \\
51 &SSB 13238& 06:41:08.61 & +09:35:42.7&  1.78   & 2.41   & member? & I     \\
52 &SSB 13244& 06:41:08.63 & +09:36:03.4& --0.10  & 0.35   & II      & II(ex)\\
53 &SSB 13427& 06:41:09.90 & +09:35:40.8&  2.08   & 2.76   & I	     & I     \\
54 &SSB 13499& 06:41:10.43 & +09:34:18.6& --0.35  & 0.06   & II,X    & II(ex)\\
55 &SSB 13567& 06:41:10.93 & +09:34:08.3& --0.58  & 0.15   & II,x    & II(ex)\\
56 &SSB 13696& 06:41:11.84 & +09:35:31.5& --0.47  & 0.43   & I,X     & I     \\
\enddata		    				  						 
%% Text for table notes should follow after the \enddata but before
%% the \end{deluxetable}. Make sure there is at least one \tablenotemark
%% in the table for each \tablenotetext.
%\tablecomments{Blablabla}
\tablenotetext{a}{for NGC 2264 from \citet{sun09} with the following symbols: + = H{alpha} emission star with X-ray emission; X = X-ray emission star; x = X-ray emission candidate; P = X-ray emission star with strong H$\alpha$ emission; for IC 348 from \citet[][disk types]{lad06} and \citet{mue07}}
\tablenotetext{b}{Sources 6 and 7 form a single blended source at most wavelengths, and are thus discussed as such.}
\tablenotetext{c}{Due to blending with nearby sources, insufficient photometry at longer wavelengths to warrant modeling and further discussion.}
\end{deluxetable}

\begin{deluxetable}{lrrrrrl}
\tabletypesize{\scriptsize}
\tablecaption{Spectral features and fit results}
\tablewidth{0pt}
\tablecolumns{7}
\tablehead{
\colhead{Source} & \colhead{Stage} & \colhead{$\tau_{6.0}$\tablenotemark{b}} & \colhead{$\tau_{9.7}$\tablenotemark{b}} & \colhead{FWHM 9.7\tablenotemark{b}} & \colhead{$N_{\rm tot}$(ice)\tablenotemark{b}}     &  \colhead{Comments\tablenotemark{c}} \\
\colhead{      } & \colhead{     } & \colhead{            } & \colhead{            } & \colhead{  [\mic]} & \colhead{[$\times 10^{18}$\scm]} &  \colhead{        } \\
}
\startdata
2 	&   I	& 1.01*   & 2.79*   & \ldots & 0.65    &\\
3 	&   I	& 1.21*   & 3.00*   & \ldots & 2.08    & \htwo\ S(1)--S(7), \neiif\ 12.8, \feiif\ 17.9, 26.0  \\
4 	&   I	& \ldots & \ldots & \ldots & $<$3.64   & \htwo\ S(1)--S(2) , \feiif\ 17.9, 24.5, 26.0, 35.3, \sif\ 25.2, \siliif\ 34.8; no SL  \\
23	&   I   & 0.49    & 1.85	 & 2.2     & 2.02    &\\
24	&   I   & 1.61*   & 1.94*	 & \ldots & \ldots   & no LL/LH\\
26	&   I	& 0.93    & 2.86  & 2.1  & 3.25        &\\
28	&   I   & $<$0.75 & 1.89*	 & 2.1     & \ldots  & no LL/LH\\
35	&   I   & $<$0.14 & $<$0.05 & \ldots & $<$0.15  &\\
40	&   I	& 0.31    & 1.36    & 2.6  & 0.85	 &\\
42	&   I	& \ldots & \ldots & \ldots & 0.78	 & PAH\\
44	&   I	& 0.52    & 1.60    & 2.1  & 1.41	 &\\
45	&   I	& 1.41    & 5.93    & 2.0  & 5.27	 &\\
48	&   I	& 1.34    & 3.08    & 2.1  & 2.65	 &\\
50	&   I	& 0.92    & 3.87    & 2.1  & 4.46	 &\\
51	&   I	& $<$1.26 & $<$1.58 & \ldots & $<$0.42   & \htwo\ S(1)--S(2)\\
53	&   I	& 1.94    & 6.27    & 1.8  & 1.72	 & \\
56	&   I	& 0.68*   & 1.28*   & 2.1  & 1.82	 & \htwo\ S(2), S(5), \neiif\ 12.8, \feiif\ 17.9, 26.0, \siliif\ 34.8\\
\hline
1 	&   II(ex)   & $<$0.07&--0.35	& 3.0  & $<$0.13   &\\
5 	&   II(ex)   & 0.16	 & 0.43   & 2.6    & 0.23  &\\
9 	&   II(ex)   & $<$0.88&$<$0.50 & \ldots & $<$2.66  & \htwo\ S(1)--S(5) \\
10	&   II(ex)   & 0.13	 & 1.48   & 2.3    & 0.30  &\\
11	&   II(ex)   & 0.22*	 & 0.83   & 2.8    & 0.75  &\\
13	&   II(ex)   & 0.20	 & 0.45   & 2.1    & 0.81  &\\
20	&   II(ex)   & $<$0.05 & $<$0.06 & \ldots & $<$0.11  & \htwo\ S(1)--S(2), \neiif\ 12.8 \\
27	&   II(ex)   & $<$1.08 & $<$0.32 & \ldots & \ldots   & no LL/LH\\
33	&   II(ex)   & $<$0.42 & $<$0.54 & \ldots & \ldots   & \htwo\ S(2)--S(7); no LL/LH \\
34	&   II(ex)   & 0.35    & 1.27	 & 2.3     & 0.56    & \htwo\ S(1)--S(3), \neiif\ 12.8, \feiif\ 17.9, 26.0\\
39	&   II(ex)   & 0.19*   & 0.38*	 & 2.3     & 0.27*   &\\
46	&   II(ex)   & 0.22*   & 0.68*	 & 1.9     & \ldots  &\\
52	&   II(ex)   & 0.51*   & 1.13*	 & 2.2     & \ldots  & no LL/LH\\
54	&   II(ex)   & 0.23*   & 0.51*	 & 2.5     & 0.88*   & \htwo\ S(1)--S(7)\\
55	&   II(ex)   & $<$0.07 & $<$0.50 & \ldots & $<$0.33  & \htwo\ S(0)--S(5), \siliif\ 34.8\\
\hline
15	&   II   & $<$0.20& 0.54   & \ldots & $<$0.54	 &\\
17	&   II   & $<$0.07 & --0.25  & 3.6 & $<$0.17	 & \htwo\ S(0)--S(3), \neiif\ 12.8 \\
18	&   II   & $<$0.08 & --1.09  & 2.8 &  \ldots 	 & no LL/LH \\
19	&   II   & $<$0.21 & $<$0.38 & \ldots & \ldots   & no LL/LH \\
21	&   II   & $<$0.11 & --0.54  &  2.8 & $<$0.63	 & \htwo\ S(0)--S(2), \neiif\ 12.8 \\
22	&   II   & $<$0.06 & --0.85  &  2.7&   \ldots 	 & \htwo\ S(2)--S(3); no LL/LH\\
29	&   II   & $<$0.08 & --0.69  & 3.0 & $<$0.18	 &\\
30	&   II   & $<$0.18 & $<$0.14 & \ldots & $<$0.54  &\\
31	&   II   & $<$0.02 & --0.53  & 3.0 & $<$0.25	 &\\
37	&   II   & $<$0.32 & --1.03    & 2.4&  \ldots 	 & no LL/LH\\
38	&   II   & $<$0.08 & --0.60  & 3.4 & $<$0.26	 &\\
47	&   II   & $<$0.08 & --0.34  & 2.8 & $<$0.09	 &\\
\hline
6+7 	&       &0.71	& 1.96   & 2.4    & 1.59       & \htwo\ S(1)--S(6), \neiif\ 12.8 \\
8 	&	& 0.23   & 1.00   & 2.6    & 1.36      &\\
12	&	& $<$0.78&$<$0.23 & \ldots & $<$0.42   &\\
14	&	& 1.11*   & 2.25*   & \ldots   & $<$1.00   & \htwo\ S(1)--S(7), \neiif\ 12.8, \neiiif\ 15.5, \feiif\ 17.9; no LL/LH \\
16	&	& $<$0.58& 0.82   & \ldots & $<$0.59   &\\
25	&	& $<$0.68 & \ldots & \ldots & $<$0.19	 & PAH \\
32	&	& 0.47    & 1.18*    & 2.1  & 1.75	 &\\
36	&	& $<$0.12 & $<$0.26 & \ldots & 0.29	 & PAH \\
41	&	& 0.60    & 1.60    & 1.9  & 1.72	 & \htwo\ S(1)--S(7), \neiif\ 12.8, \feiif\ 5.3, 17.9, 26.0; part of microcluster \\
43	&	& 0.97    & 2.66    & 2.4  & 2.19	 & \htwo\ S(3)--S(7), \feiif\ 5.3, 17.9; part of microcluster\\
49	&	& 0.19    & 0.77    & 2.1  & \ldots	 & no LL/LH\\
\enddata
\tablecomments{Silicate optical depths and ice column densities are generally estimated to be accurate to $\sim25$\% except for some cases,
marked by asterisks, where the error is estimated to be $\sim50$\%.}
\tablenotetext{a}{(\ldots) = no silicate band or no data for FWHM fit; for ices: (\ldots) = no spectral data (e.g., SL or SH/LH only)}
\tablenotetext{b}{($cotwo:\co:\water$)}
\tablenotetext{c}{H$_2$ transitions are H$_2$(0--0) transitions.}
\label{tab_icefit}
\end{deluxetable}

%\begin{minipage}{20cm}
\begin{deluxetable}{l cc cc c cc cc c cc cc cc c}
\rotate
\tabletypesize{\tiny}
\tablewidth{0pt}
\tablecaption{Ranges of parameters from the modeling with the R07 SED fitting tool.\label{tab:sedmodeling}}
\tablehead{Source & \multicolumn{2}{c}{External A$_{\rm V}$} &\multicolumn{2}{c}{Total A$_{\rm V}$} & X-ray A$_v$ &
\multicolumn{2}{c}{L$_{\star}$} &\multicolumn{2}{c}{T$_{\star}$} & Sp. Type &\multicolumn{2}{c}{M$_{\rm env}$} &\multicolumn{2}{c}{$\dot{M}_{\rm
infall}$} &\multicolumn{2}{c}{M$_{\rm disk}$} & Stage\tablenotemark{a} \\
 & \multicolumn{2}{c}{(mag)} & \multicolumn{2}{c}{(mag)} & (mag) & \multicolumn{2}{c}{(L$_\odot$)} & \multicolumn{2}{c}{(K)} & & \multicolumn{2}{c}{(M$_\odot$)}& \multicolumn{2}{c}{(M$_\odot$/yr)} & \multicolumn{2}{c}{(M$_\odot$)} \\
  & min & max & min & max &  & min & max & min & max &  & min & max & min & max & min & max }
\startdata
2               &       36.8 &       40.0 &       52.9 &       78.5 &               &        0.1 &        0.5 &  2867 &  3106 &          &     3.3$\times10^{-3}$&     1.4$\times10^{-1}$&     5.0$\times10^{-7}$&     1.2$\times10^{-5}$&     3.2$\times10^{-6}$&     1.9$\times10^{-3}$& I \\
3               &       20.8 &       39.7 &       50.2 &       52.0 &               &        0.2 &        0.2 &  2960 &  2960 &          &     4.1$\times10^{-2}$&     4.1$\times10^{-2}$&     4.6$\times10^{-6}$&     4.6$\times10^{-6}$&     6.5$\times10^{-3}$&     6.5$\times10^{-3}$& I \\
4               &        0.0 &       40.0 &      481.9 &      700.2 &               &        1.2 &        2.5 &  2740 &  3537 &          &     1.7&     4.4&     5.0$\times10^{-5}$&     1.1$\times10^{-4}$&     7.7$\times10^{-5}$&     1.9$\times10^{-3}$& I \\
23              \tablenotemark{b}&       37.0 &       40.0 &       55.9 &       65.4 &               &        1.6 &        2.3 &  2637 &  2894 &          &     9.1$\times10^{-3}$&     4.3$\times10^{-2}$&     7.2$\times10^{-7}$&     1.6$\times10^{-6}$&     2.5$\times10^{-4}$&     1.5$\times10^{-2}$& I \\
24              &        7.3 &       40.0 &       43.7 &      462.6 &               &        0.4 &        3.8 &  2603 &  3886 &          &     4.2$\times10^{-3}$&     8.2$\times10^{-1}$&     6.5$\times10^{-7}$&     6.8$\times10^{-5}$&     7.7$\times10^{-6}$&     1.6$\times10^{-2}$& I \\
26              &        0.2 &        0.2 &       77.4 &       77.4 &               &        1.3 &        1.3 &  2601 &  2601 &          &     1.4$\times10^{-1}$&     1.4$\times10^{-1}$&     5.2$\times10^{-6}$&     5.2$\times10^{-6}$&     7.6$\times10^{-3}$&     7.6$\times10^{-3}$& I \\
28              &       34.2 &       40.0 &       50.0 &       70.8 &               &        0.4 &        0.9 &  2643 &  3193 &          &     4.2$\times10^{-3}$&     9.7$\times10^{-2}$&     6.8$\times10^{-7}$&     5.8$\times10^{-6}$&     1.2$\times10^{-4}$&     6.4$\times10^{-3}$& I \\
35              &        0.9 &        3.3 &        4.5 &       15.2 &               &        5.9 &       16.8 &  3387 &  4470 &          &     1.1$\times10^{-2}$&     1.4$\times10^{-1}$&     1.4$\times10^{-6}$&     4.2$\times10^{-6}$&     1.5$\times10^{-3}$&     3.6$\times10^{-2}$& I \\
40              &       23.7 &       36.9 &       30.8 &       37.1 &               &       20.6 &      117.5 &  4293 &  5867 &          &     3.0$\times10^{-3}$&     5.3$\times10^{-1}$&     1.4$\times10^{-7}$&     1.2$\times10^{-5}$&     4.5$\times10^{-5}$&     6.8$\times10^{-2}$& I \\
42              &       10.6 &       23.0 &       19.0 &       24.5 &16.85$\pm$ 1.45&       21.2 &      114.1 &  4222 &  5279 &          &     2.1$\times10^{-1}$&     1.2$\times10^{1}$&     5.0$\times10^{-6}$&     2.6$\times10^{-4}$&     7.4$\times10^{-5}$&     1.2$\times10^{-1}$& I \\
44              &       21.8 &       38.3 &       28.6 &       43.1 &               &        2.9 &       19.1 &  2803 &  4700 &          &     7.4$\times10^{-3}$&     1.8&     1.4$\times10^{-6}$&     2.6$\times10^{-5}$&     1.0$\times10^{-4}$&     8.2$\times10^{-2}$& I \\
45              &       20.0 &       40.0 &      124.0 &      192.3 &               &        3.1 &        6.9 &  3039 &  3957 &          &     3.6$\times10^{-2}$&     6.5$\times10^{-1}$&     4.4$\times10^{-6}$&     3.5$\times10^{-5}$&     2.7$\times10^{-3}$&     3.9$\times10^{-2}$& I \\
48              &       28.5 &       40.0 &       77.3 &      104.4 &               &        4.9 &       15.8 &  3449 &  4630 &          &     1.9$\times10^{-2}$&     9.8&     5.5$\times10^{-6}$&     1.4$\times10^{-4}$&     6.3$\times10^{-5}$&     3.1$\times10^{-2}$& I \\
50              &       17.0 &       40.0 &       68.8 &      620.6 &50.50$\pm$ 5.75&        3.7 &       15.5 &  2896 &  4351 &          &     6.5$\times10^{-2}$&     3.6&     5.4$\times10^{-6}$&     9.9$\times10^{-5}$&     1.5$\times10^{-4}$&     3.5$\times10^{-2}$& I \\
51              &        0.0 &       40.0 &      476.3 &     2383.3 &               &        8.9 &       24.1 &  2651 &  4197 &          &     1.2$\times10^{-1}$&     5.9$\times10^{-1}$&     1.2$\times10^{-5}$&     1.3$\times10^{-4}$&     4.7$\times10^{-4}$&     5.6$\times10^{-2}$& I \\
53              &       27.0 &       40.0 &      153.4 &      164.2 &               &       18.7 &       34.6 &  4245 &  4726 &          &     1.3$\times10^{-1}$&     2.2$\times10^{-1}$&     4.2$\times10^{-5}$&     6.6$\times10^{-5}$&     5.8$\times10^{-3}$&     9.4$\times10^{-2}$& I \\
56              &       10.8 &       27.5 &       27.2 &       37.8 &               &       16.6 &       22.3 &  4202 &  4385 &          &     7.3$\times10^{-1}$&     1.8&     2.4$\times10^{-5}$&     5.4$\times10^{-5}$&     2.1$\times10^{-3}$&     1.8$\times10^{-2}$& I \\
\hline
1               &       13.3 &       15.8 &       13.3 &       15.8 &               &        5.4 &       28.1 &  5013 &  9888 &          &     1.3$\times10^{-8}$&     2.1$\times10^{-4}$&        0    &        0    &     5.8$\times10^{-6}$&     2.9$\times10^{-2}$& II(ex) \\
5               &       15.0 &       18.4 &       15.0 &       18.4 &               &        8.8 &       31.7 &  4197 &  7183 &      M0.5&     1.2$\times10^{-6}$&     3.6$\times10^{-2}$&        0    &     8.3$\times10^{-7}$&     3.0$\times10^{-5}$&     1.1$\times10^{-1}$& II(ex) \\
9               &        8.7 &       10.1 &        8.7 &       10.1 &               &        0.0 &        0.0 &  3073 &  3074 &          &     1.9$\times10^{-9}$&     2.1$\times10^{-8}$&        0    &        0    &     6.7$\times10^{-6}$&     2.0$\times10^{-5}$& II(ex) \\
10              \tablenotemark{b}&        1.4 &       16.0 &       16.0 &       40.2 &               &        1.3 &       14.1 &  2585 &  4504 &          &     7.4$\times10^{-4}$&     5.3$\times10^{-2}$&     1.1$\times10^{-7}$&     1.4$\times10^{-6}$&     4.8$\times10^{-3}$&     5.5$\times10^{-2}$& II(ex) \\
11              &       24.7 &       30.3 &       24.7 &       30.3 &               &        1.2 &        6.6 &  3425 &  5714 &          &     2.8$\times10^{-9}$&     4.2$\times10^{-3}$&        0    &     2.7$\times10^{-7}$&     2.3$\times10^{-4}$&     3.4$\times10^{-2}$& II(ex) \\
13              &       22.7 &       34.0 &       23.9 &       34.2 &               &        5.8 &       42.0 &  3936 & 10731 &          &     3.8$\times10^{-9}$&     4.4$\times10^{-1}$&        0    &     1.1$\times10^{-5}$&     1.4$\times10^{-5}$&     1.1$\times10^{-1}$& II(ex) \\
20              &        8.7 &       11.3 &        9.8 &       11.4 & 4.90$\pm$ 0.90&       71.5 &      181.5 &  5724 & 13452 &          &     1.1$\times10^{-8}$&     4.9$\times10^{-2}$&        0    &     2.0$\times10^{-7}$&     3.8$\times10^{-5}$&     1.1$\times10^{-1}$& II(ex) \\
27              &        0.0 &       14.4 &       12.6 &       42.0 &               &        0.2 &        0.9 &  2684 &  3409 &          &     5.9$\times10^{-5}$&     2.5$\times10^{-1}$&     2.9$\times10^{-9}$&     2.4$\times10^{-5}$&     1.5$\times10^{-6}$&     8.7$\times10^{-3}$& II(ex) \\
33              &        6.8 &       31.1 &       16.1 &       49.7 &               &        0.4 &        3.9 &  2654 &  5415 &          &     6.5$\times10^{-10}$&     8.8$\times10^{-1}$&        0    &     4.3$\times10^{-5}$&     1.7$\times10^{-6}$&     4.3$\times10^{-2}$& II(ex) \\
34              &        0.9 &       40.0 &       24.5 &       56.6 &               &        1.1 &       21.3 &  2585 &  7176 &          &     1.6$\times10^{-6}$&     1.1$\times10^{1}$&        0    &     1.6$\times10^{-4}$&     1.4$\times10^{-5}$&     1.1$\times10^{-1}$& II(ex) \\
39              &       14.6 &       24.8 &       18.8 &       27.1 &               &       20.6 &      164.6 &  4335 & 13713 &          &     2.3$\times10^{-8}$&     2.0$\times10^{-1}$&        0    &     1.9$\times10^{-6}$&     1.8$\times10^{-5}$&     2.1$\times10^{-1}$& II(ex) \\
46              &       15.1 &       30.1 &       23.1 &       33.3 &               &        1.6 &       13.7 &  3280 &  5093 &          &     4.6$\times10^{-9}$&     2.1&        0    &     3.4$\times10^{-5}$&     3.3$\times10^{-6}$&     5.5$\times10^{-2}$& II(ex) \\
52              &        5.4 &       32.6 &       23.2 &       40.0 &               &        0.5 &       49.9 &  2716 &  9658 &          &     2.1$\times10^{-6}$&     5.1$\times10^{-1}$&        0    &     2.7$\times10^{-5}$&     7.8$\times10^{-8}$&     1.8$\times10^{-2}$& II(ex) \\
54              &       14.2 &       22.8 &       14.2 &       23.0 &               &        5.6 &       33.6 &  4041 & 10334 &          &     1.4$\times10^{-8}$&     4.7$\times10^{-1}$&        0    &     1.9$\times10^{-5}$&     1.0$\times10^{-5}$&     1.1$\times10^{-1}$& II(ex) \\
55              &        7.3 &       22.6 &       16.8 &       29.9 &               &        2.7 &       14.8 &  3776 &  7710 &          &     1.9$\times10^{-9}$&     7.3$\times10^{-1}$&        0    &     1.9$\times10^{-5}$&     3.3$\times10^{-6}$&     1.1$\times10^{-1}$& II(ex) \\
\hline
15              &        4.3 &        9.4 &        9.4 &       33.6 &               &        0.1 &        0.3 &  2883 &  3275 &     M5.75&     7.3$\times10^{-10}$&     6.9$\times10^{-3}$&        0    &     9.6$\times10^{-7}$&     2.0$\times10^{-5}$&     7.9$\times10^{-3}$& II \\
17              &        5.1 &        7.0 &        5.1 &        7.0 &               &       25.2 &       86.4 &  4688 & 12157 &        K6&     2.3$\times10^{-8}$&     1.4$\times10^{-3}$&        0    &     9.5$\times10^{-8}$&     4.6$\times10^{-4}$&     1.1$\times10^{-1}$& II \\
18              &        0.6 &        2.4 &        1.1 &        2.4 & 0.00$\pm$ 0.10&        3.9 &        8.5 &  4874 &  5912 &     K4IVe&     6.5$\times10^{-9}$&     1.8$\times10^{-5}$&        0    &        0    &     9.7$\times10^{-6}$&     1.6$\times10^{-2}$& II \\
19              &        0.2 &        4.0 &        0.2 &        4.0 &               &        0.9 &        9.0 &  3138 &  7468 &        K7&     2.0$\times10^{-9}$&     1.2$\times10^{-2}$&        0    &     3.5$\times10^{-6}$&     8.6$\times10^{-6}$&     3.2$\times10^{-2}$& II \\
21              &        0.0 &        4.9 &        0.5 &        5.5 &               &        0.6 &        6.3 &  2790 &  4990 &          &     1.0$\times10^{-8}$&     4.3$\times10^{-1}$&        0    &     4.2$\times10^{-5}$&     1.6$\times10^{-5}$&     5.9$\times10^{-2}$& II \\
22              &        0.0 &        4.7 &        0.0 &        4.7 & 0.10$\pm$ 0.30&        1.4 &       48.5 &  3425 & 10610 &  K6/K1IVe&     5.2$\times10^{-9}$&     1.1$\times10^{-2}$&        0    &     4.4$\times10^{-7}$&     5.0$\times10^{-7}$&     4.3$\times10^{-2}$& II \\
29              &        3.4 &        5.6 &        3.4 &        5.6 & 4.40$\pm$ 0.75&       13.3 &       60.0 &  4549 & 11533 &        K3&     8.4$\times10^{-9}$&     6.5$\times10^{-3}$&        0    &     5.5$\times10^{-9}$&     2.9$\times10^{-6}$&     5.9$\times10^{-2}$& II \\
30              &        0.0 &        6.8 &        1.8 &        6.8 &               &        2.2 &      166.6 &  3538 & 13762 &        K3&     1.3$\times10^{-8}$&     5.7$\times10^{-1}$&        0    &     2.8$\times10^{-5}$&     1.4$\times10^{-8}$&     3.7$\times10^{-2}$& II \\
31              &        0.0 &        9.5 &        4.8 &       13.7 & 1.45$\pm$ 0.45&        2.2 &      151.2 &  3514 & 13523 &      M3.5&     1.0$\times10^{-8}$&     2.3$\times10^{-1}$&        0    &     5.5$\times10^{-6}$&     5.6$\times10^{-8}$&     9.4$\times10^{-3}$& II \\
37              &        0.0 &        3.7 &        2.5 &       13.3 & 0.50$\pm$ 0.40&        3.6 &        6.4 &  3930 &  5046 &          &     4.5$\times10^{-9}$&     7.5$\times10^{-3}$&        0    &     1.6$\times10^{-6}$&     4.3$\times10^{-5}$&     1.9$\times10^{-2}$& II \\
38              &        0.1 &        2.9 &        0.6 &        2.9 & 0.15$\pm$ 0.15&        8.7 &       31.9 &  4264 &  5978 &        K1&     7.8$\times10^{-9}$&     4.7$\times10^{-1}$&        0    &     1.9$\times10^{-5}$&     2.5$\times10^{-5}$&     1.1$\times10^{-1}$& II \\
47              &        6.2 &        7.5 &        6.2 &        7.5 & 9.25$\pm$ 1.30&       84.9 &      759.7 &  7076 & 17926 &        G6&     3.7$\times10^{-8}$&     6.0$\times10^{-1}$&        0    &     9.7$\times10^{-9}$&     1.1$\times10^{-6}$&     1.7$\times10^{-1}$& II \\
\enddata
\tablenotetext{a}{The fit results listed here, including the stages listed here were determined using an extinction law by M. Wolff, as described in the text. Using the two extinction laws derived by \citet{mcc09} leads to the following differences: For the low-$A_K$ law, sources 23, 31, 40, and 44 become stage II(ex), and sources 10 and 27 become stage I; when using the high-$A_K$ law, almost the same changes occur, but source 44 remains a stage~I source.}
\tablenotetext{b}{The parameters for these sources should be treated with caution as the silicate absorption feature is not fit by the models}
\tablecomments{For each parameter, the minimum and maximum value are those of the models that provide a good fit ($\chi^2 - \chi^2_{\rm best} < 3 \times n_{\rm data}$). The X-ray $A_V$ has been estimated from the column densities that were determined from spectral fits to NGC~2264 sources by the ANCHORS project (cxc.harvard.edu/ANCHORS), according to $N_H$[cm$^{-2}$]$\approx 2\times 10^{21}\times A_V$[mag] \citep{ryt96,vuo03}. Spectral types are from SIMBAD.}
\end{deluxetable}
%\end{minipage}

\begin{deluxetable}{lrcccc}
\tabletypesize{\scriptsize}
\tablecaption{Data for gaseous infrared transitions observed in this work}
\tablewidth{0pt}
\tablecolumns{6}
\tablehead{
\colhead{Line ID} & \colhead{$\lambda_{\rm vac}$} & \colhead{IE\tablenotemark{a}} & \colhead{$E_{ i}/k_B$} & \colhead{$A_{ ij}$} & \colhead{$n_{{\rm crit,}i}$\tablenotemark{b}} \\
\colhead{} & \colhead{[\mic]} & \colhead{[eV]} & \colhead{[K]} & \colhead{[s$^{-1}$]} & \colhead{[cm$^{-3}$]} \\
}
\startdata
\htwo\ 0--0 S(0) &      28.22 &          \ldots &        510 &    2.9E-11 &          1 \\
\htwo\ 0--0 S(1) &      17.03 &          \ldots &       1015 &    4.8E-10 &         13 \\
\htwo\ 0--0 S(2) &      12.28 &           \ldots &       1682 &    2.8E-09 &         77 \\
\htwo\ 0--0 S(3) &       9.66 &          \ldots &       2504 &    9.8E-09 &        279 \\
\htwo\ 0--0 S(4) &       8.03 &        \ldots &       3474 &    2.6E-08 &        766 \\
\htwo\ 0--0 S(5) &       6.91 &         \ldots &       4586 &    5.9E-08 &       1781 \\
\htwo\ 0--0 S(6) &       6.11 &         \ldots &       5830 &    1.1E-07 &       3614 \\
\htwo\ 0--0 S(7) &       5.51 &          \ldots &       7196 &    2.0E-07 &       6612 \\
           &            &                       &            &            &            \\
   \feiif\ &      25.99 &             7.9 &        554 &    2.1E-03 &       7665 \\
           &      35.35 &            7.9 &        961 &    1.6E-03 &       4274 \\
           &       5.34 &            7.9 &       2694 &    6.4E-05 &        297 \\
           &      17.94 &             7.9 &       3496 &    5.9E-03 &      16185 \\
           &      24.52 &              7.9 &       4083 &    3.9E-03 &      10133 \\
           &            &                      &            &            &            \\
   \neiif\ &      12.81 &            21.6 &       1123 &    8.6E-03 &     311319 \\
  \neiiif\ &      15.56 &            41.0 &        925 &    6.0E-03 &     113670 \\
           &            &                     &            &            &            \\
     \sif\ &      25.25 &           \ldots &        570 &    1.4E-03 &      93131 \\
           &            &                      &            &            &            \\
  \siliif\ &      34.82 &             8.2 &        413 &    2.1E-04 &        784 \\
\enddata
\tablenotetext{a}{ionization energy needed to reach the respective ionization stage}
\tablenotetext{b}{critical density of initial energy level $i$: using H--\htwo\ collision rates for \htwo\
\citep{wra07}, electron collision rates for atomic lines (Fe: \citealp{ram07}, NeII: \citealp{gri01},  NeIII: \citealp{bel01}, SI: \citealp{tay04}, SiII: \citealp{duf94}), and a temperature of 2000\,K}
\label{tab;specdata}
\end{deluxetable}
%\clearpage

\clearpage

\begin{figure*}
%\begin{center}
\includegraphics*[width=1.6in, angle=90, bb = 55 36 577 705]{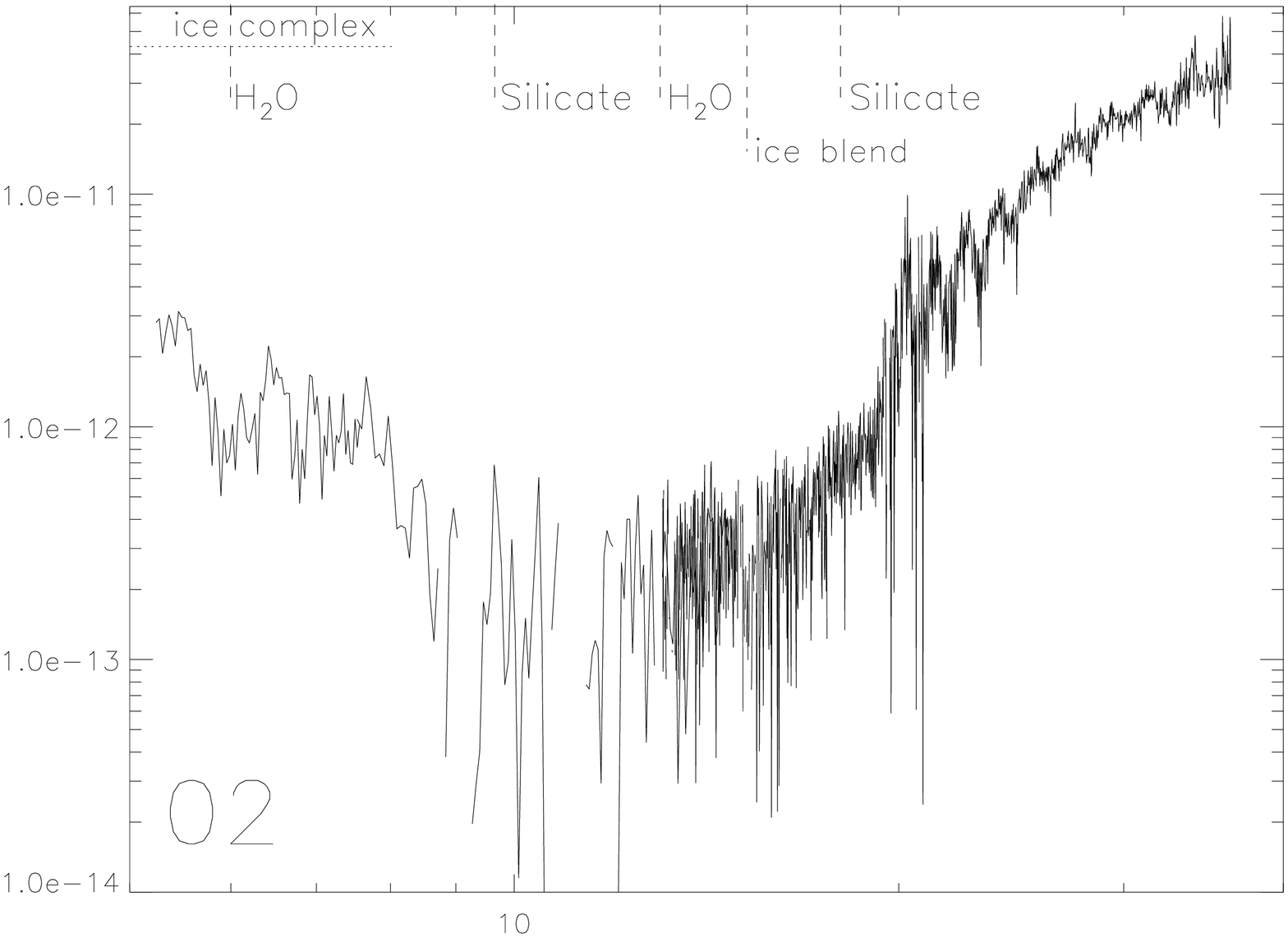}
\includegraphics*[width=1.6in, angle=90, bb = 55 36 577 705]{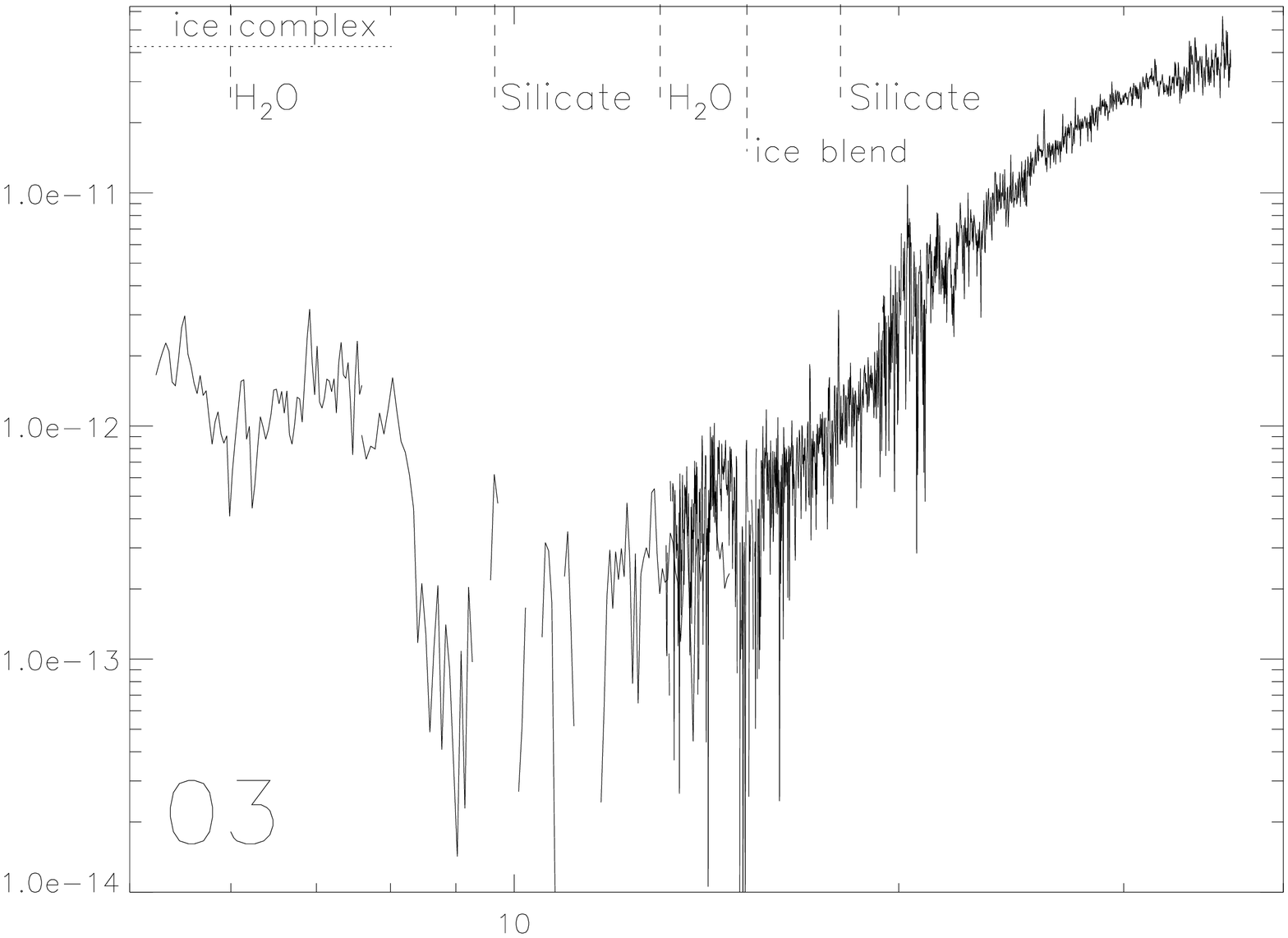}
\includegraphics*[width=1.6in, angle=90, bb = 55 36 577 705]{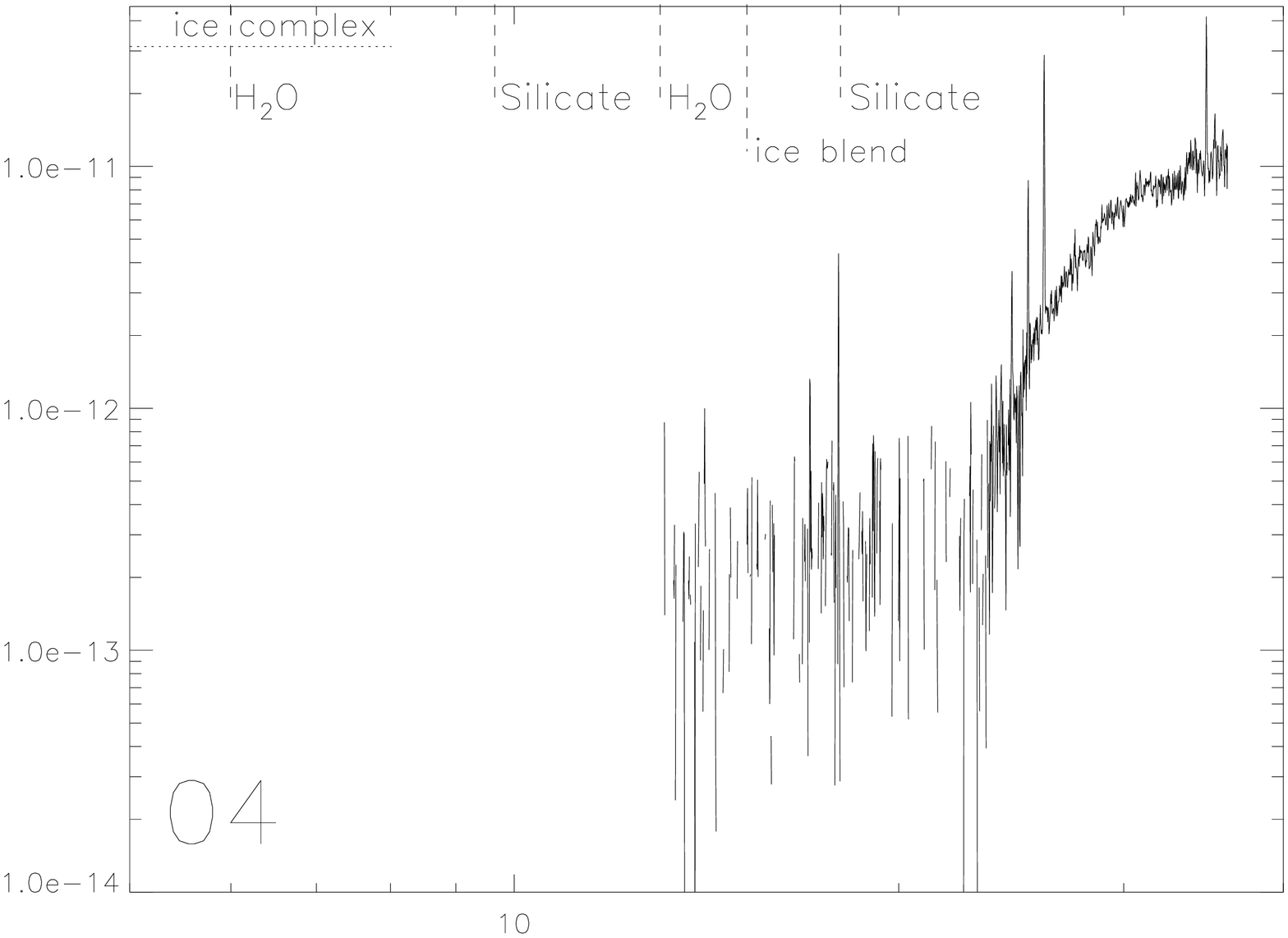}

\includegraphics*[width=1.6in, angle=90, bb = 55 36 577 705]{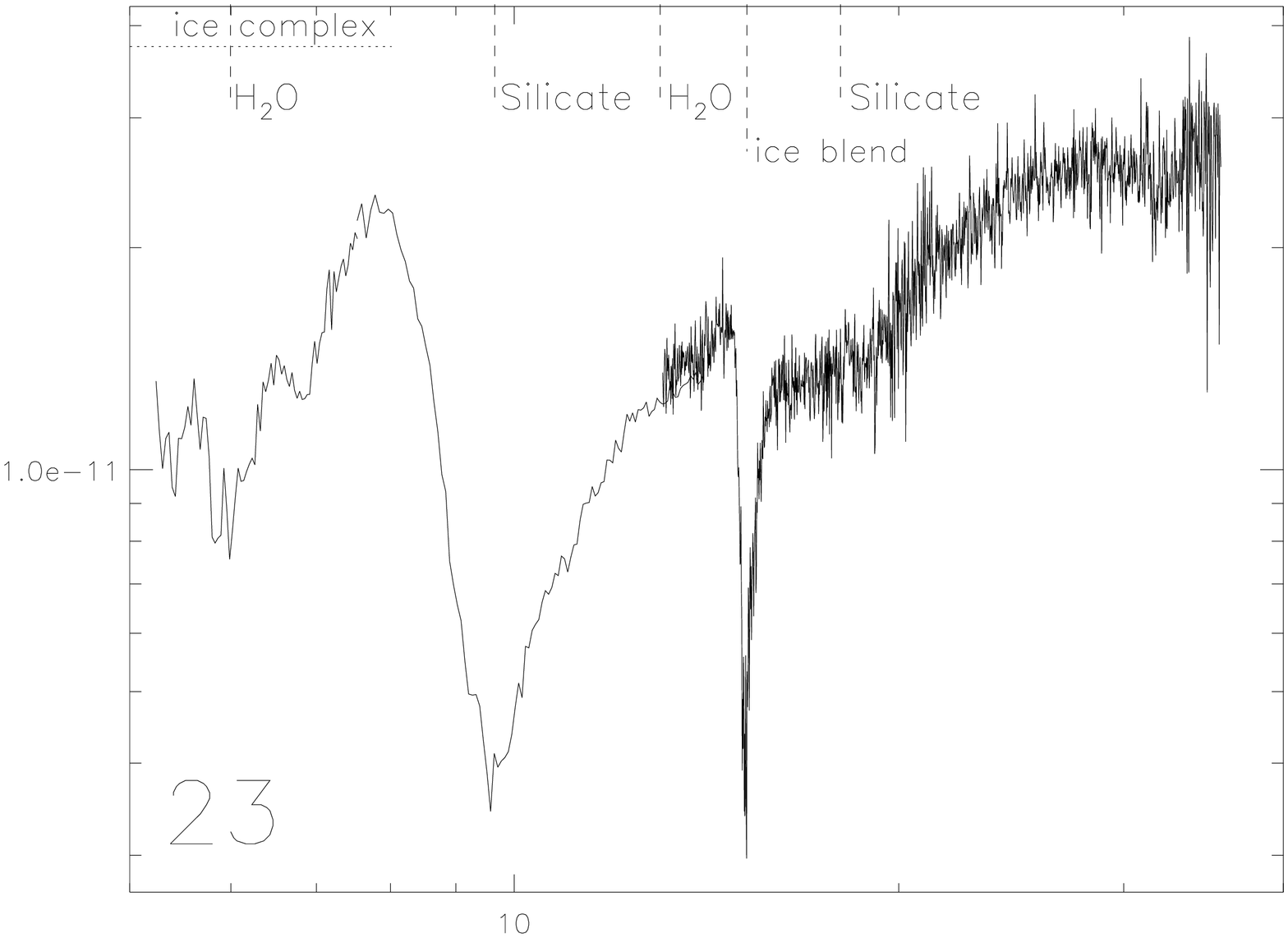}
\includegraphics*[width=1.6in, angle=90, bb = 55 36 577 705]{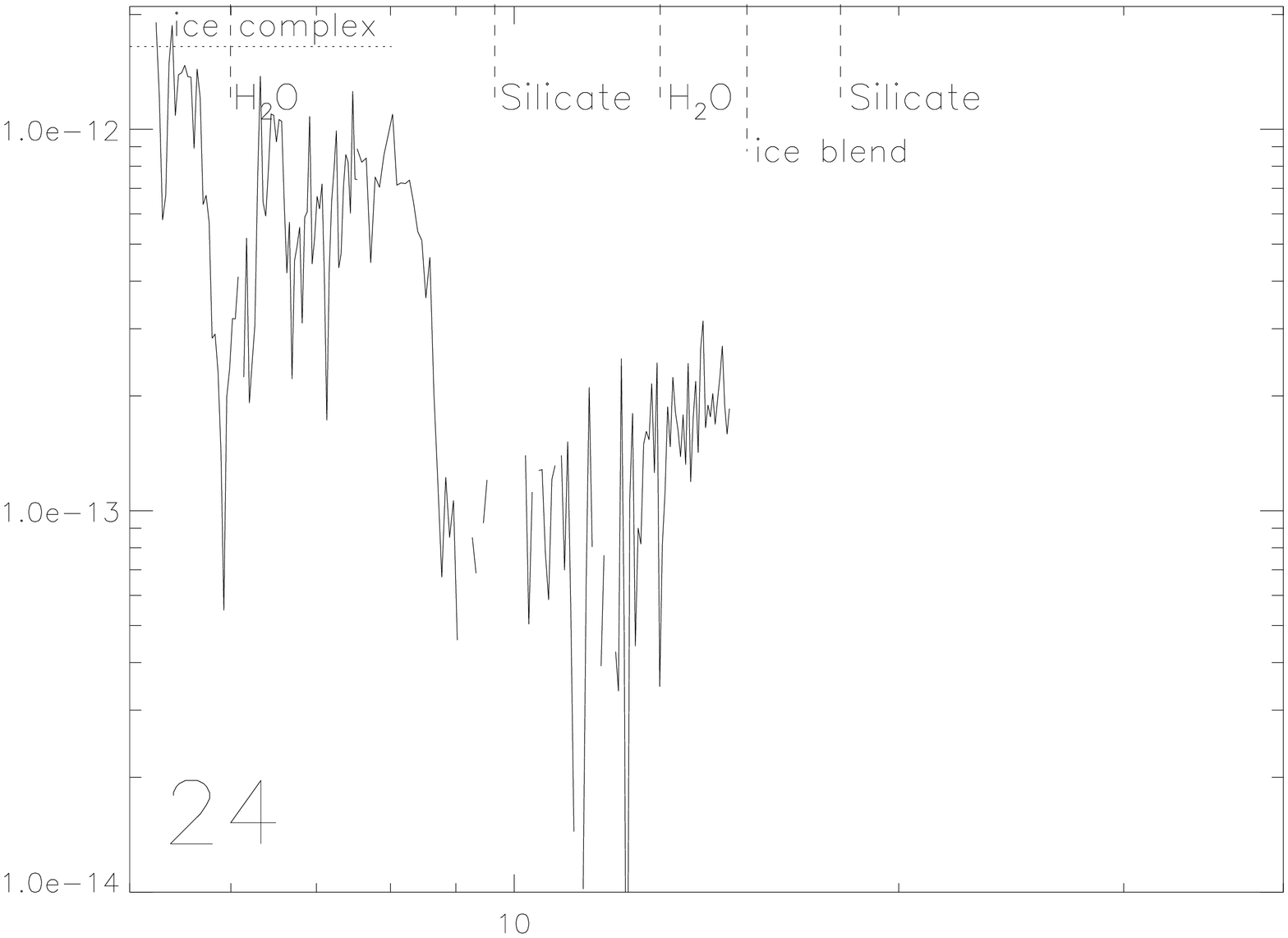}
\includegraphics*[width=1.6in, angle=90, bb = 55 36 577 705]{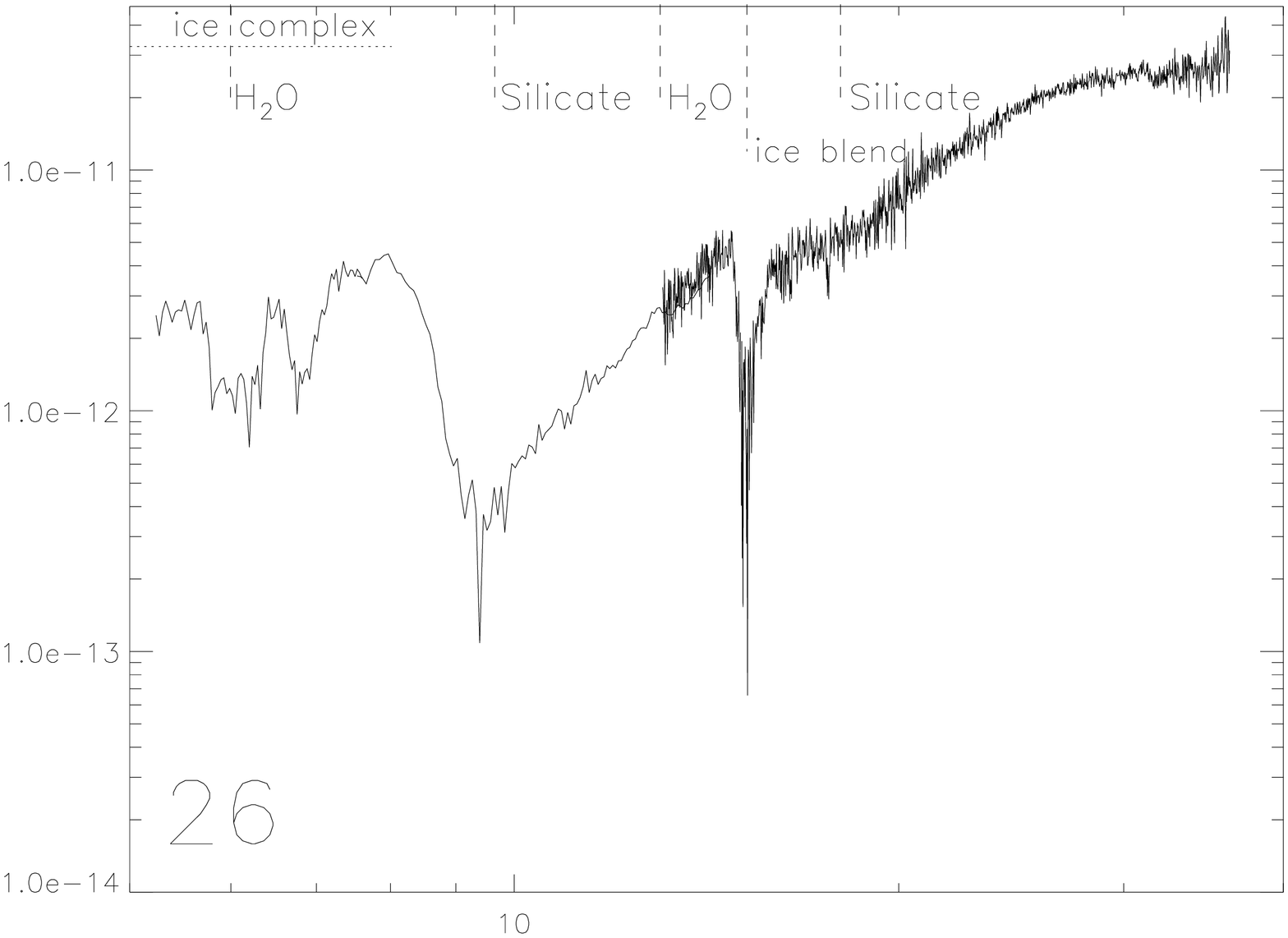}

\includegraphics*[width=1.6in, angle=90, bb = 55 36 577 705]{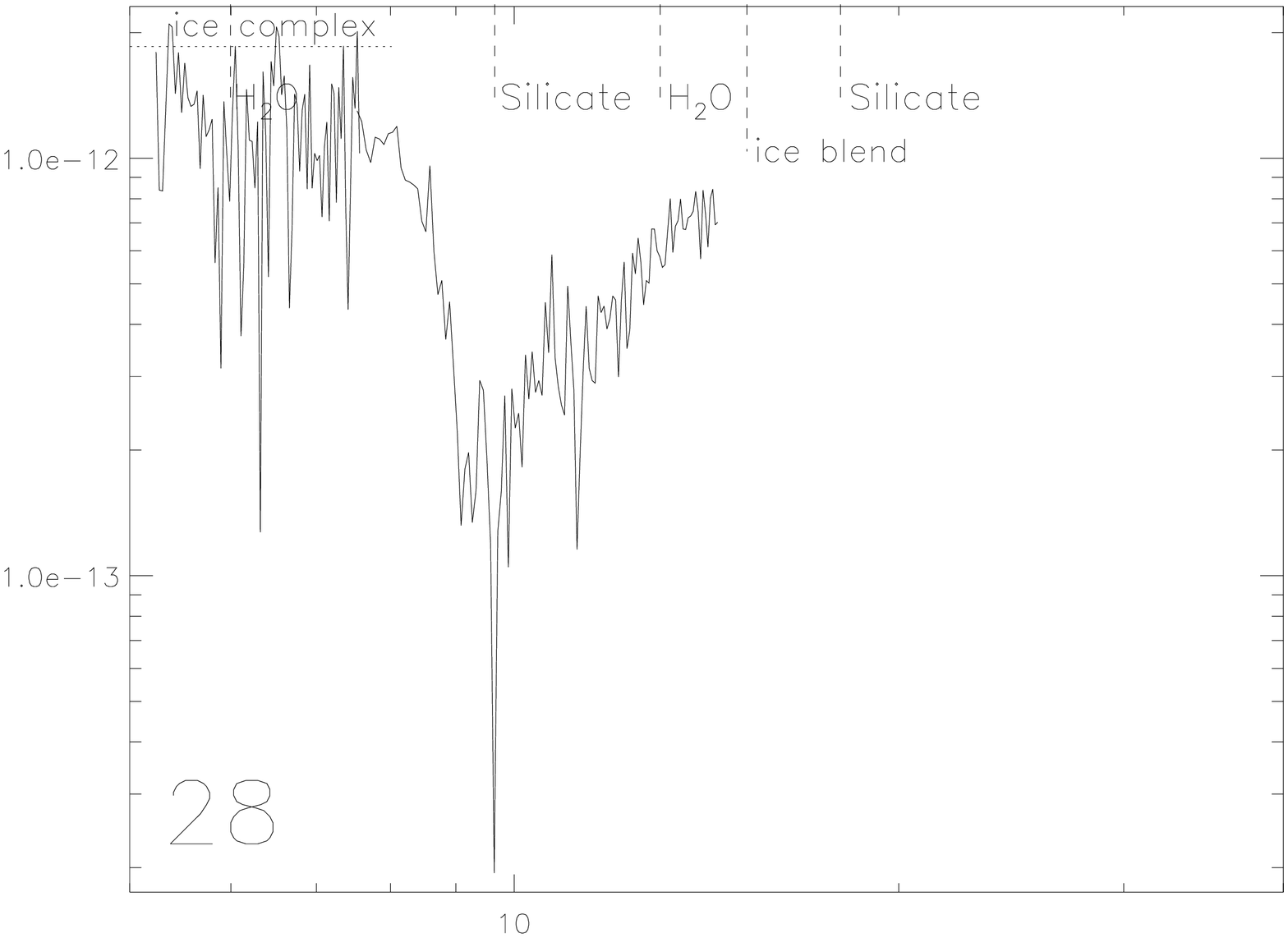}
\includegraphics*[width=1.6in, angle=90, bb = 55 36 577 705]{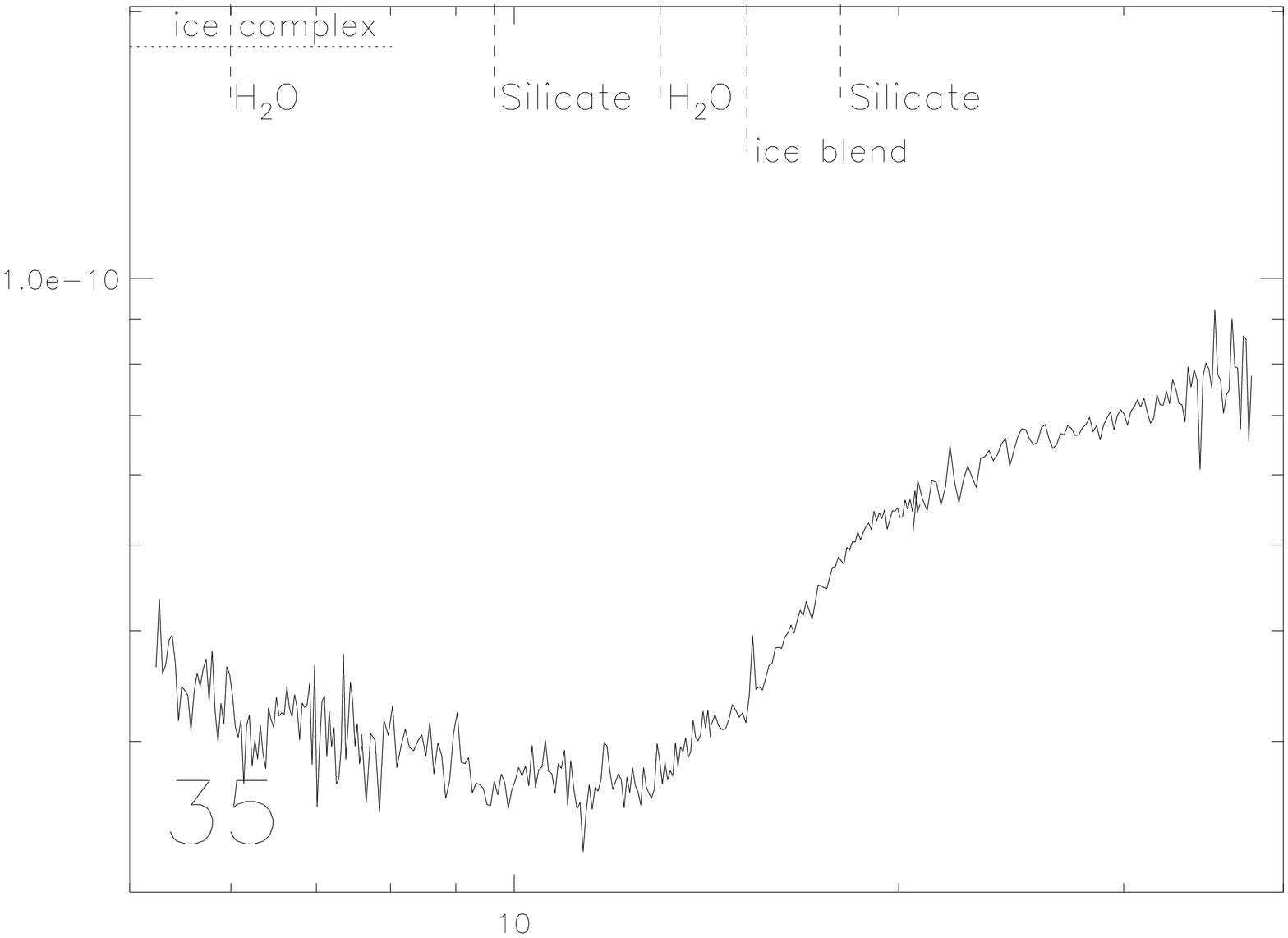}
\includegraphics*[width=1.6in, angle=90, bb = 55 36 577 705]{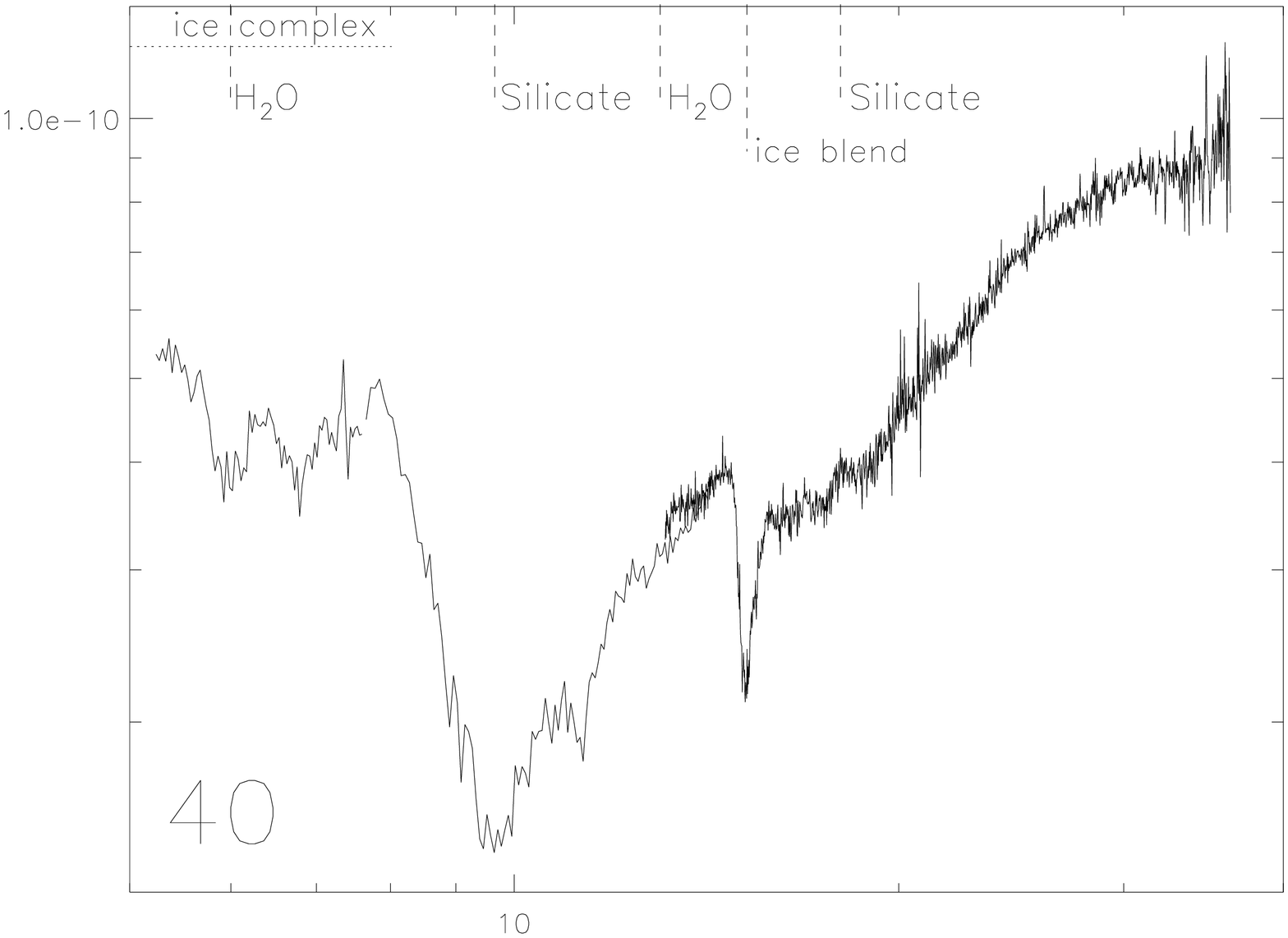}

\includegraphics*[width=1.6in, angle=90, bb = 55 36 577 705]{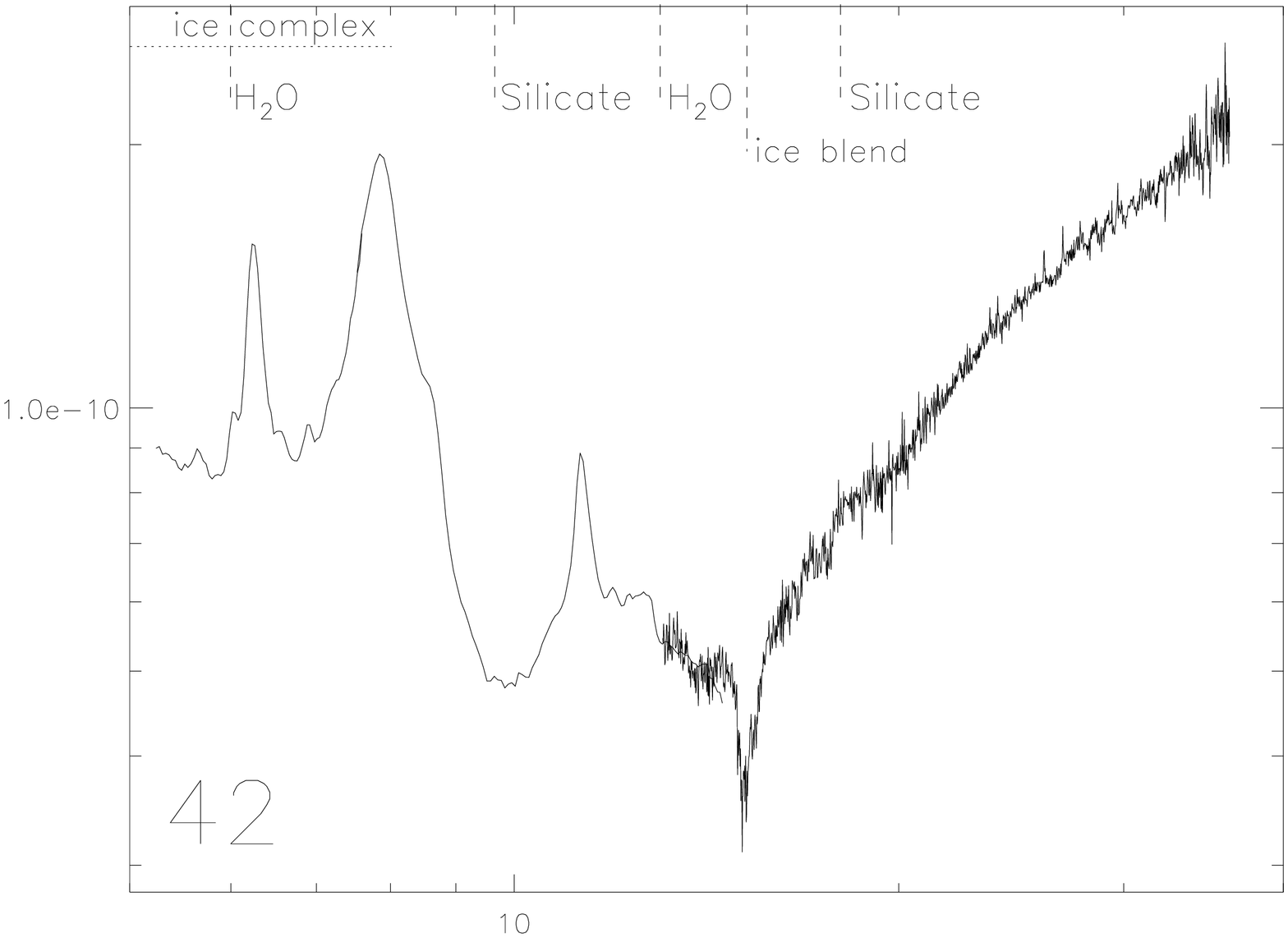}
\includegraphics*[width=1.6in, angle=90, bb = 55 36 577 705]{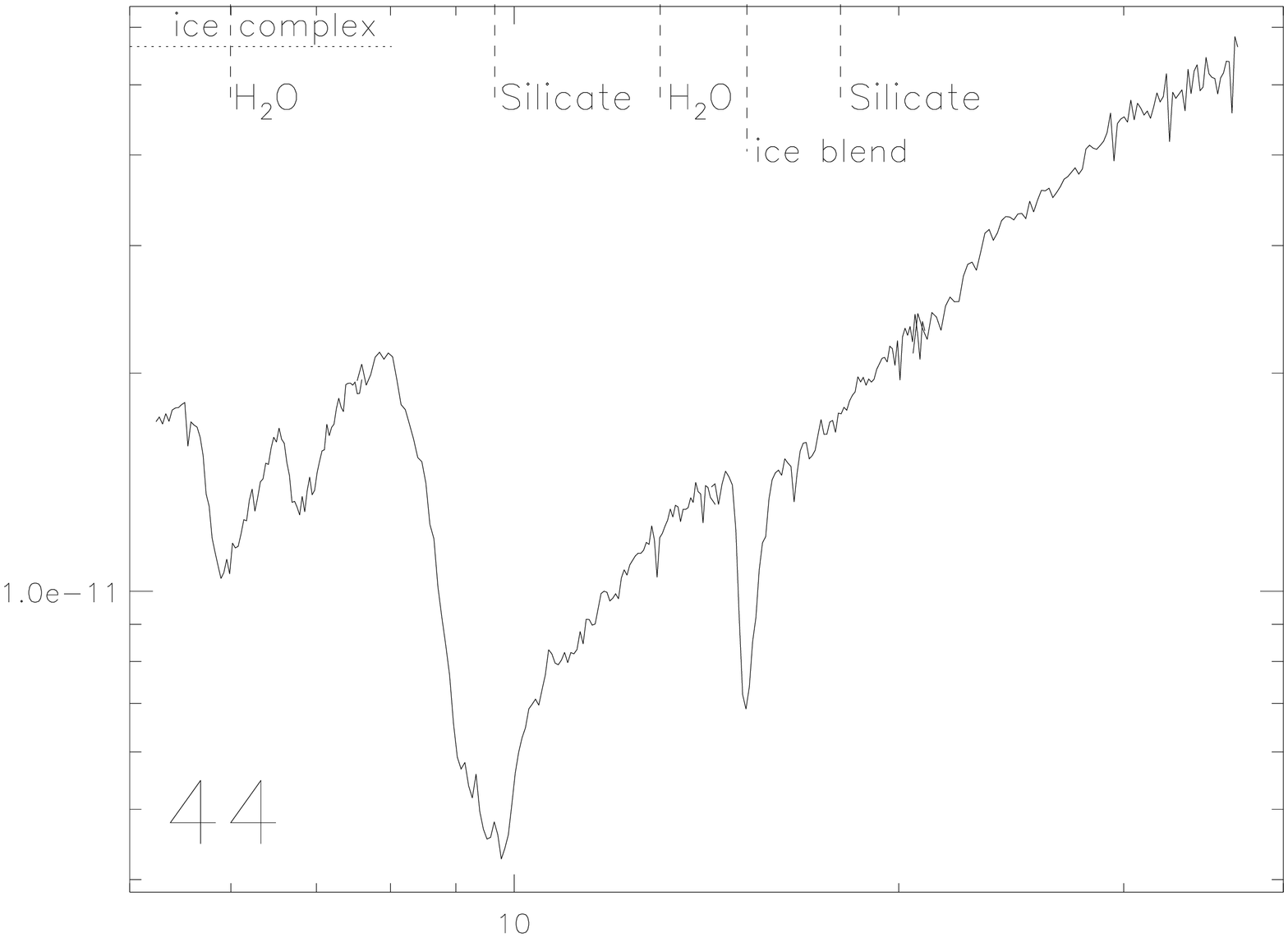}
\includegraphics*[width=1.6in, angle=90, bb = 55 36 577 705]{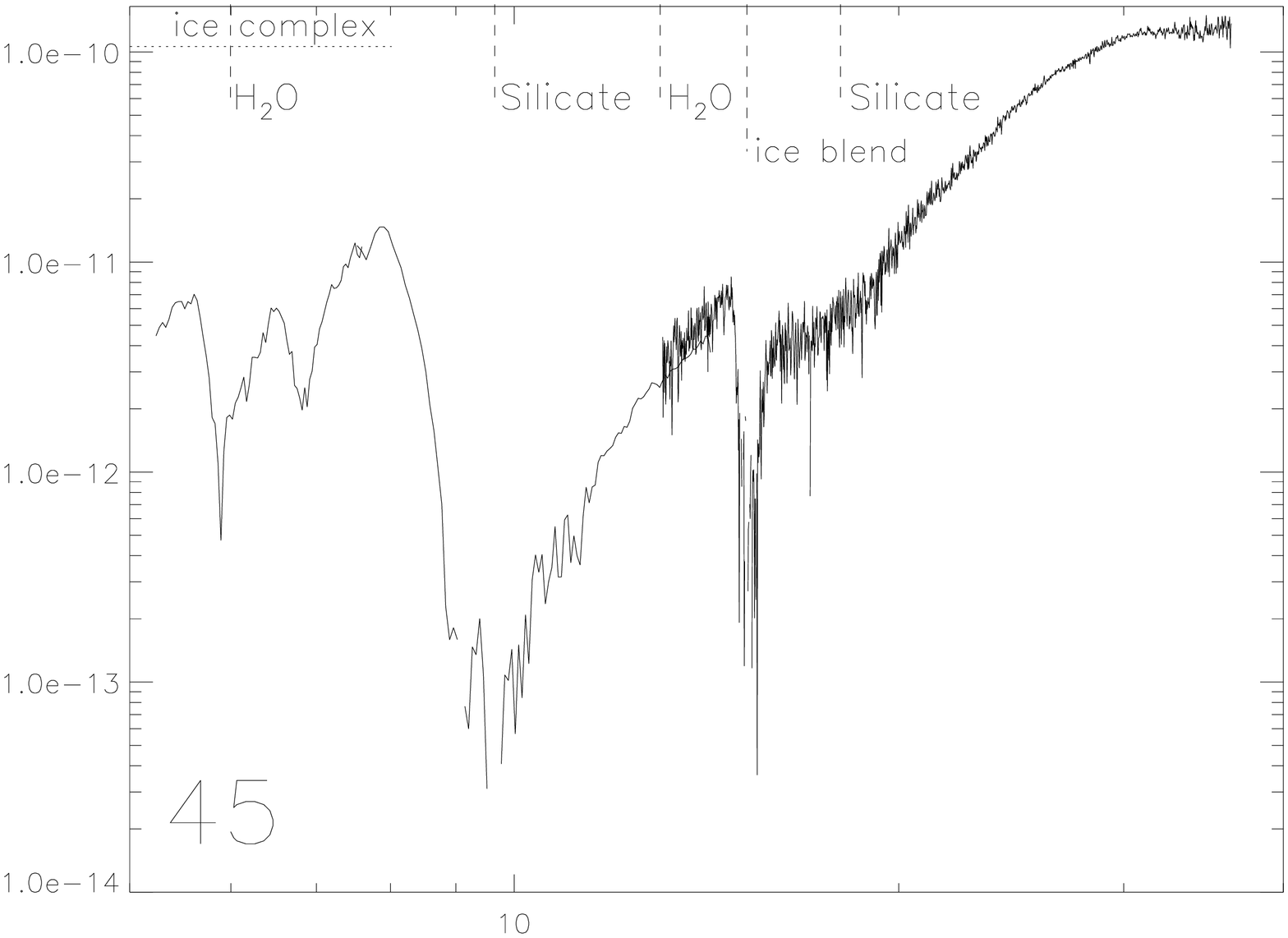}

\includegraphics*[width=1.6in, angle=90, bb = 55 36 577 705]{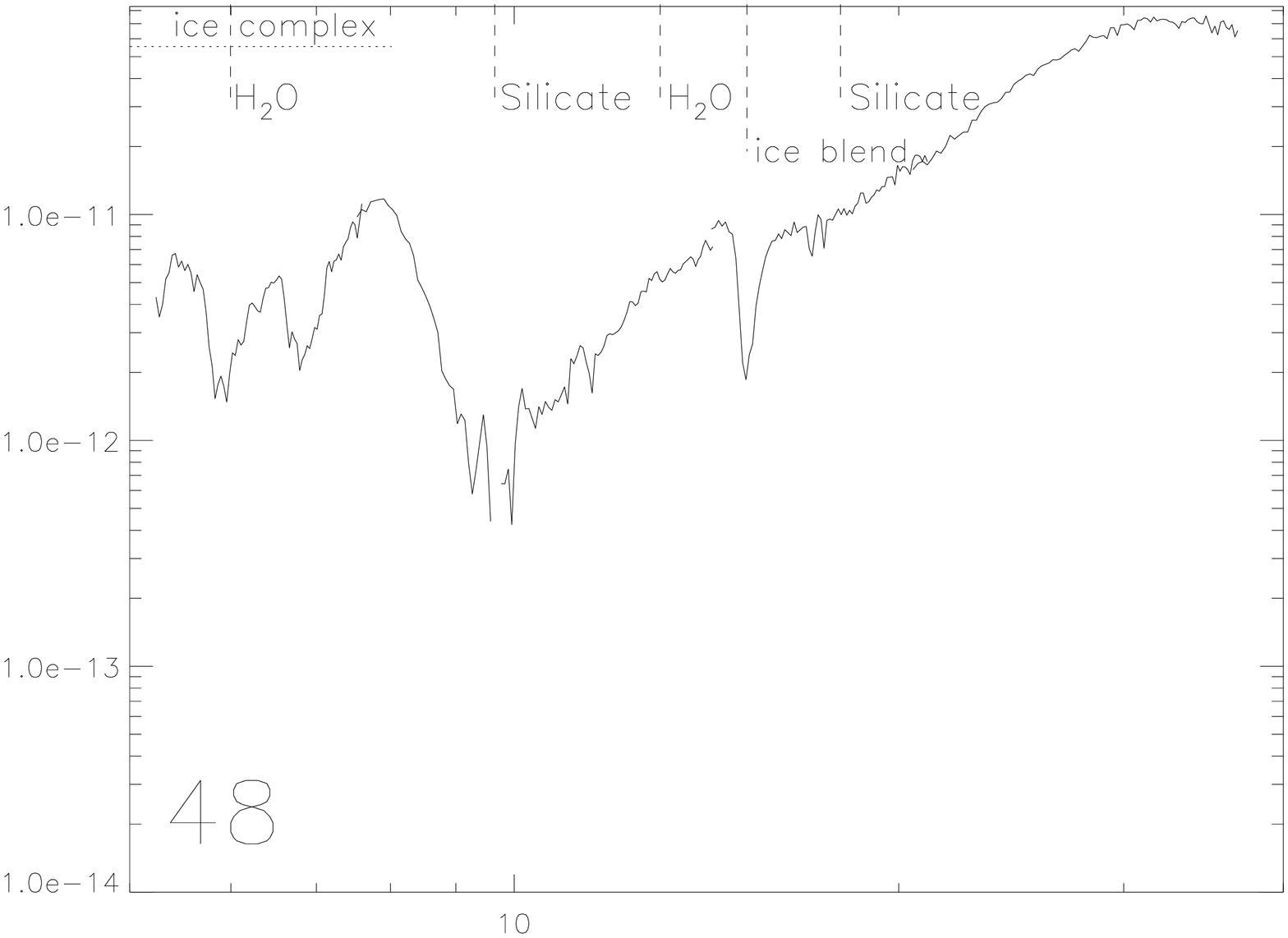}
\includegraphics*[width=1.6in, angle=90, bb = 55 36 577 705]{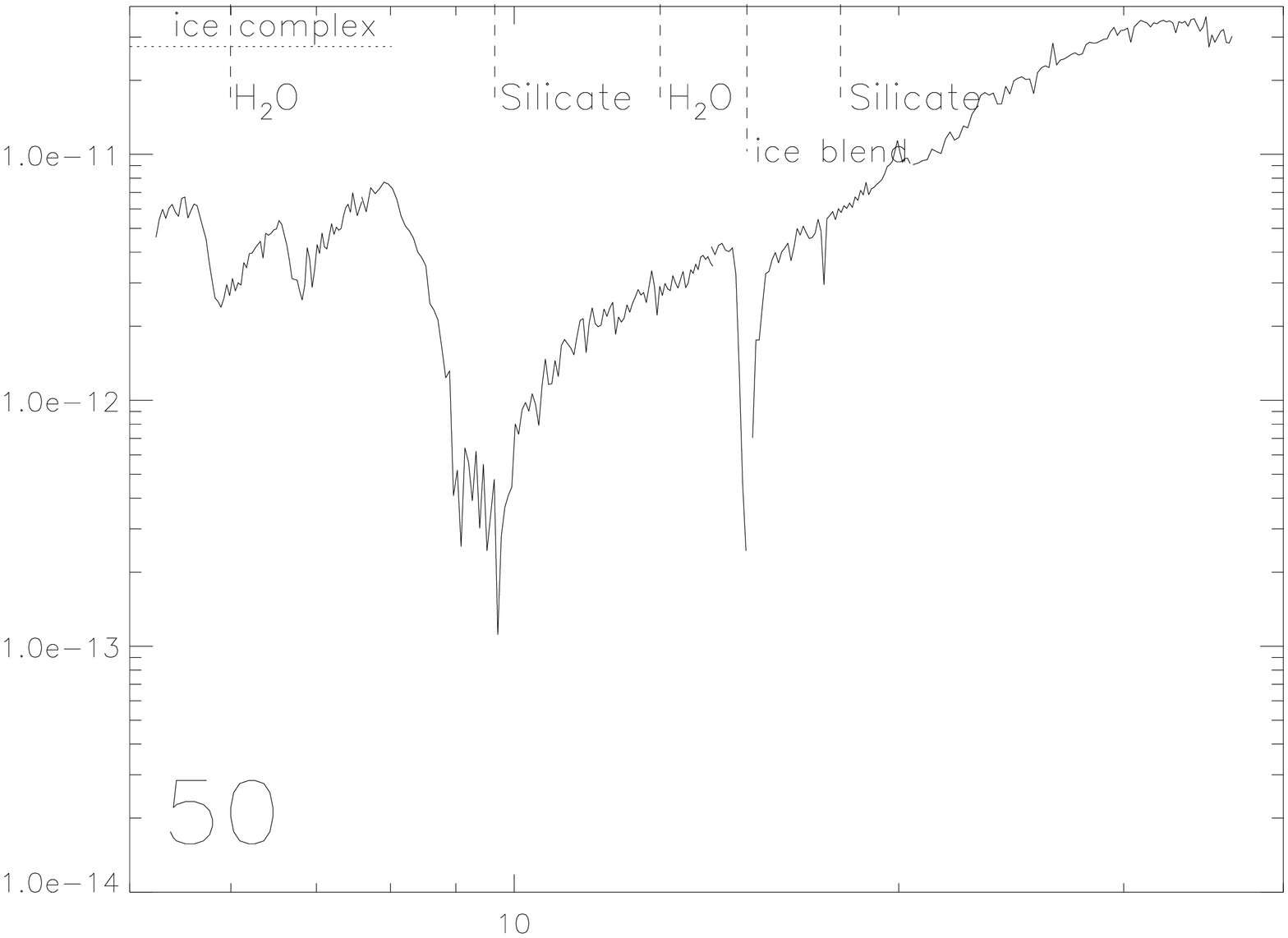}
\includegraphics*[width=1.6in, angle=90, bb = 55 36 577 705]{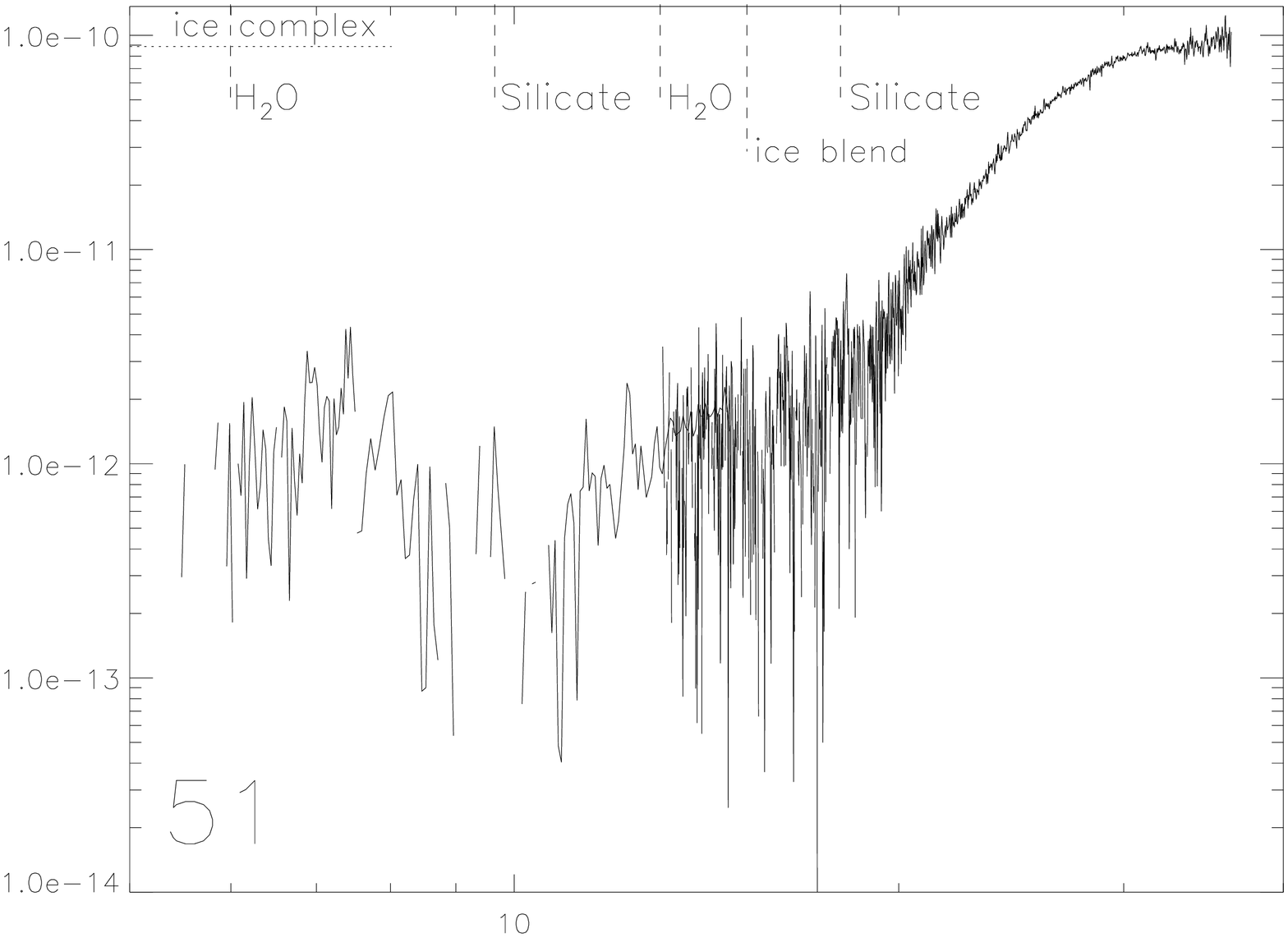}
\caption{IRS spectra of the 17 identified genuine protostars (stage~I sources), in $\lambda F_\lambda$, with orders of magnitude indicated (in erg\,s$^{-1}$\,cm$^{-2}$). \label{fig_irsplots1}}
%\end{center}
\end{figure*}

\addtocounter{figure}{-1}

\begin{figure*}
\includegraphics*[width=1.6in, angle=90, bb = 55 36 577 705]{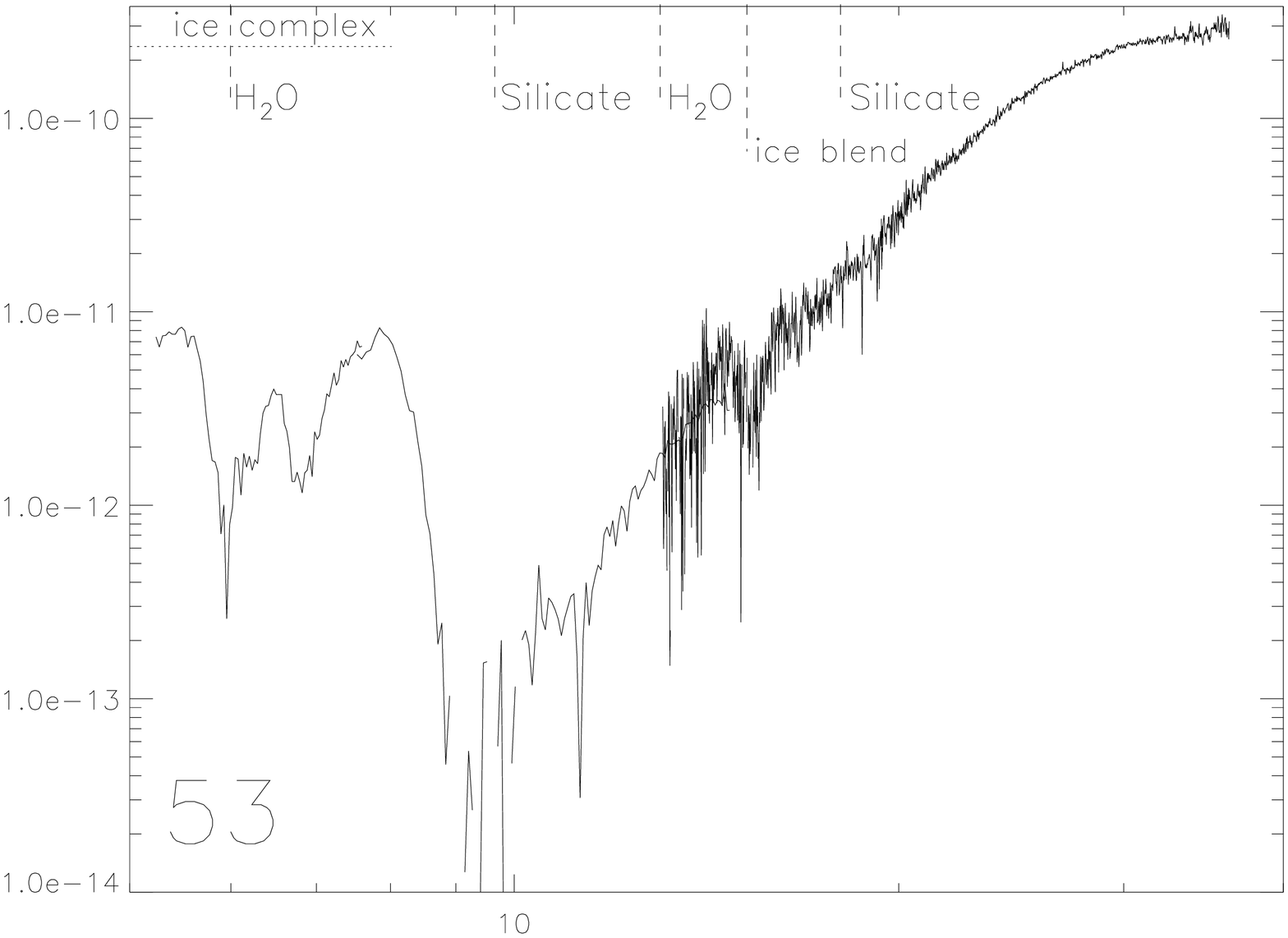}
\includegraphics*[width=1.6in, angle=90, bb = 55 36 577 705]{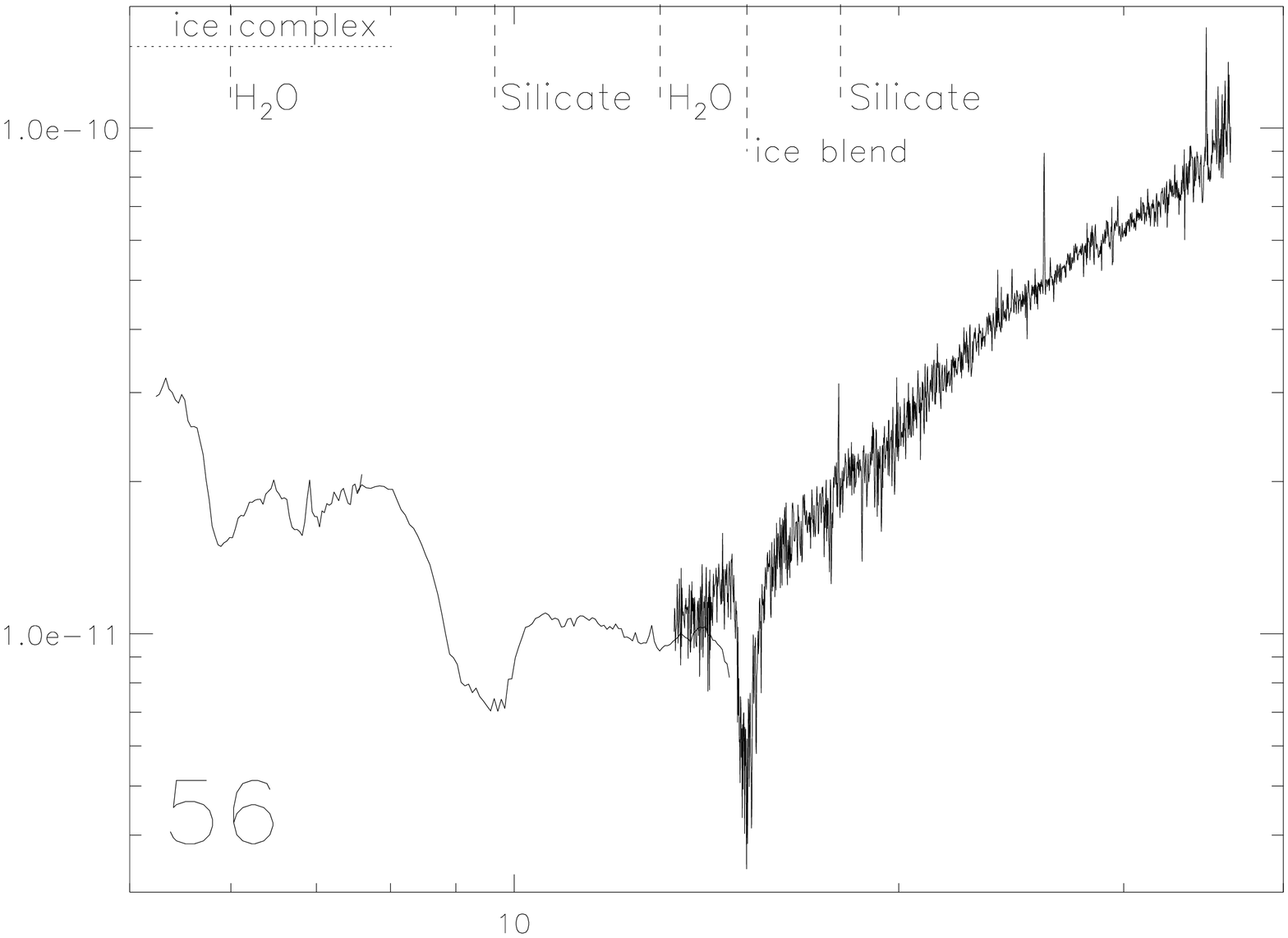}
\caption{continued}
%\end{center}
\end{figure*}

\begin{figure*}
%\begin{center}
\includegraphics*[width=1.6in, angle=90, bb = 55 36 577 705]{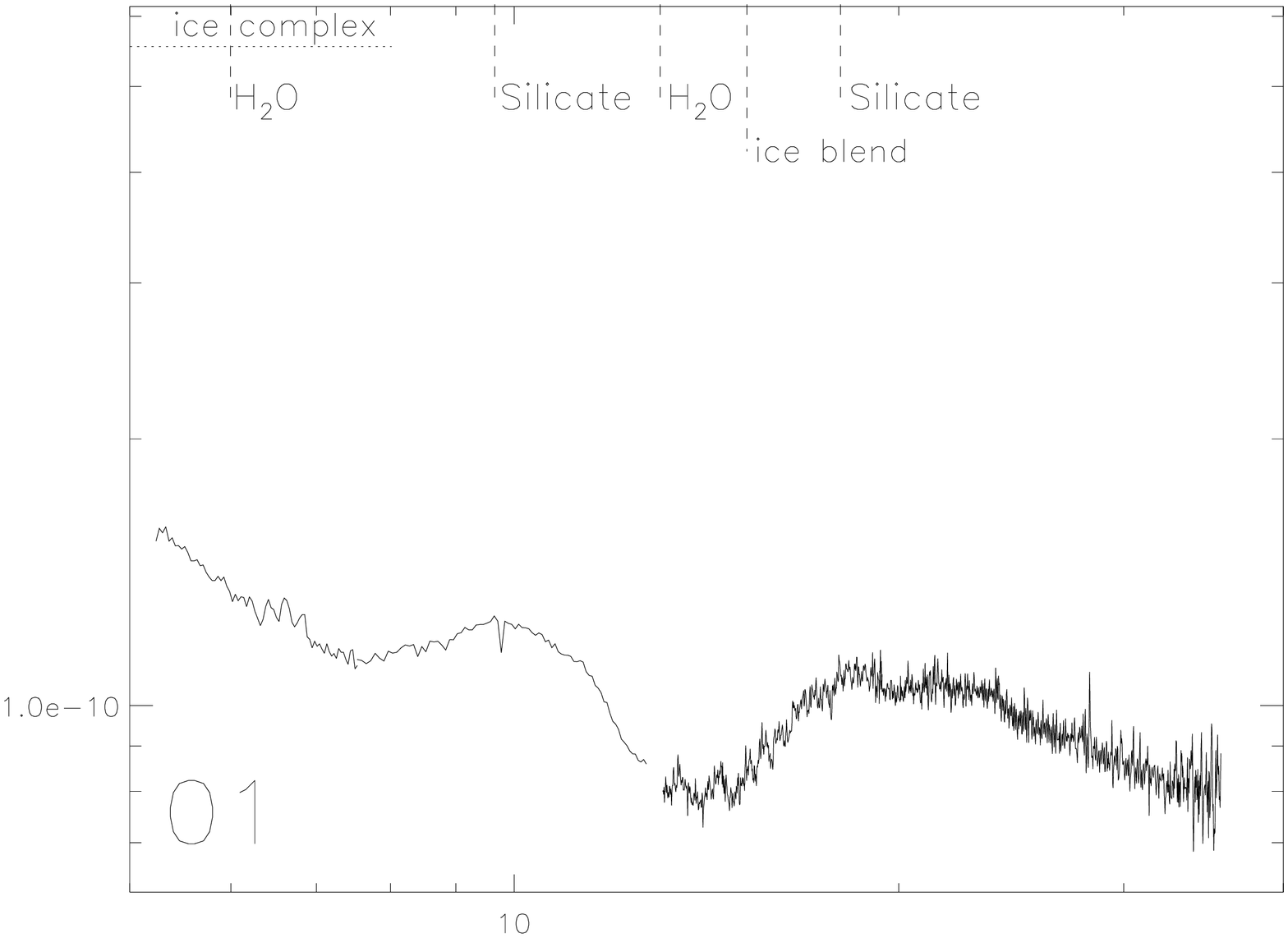}
\includegraphics*[width=1.6in, angle=90, bb = 55 36 577 705]{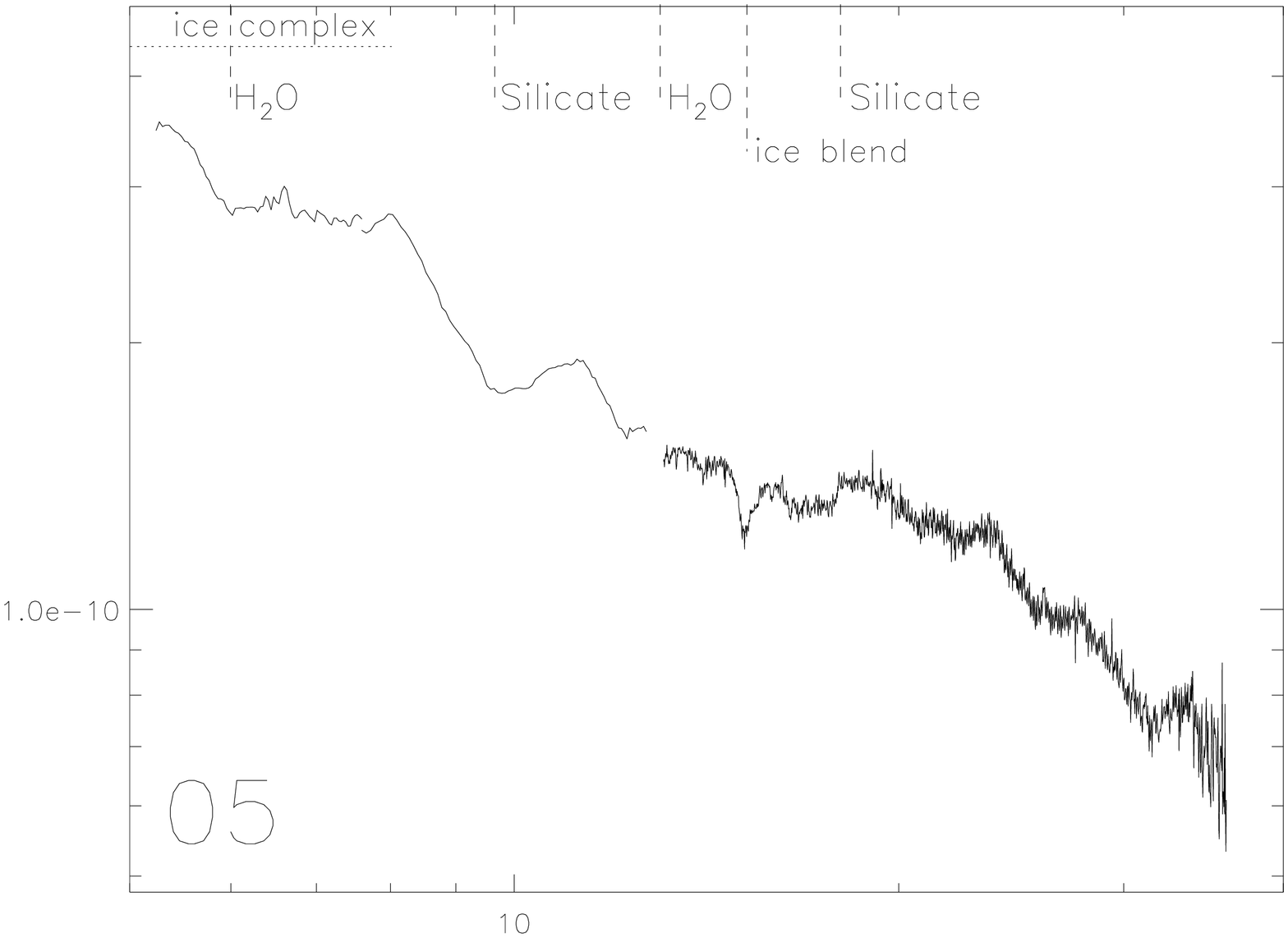}
\includegraphics*[width=1.6in, angle=90, bb = 55 36 577 705]{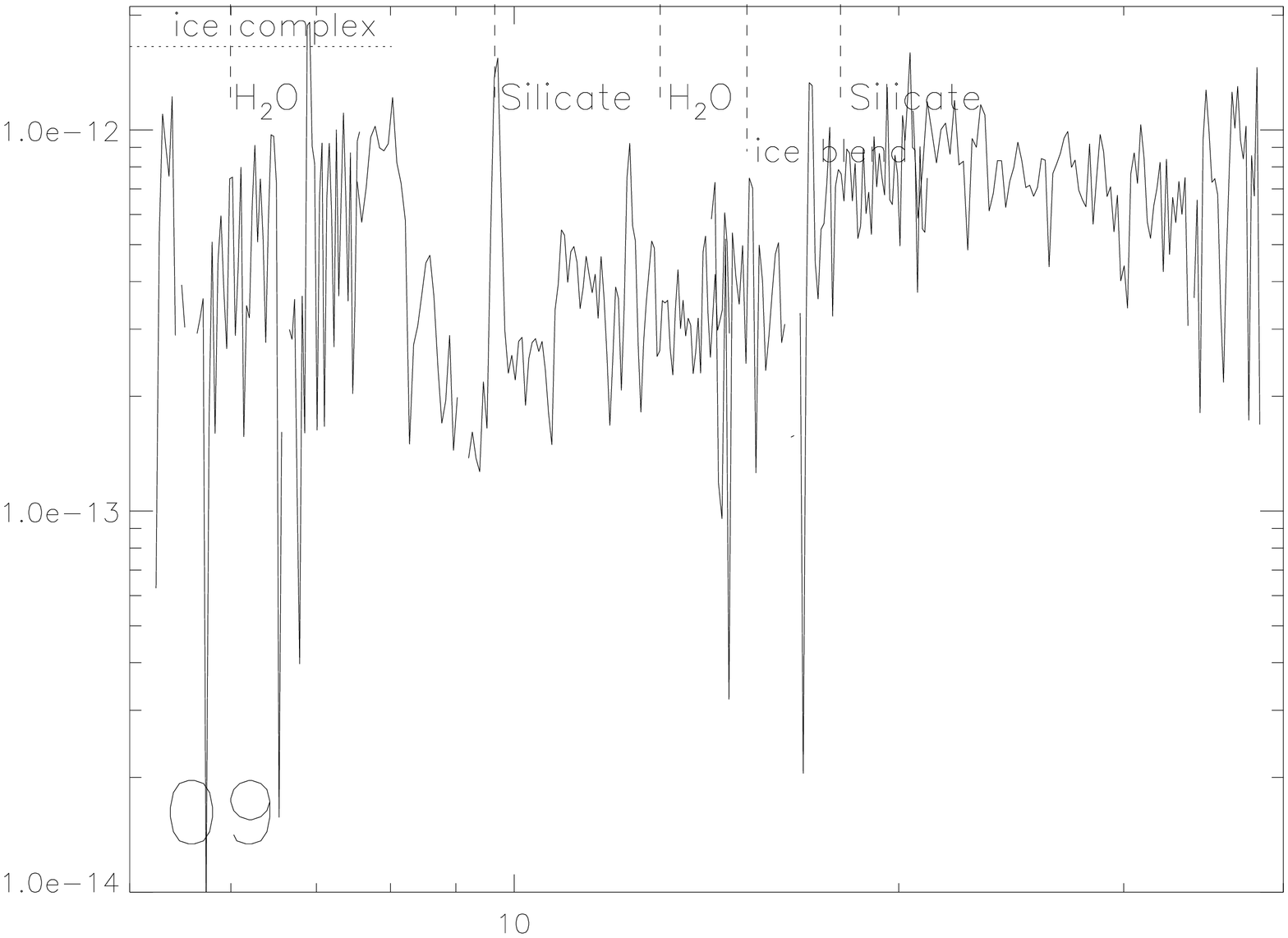}

\includegraphics*[width=1.6in, angle=90, bb = 55 36 577 705]{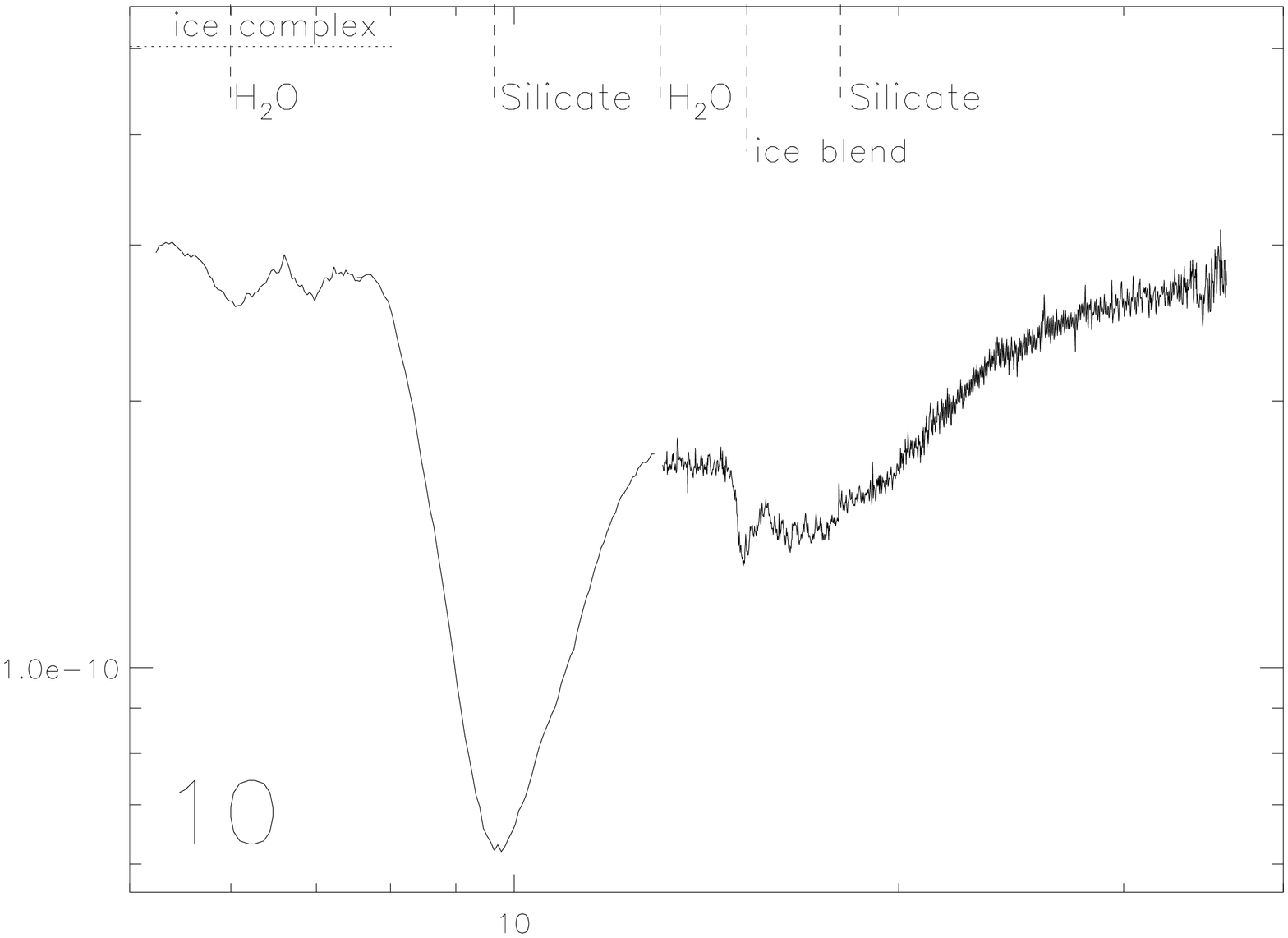}
\includegraphics*[width=1.6in, angle=90, bb = 55 36 577 705]{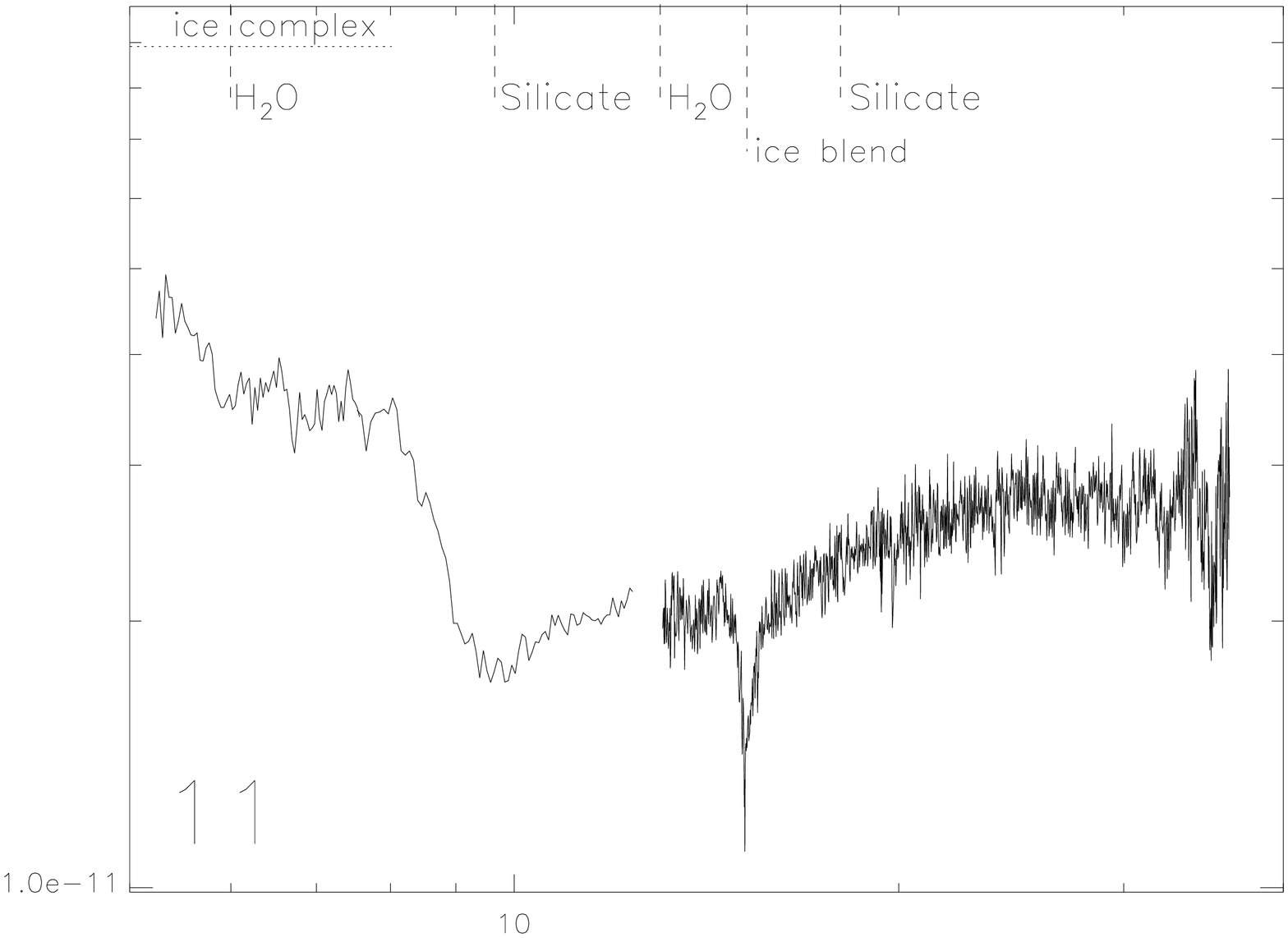}
\includegraphics*[width=1.6in, angle=90, bb = 55 36 577 705]{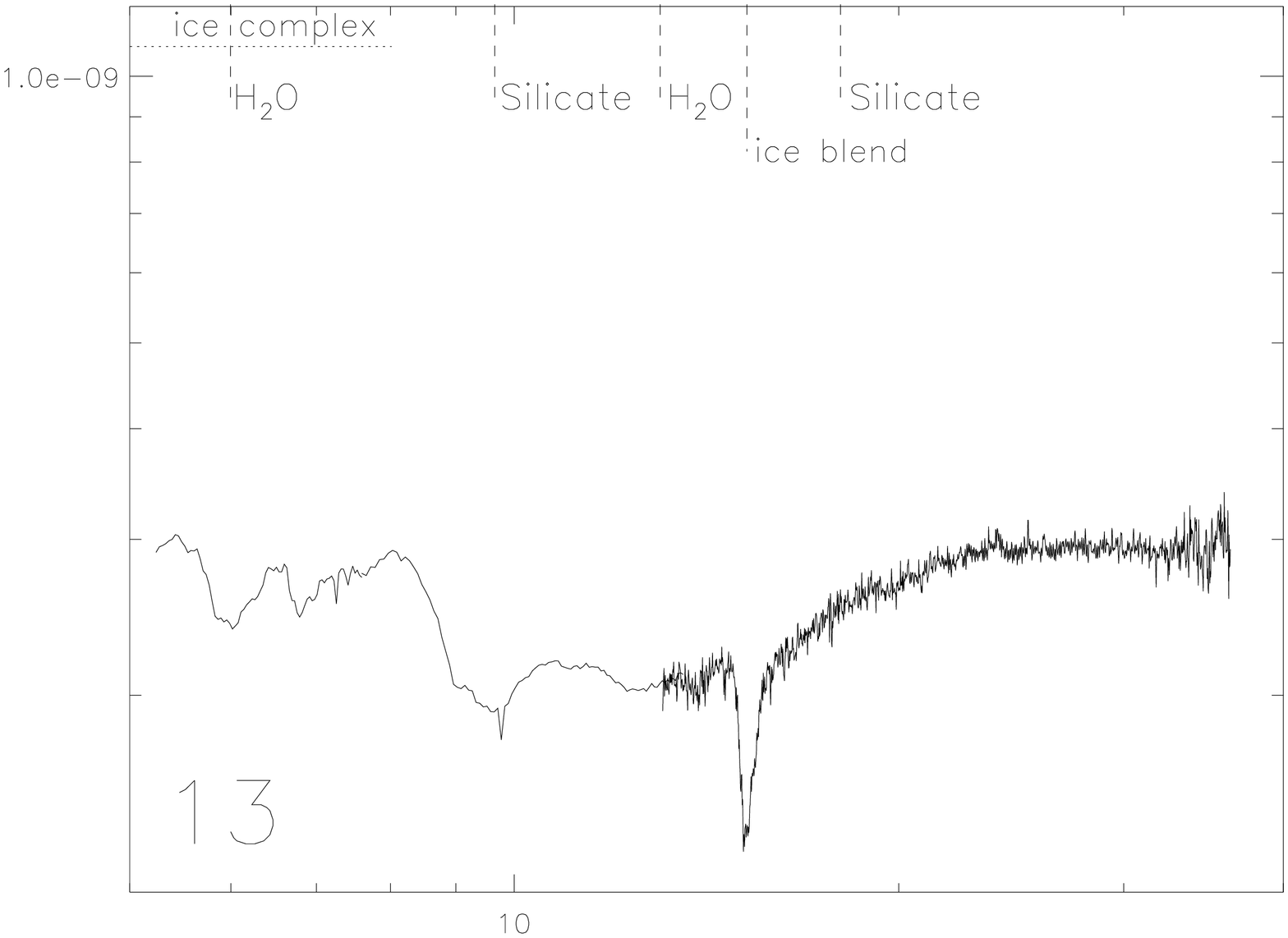}

\includegraphics*[width=1.6in, angle=90, bb = 55 36 577 705]{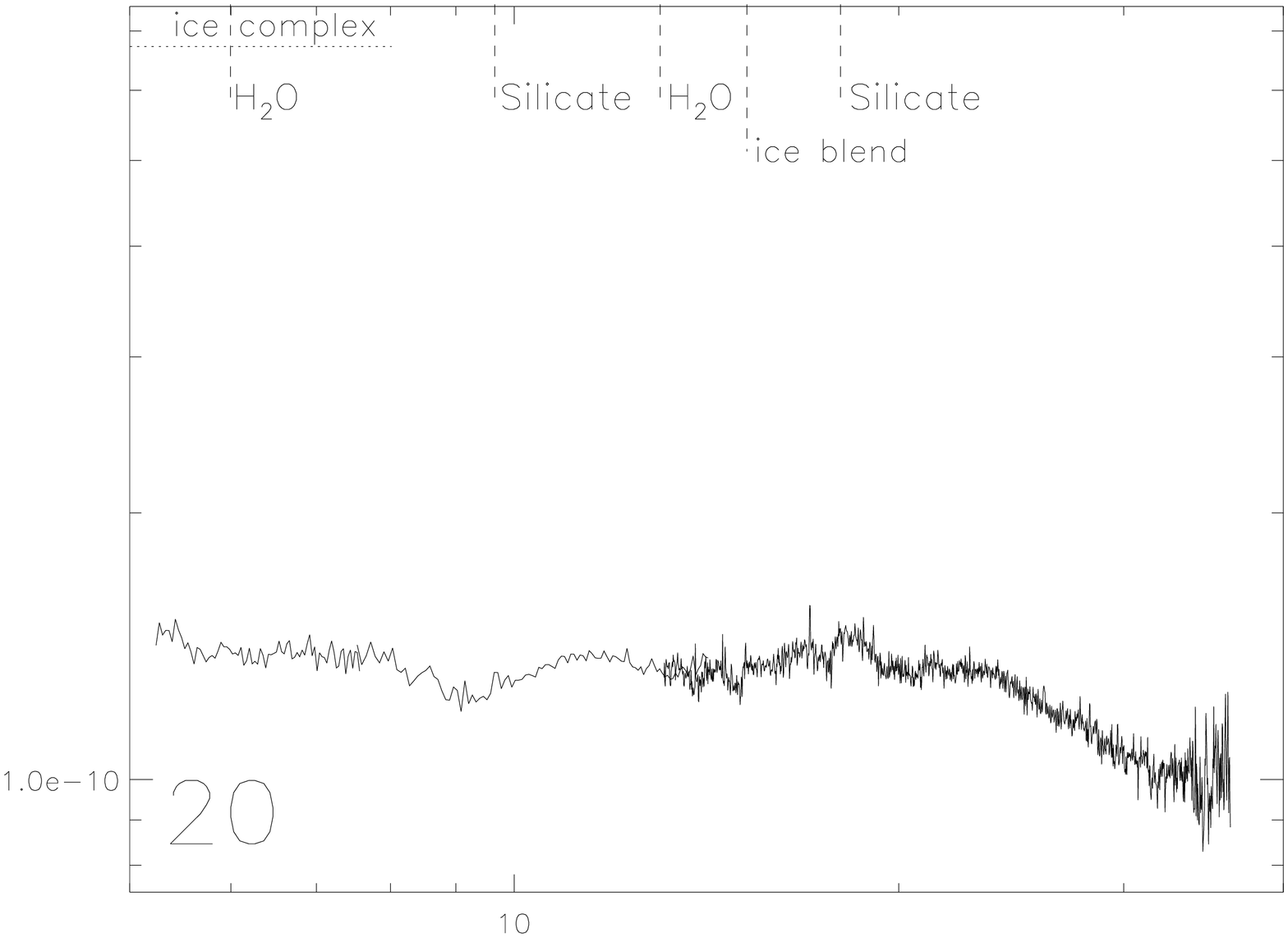}
\includegraphics*[width=1.6in, angle=90, bb = 55 36 577 705]{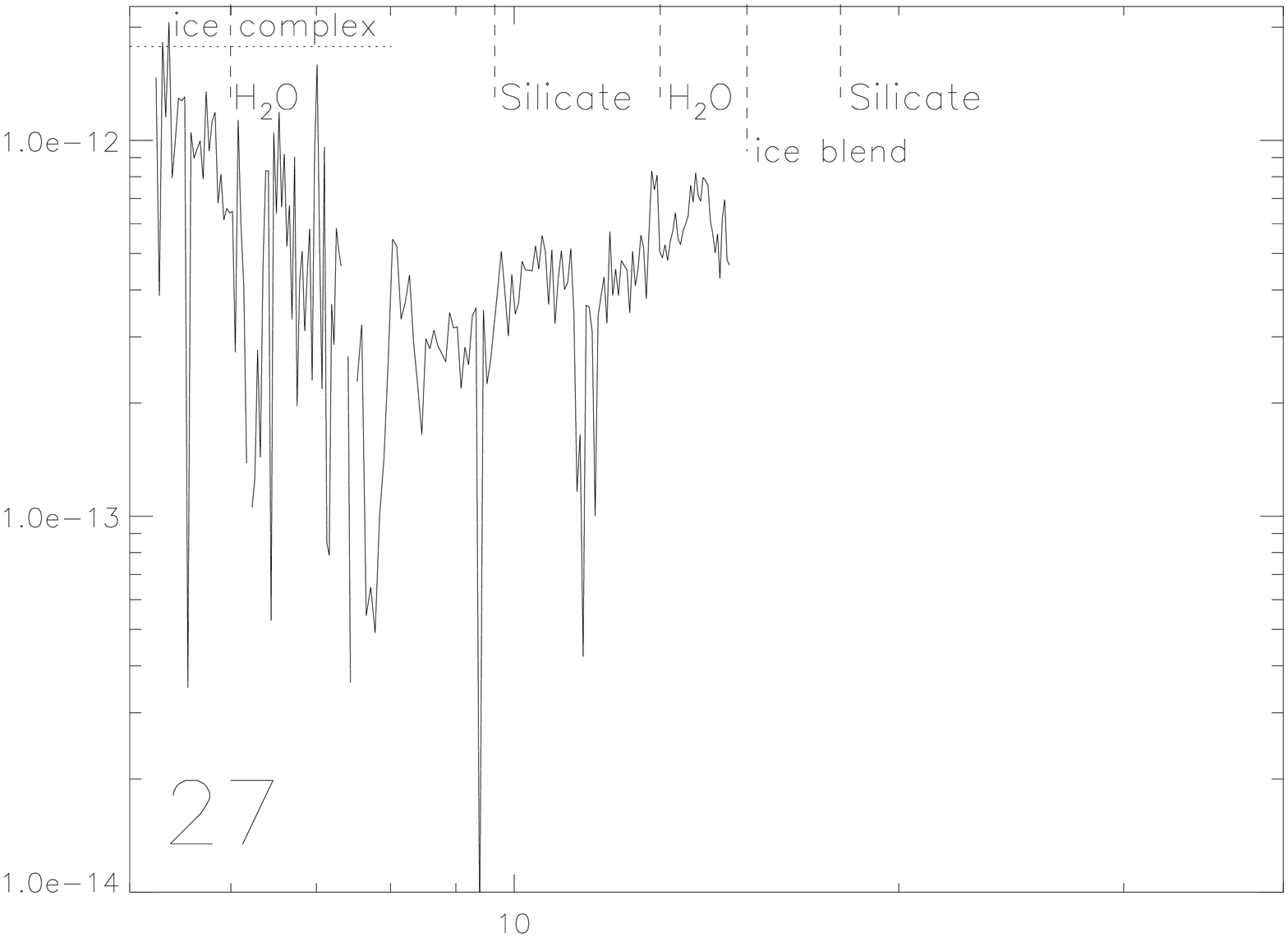}
\includegraphics*[width=1.6in, angle=90, bb = 55 36 577 705]{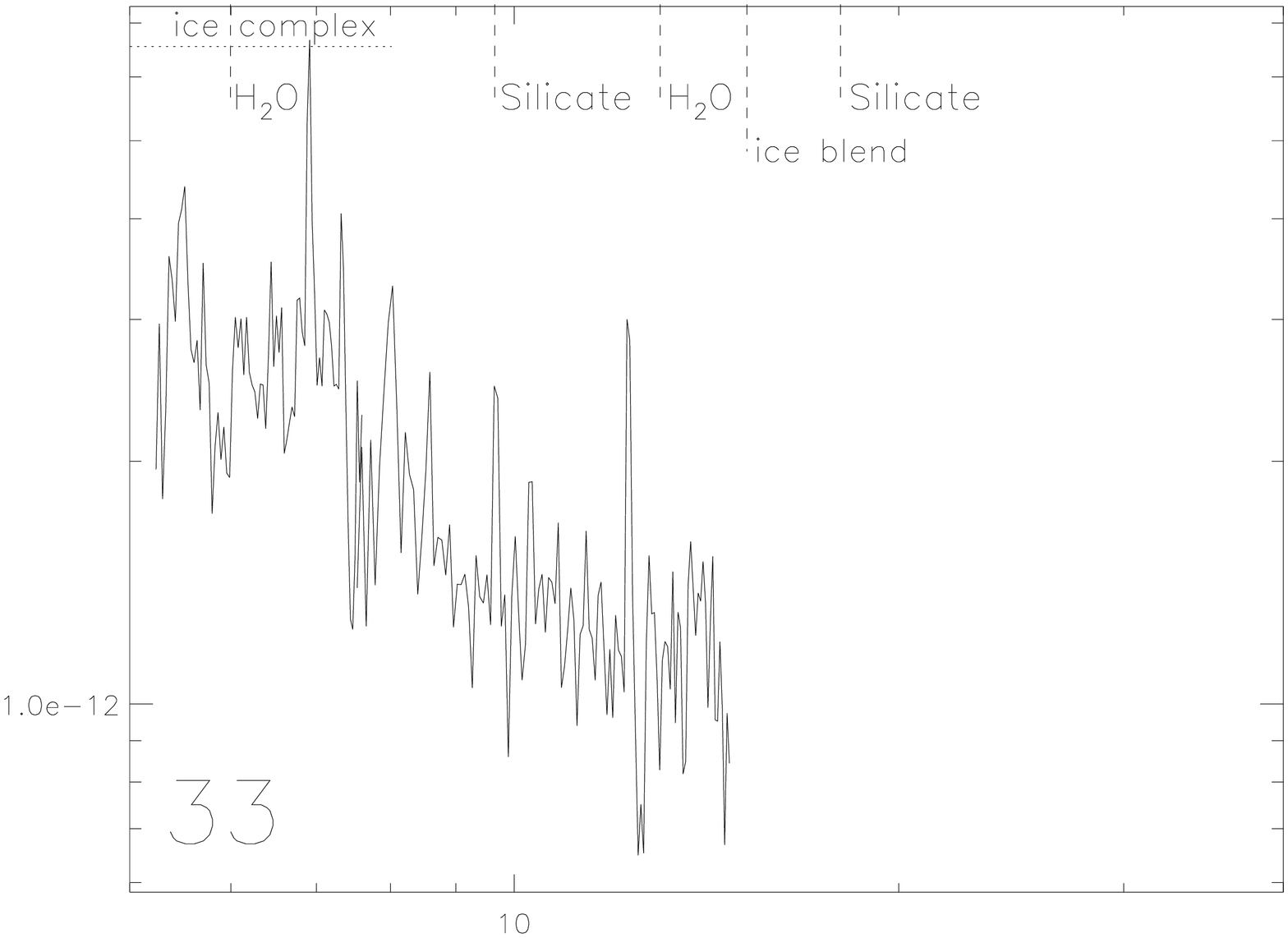}

\includegraphics*[width=1.6in, angle=90, bb = 55 36 577 705]{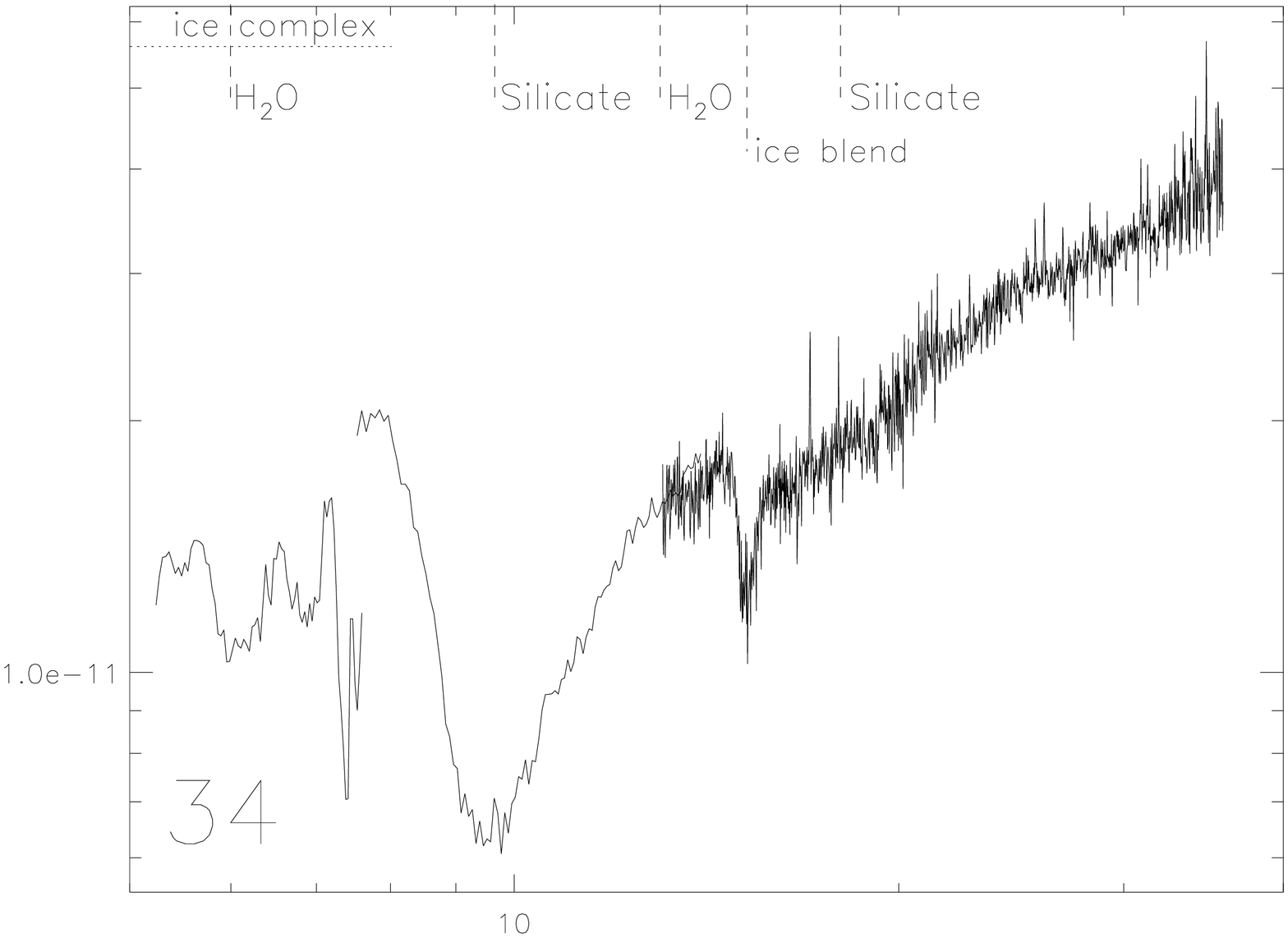}
\includegraphics*[width=1.6in, angle=90, bb = 55 36 577 705]{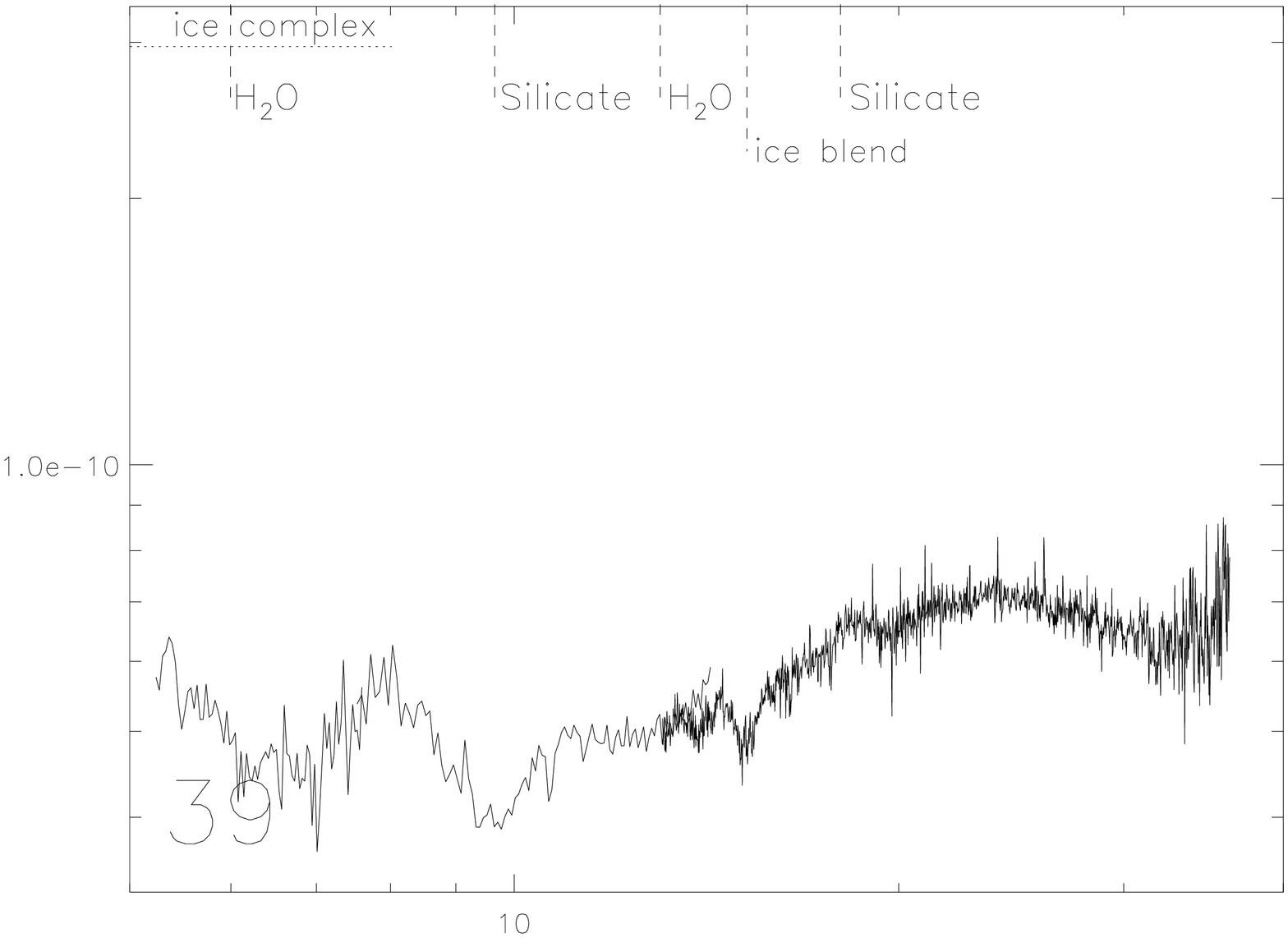}
\includegraphics*[width=1.6in, angle=90, bb = 55 36 577 705]{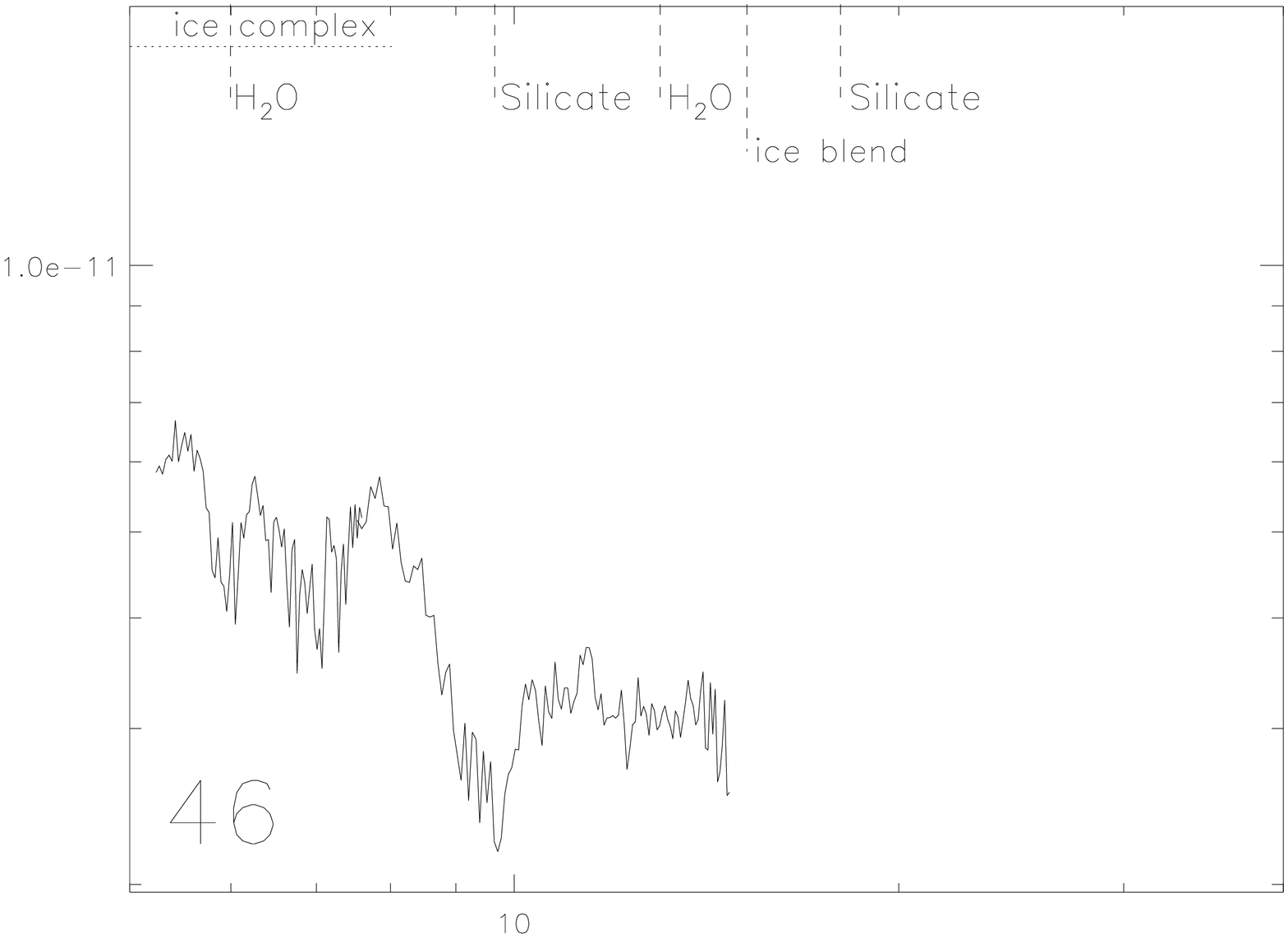}

\includegraphics*[width=1.6in, angle=90, bb = 55 36 577 705]{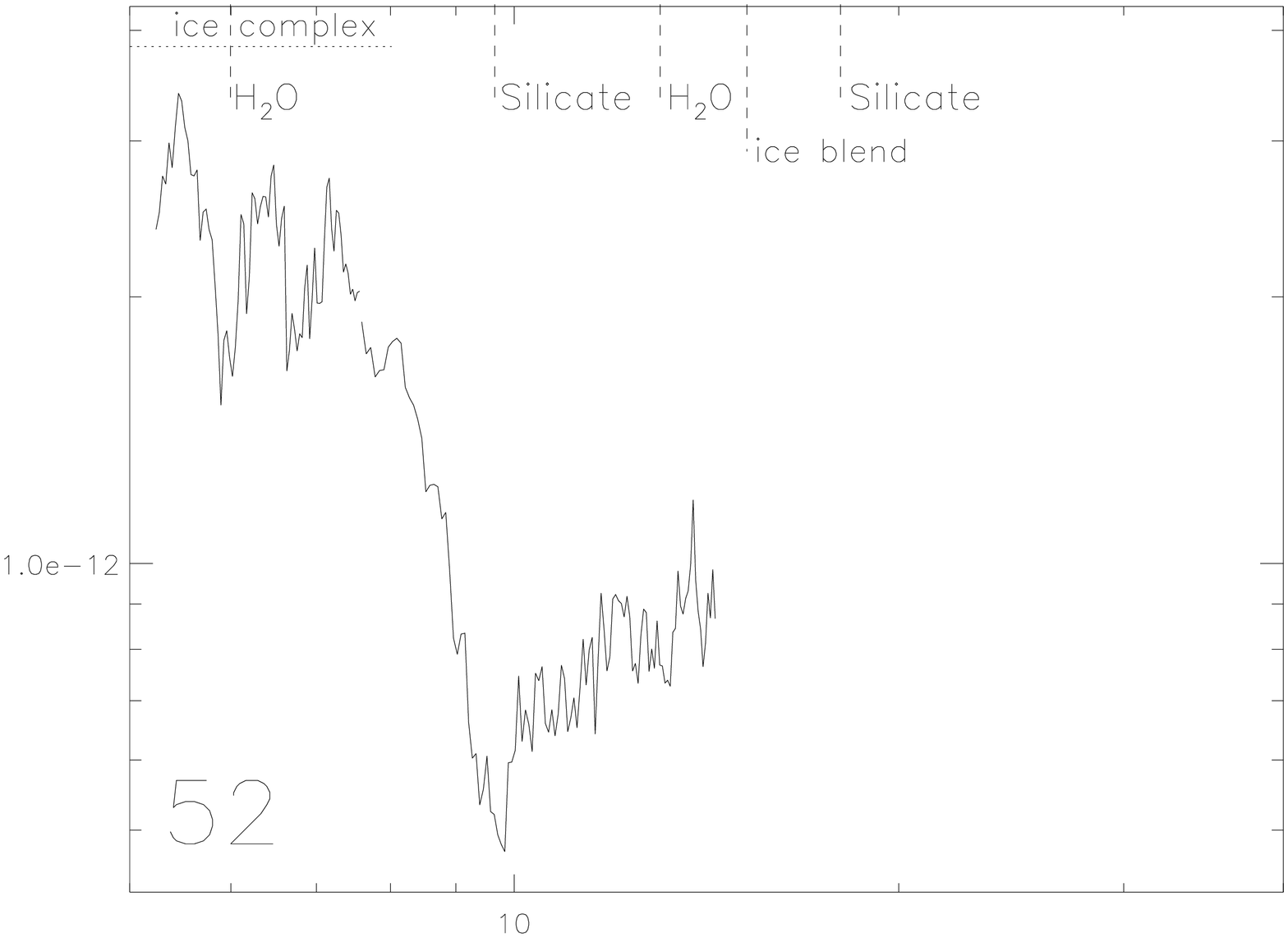}
\includegraphics*[width=1.6in, angle=90, bb = 55 36 577 705]{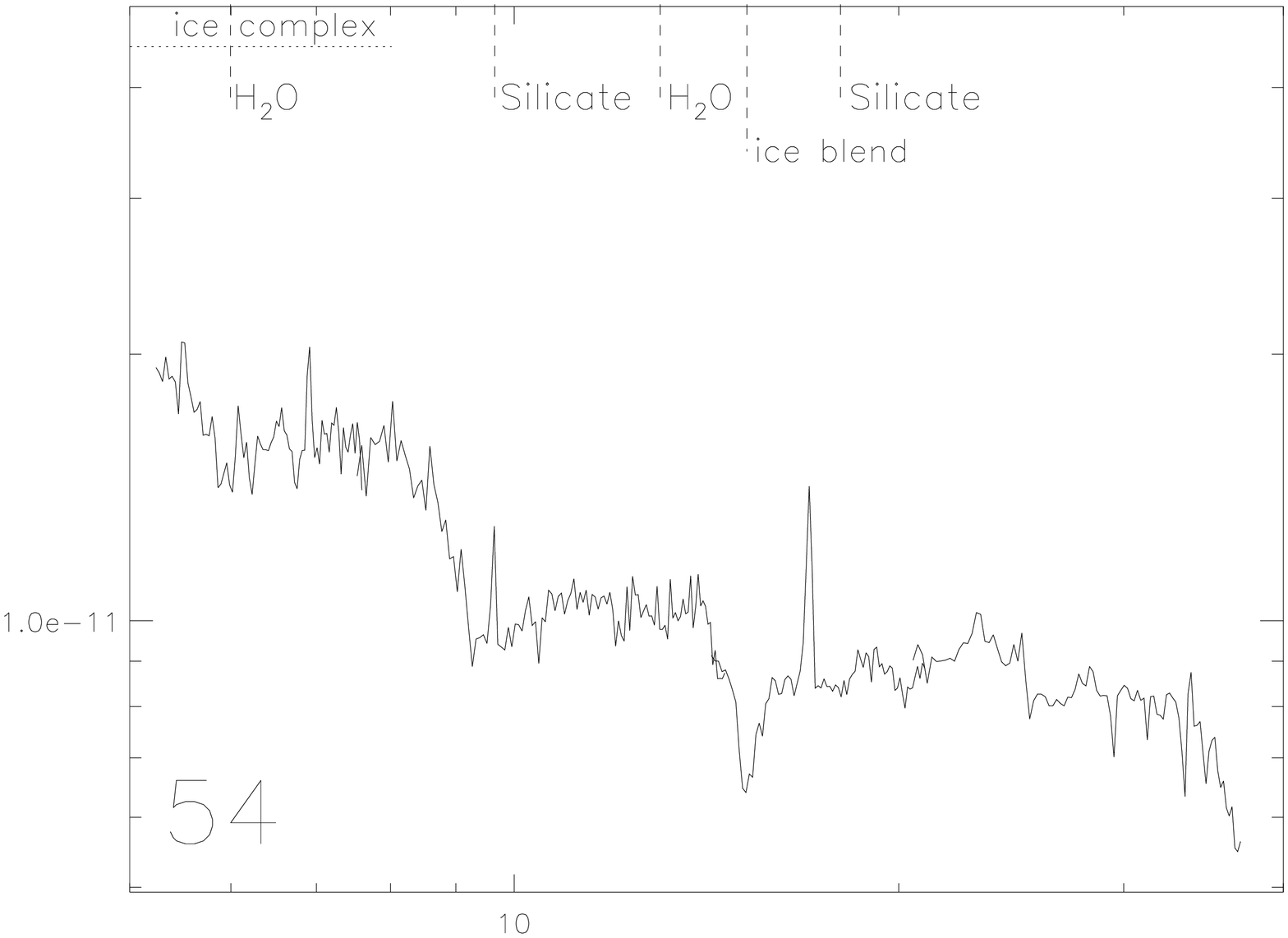}
\includegraphics*[width=1.6in, angle=90, bb = 55 36 577 705]{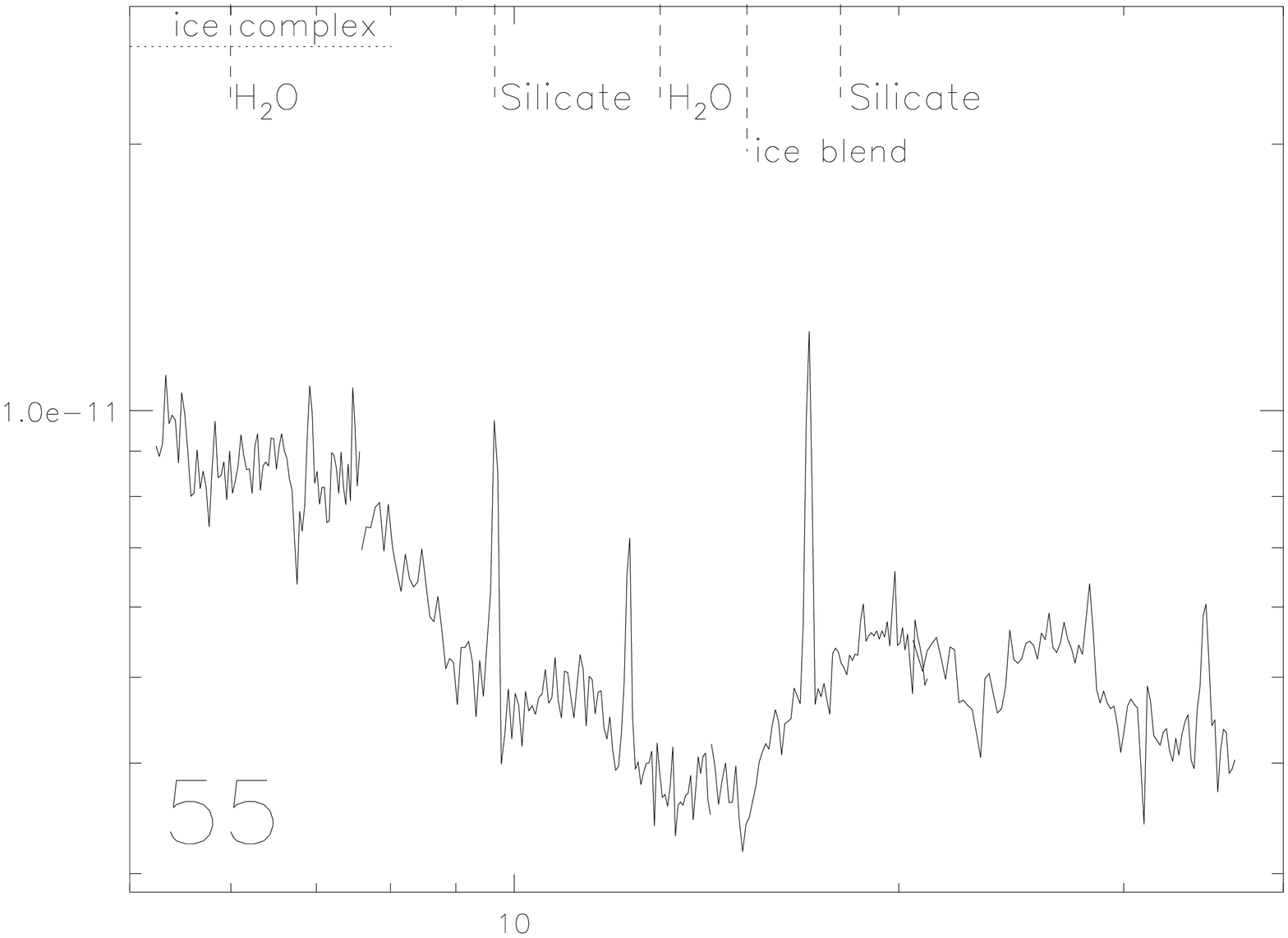}
%{Fig.~\ref{fig_irsplots} --  cntd.}
\caption{IRS spectra of the 15 identified disk sources with high external foreground extinction (stage II(ex) sources). \label{fig_irsplots2}}
%\end{center}
\end{figure*}

\begin{figure*}
\includegraphics*[width=1.6in, angle=90, bb = 55 36 577 705]{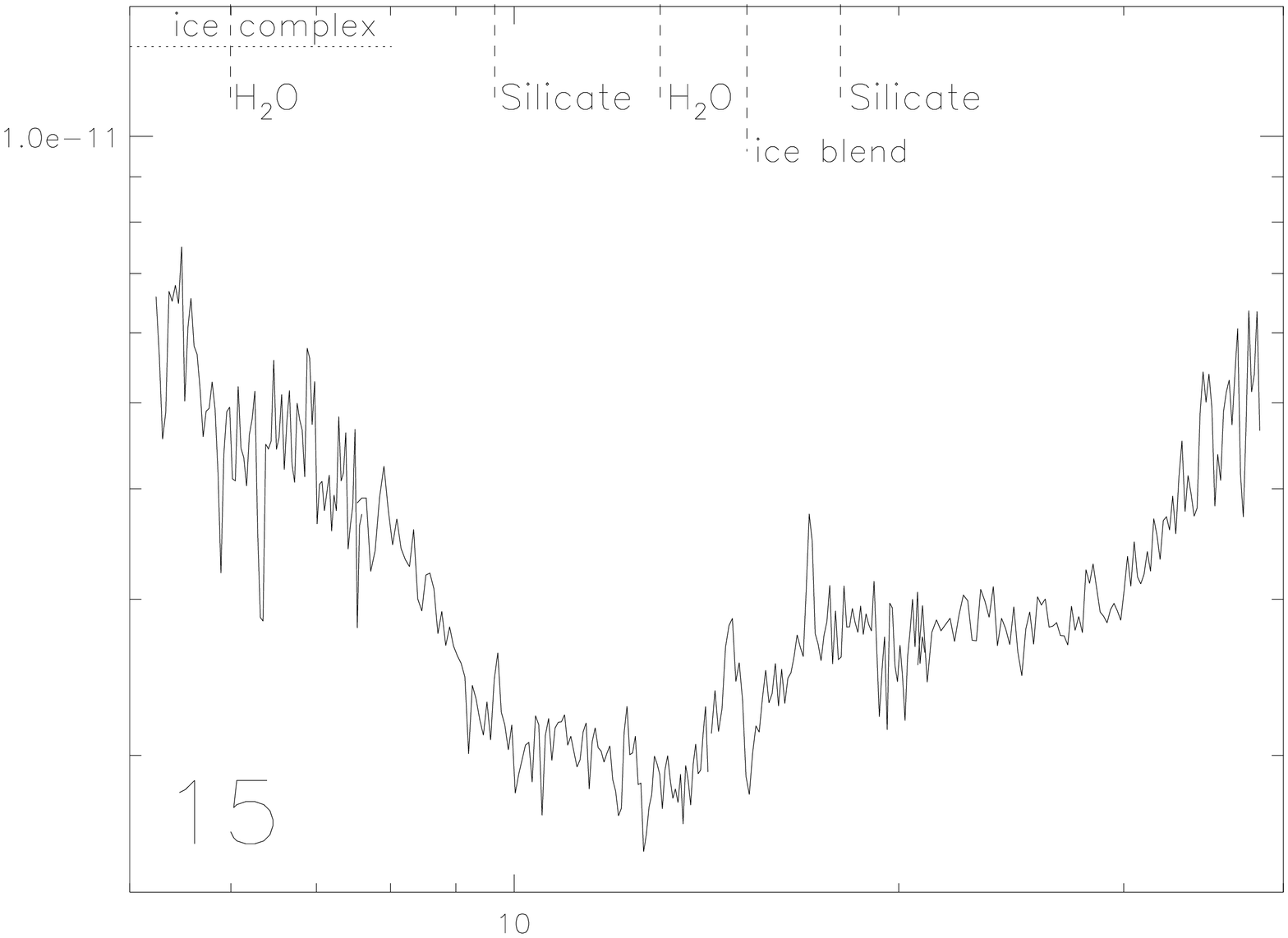}
\includegraphics*[width=1.6in, angle=90, bb = 55 36 577 705]{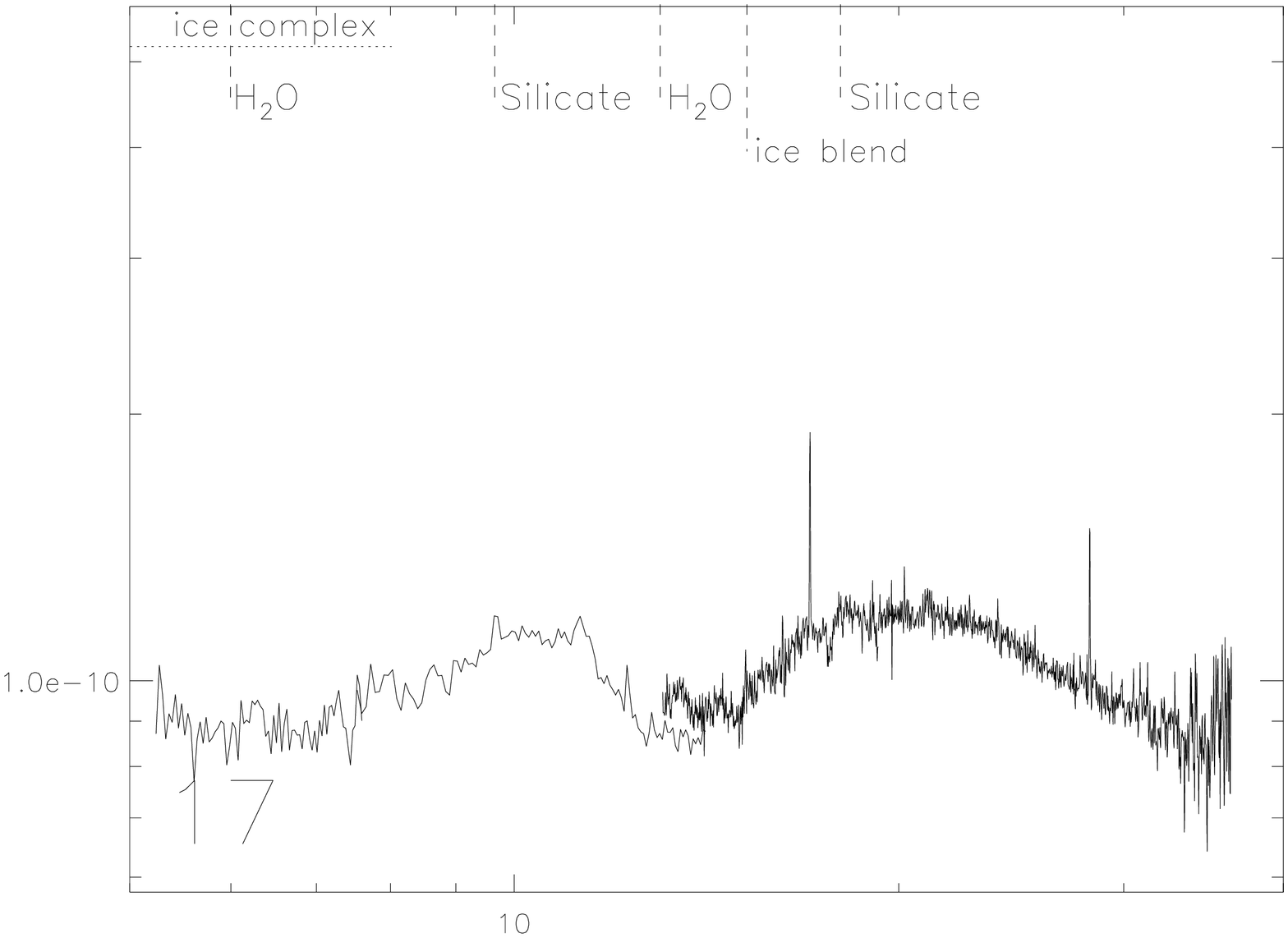}
\includegraphics*[width=1.6in, angle=90, bb = 55 36 577 705]{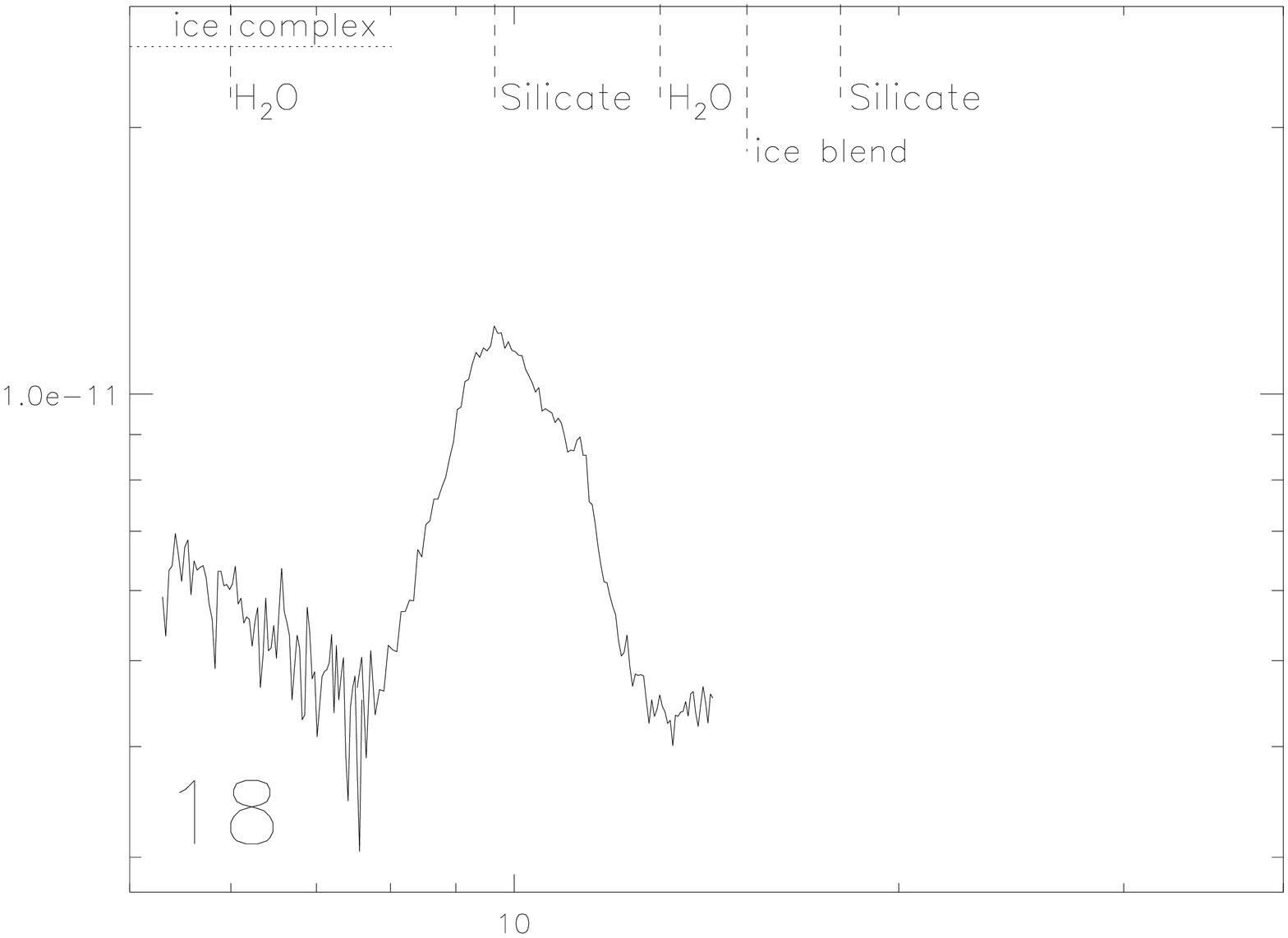}

\includegraphics*[width=1.6in, angle=90, bb = 55 36 577 705]{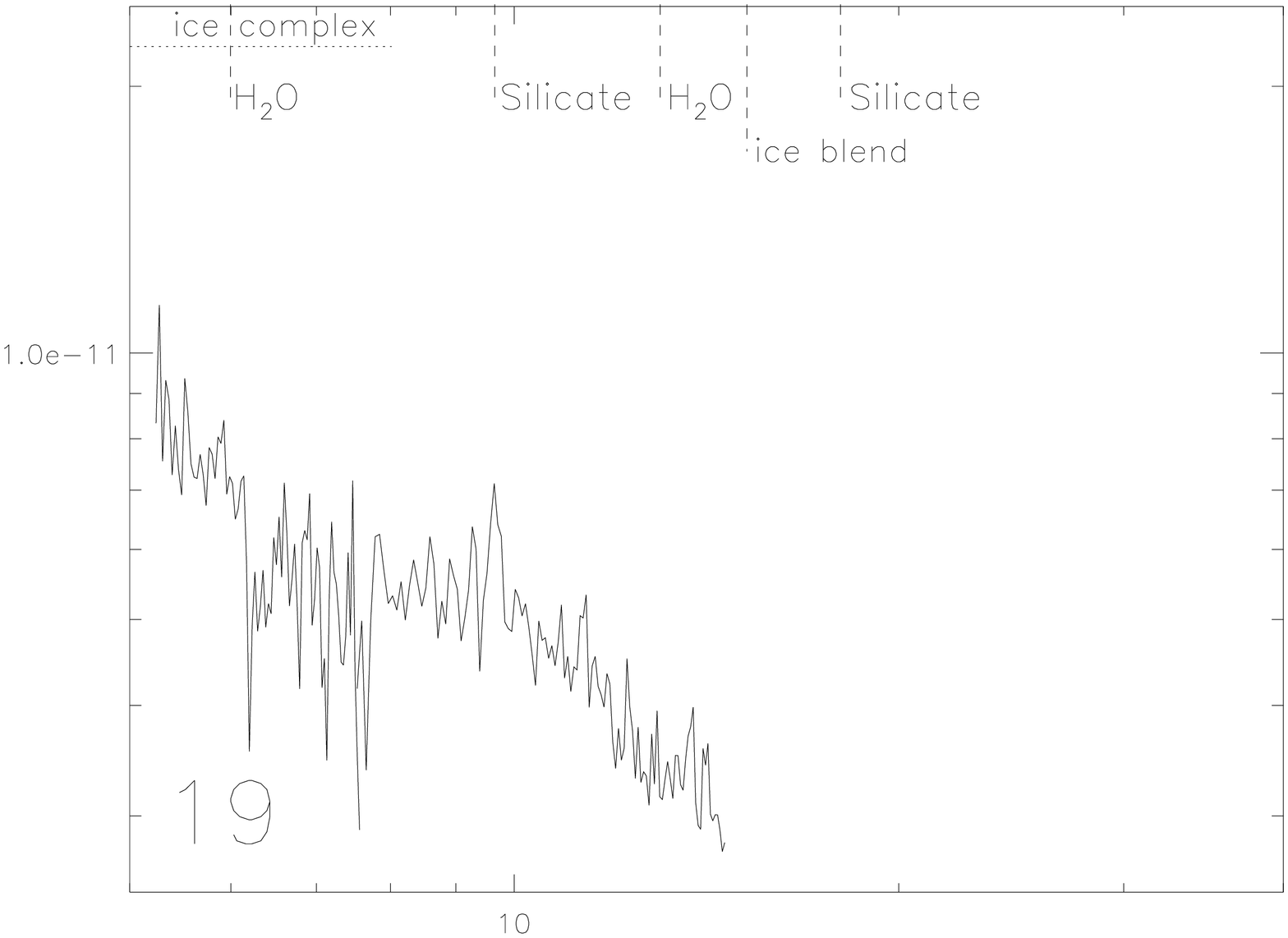}
\includegraphics*[width=1.6in, angle=90, bb = 55 36 577 705]{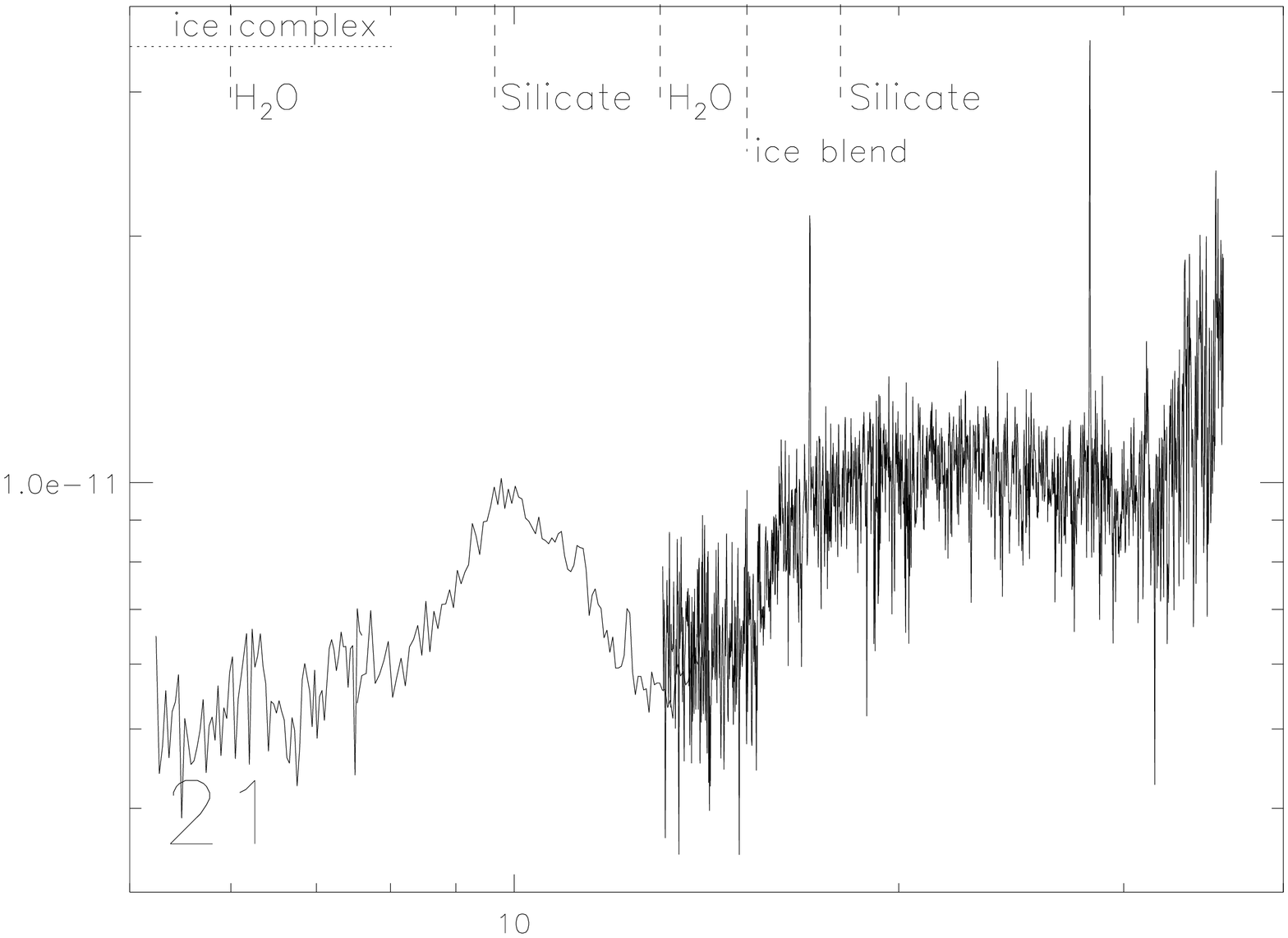}
\includegraphics*[width=1.6in, angle=90, bb = 55 36 577 705]{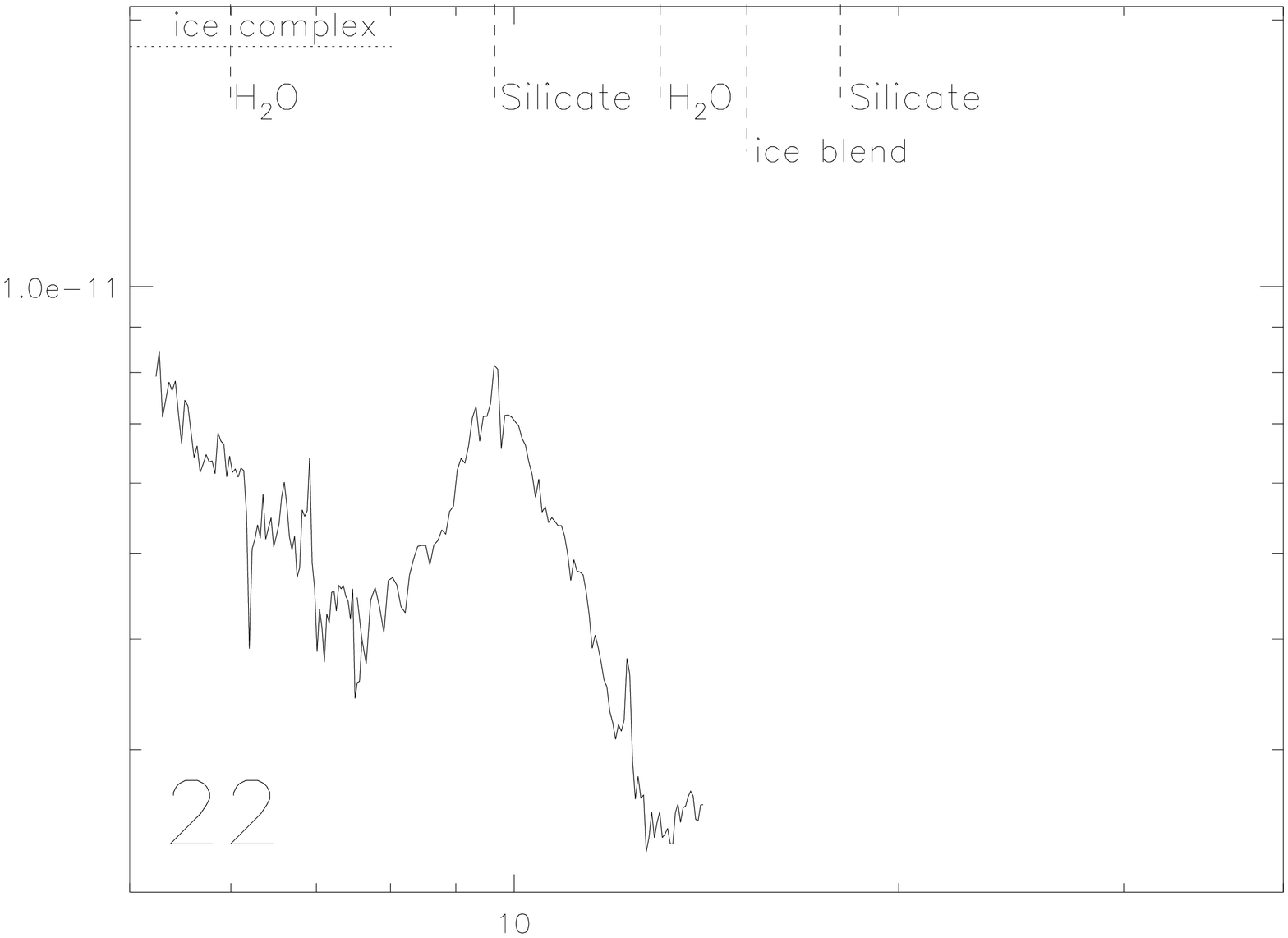}

\includegraphics*[width=1.6in, angle=90, bb = 55 36 577 705]{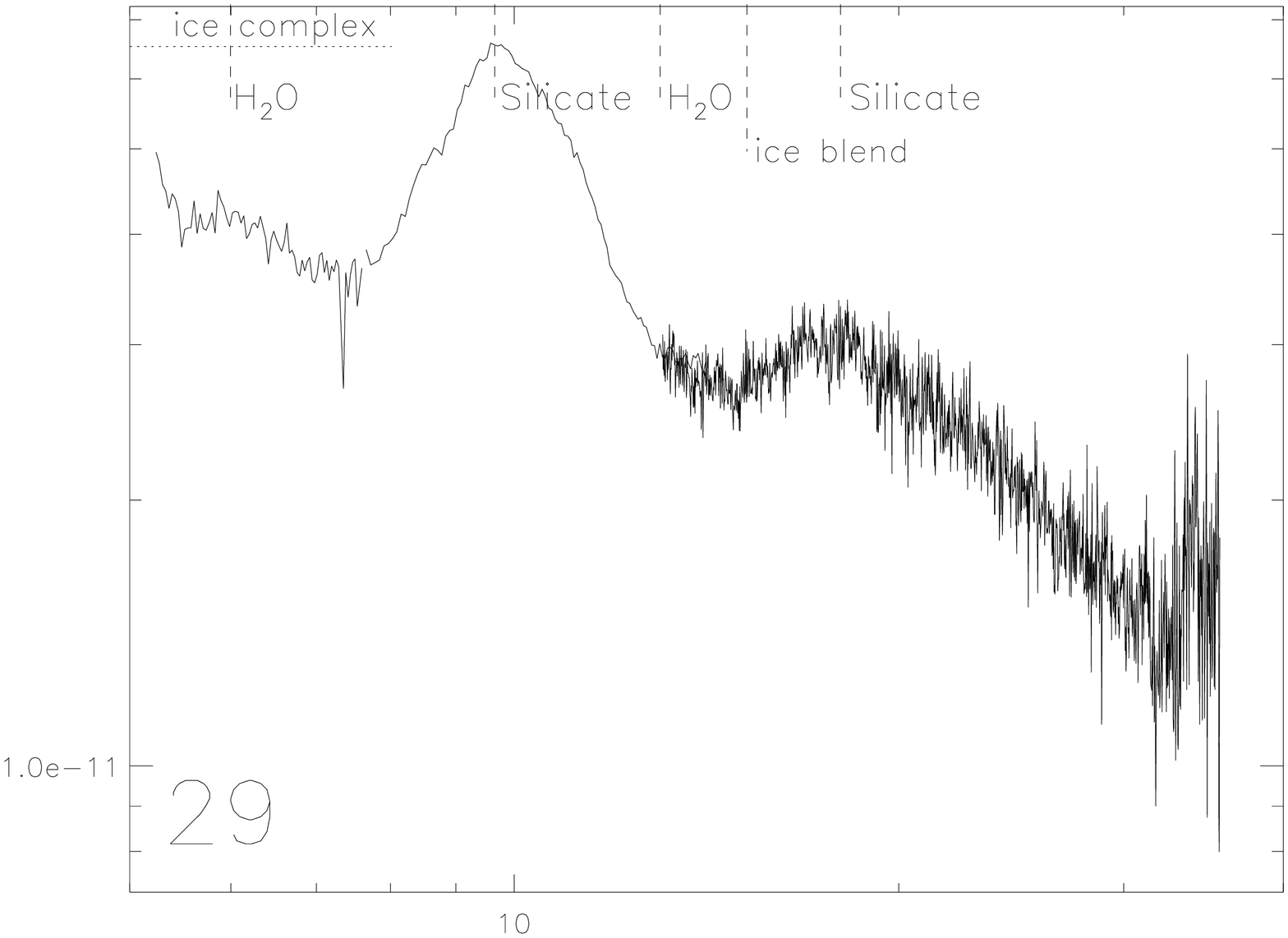}
\includegraphics*[width=1.6in, angle=90, bb = 55 36 577 705]{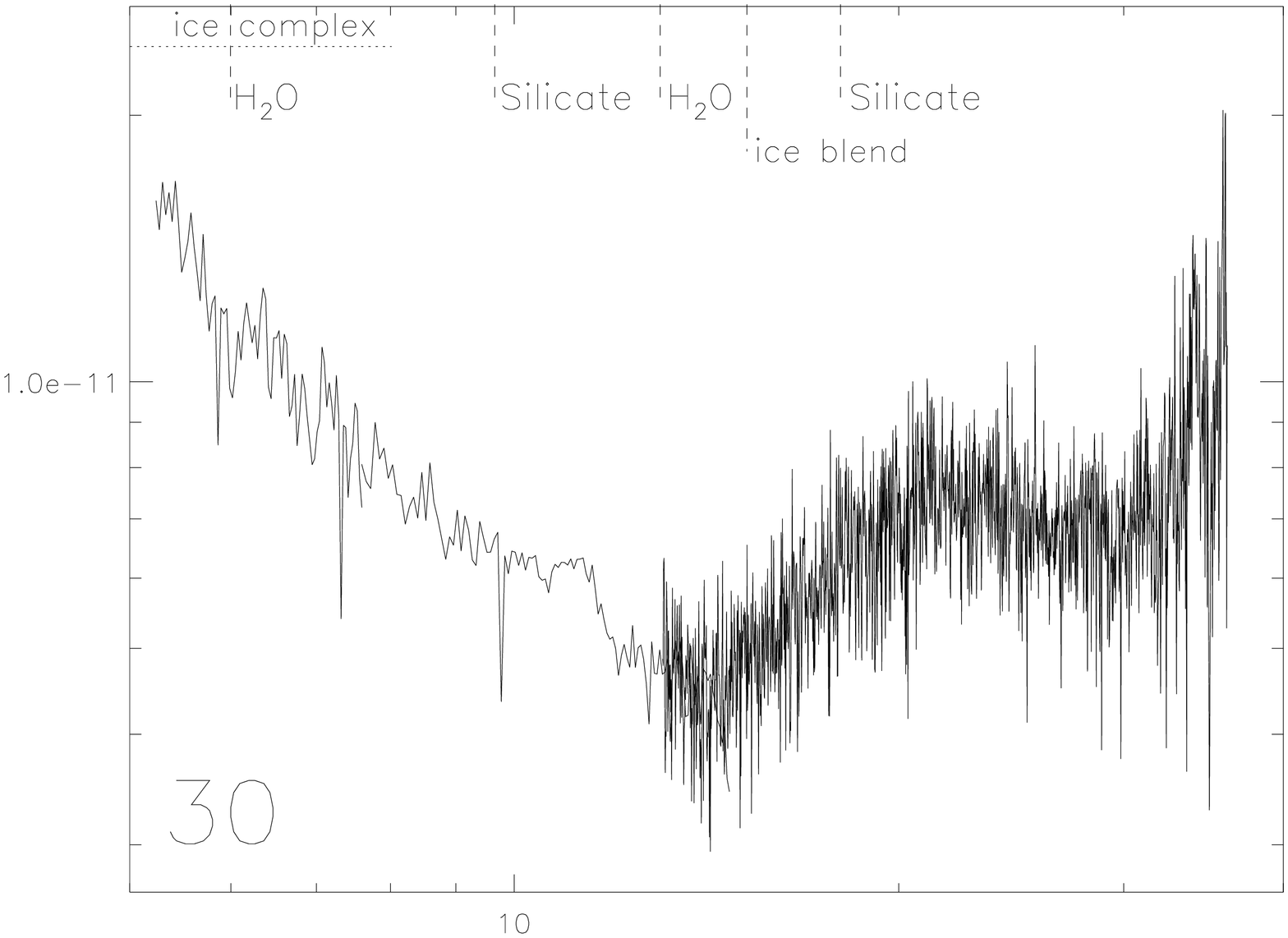}
\includegraphics*[width=1.6in, angle=90, bb = 55 36 577 705]{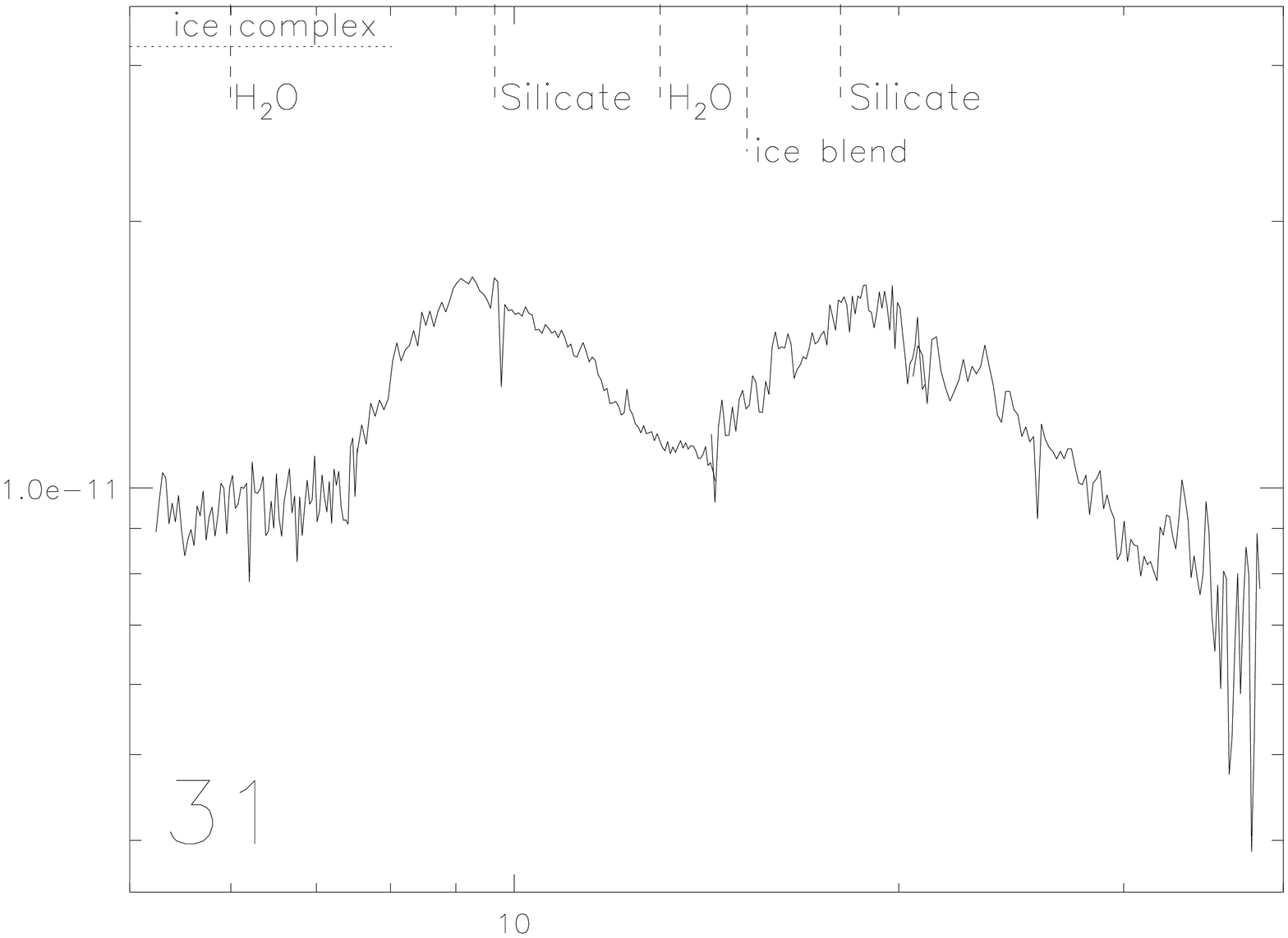}

\includegraphics*[width=1.6in, angle=90, bb = 55 36 577 705]{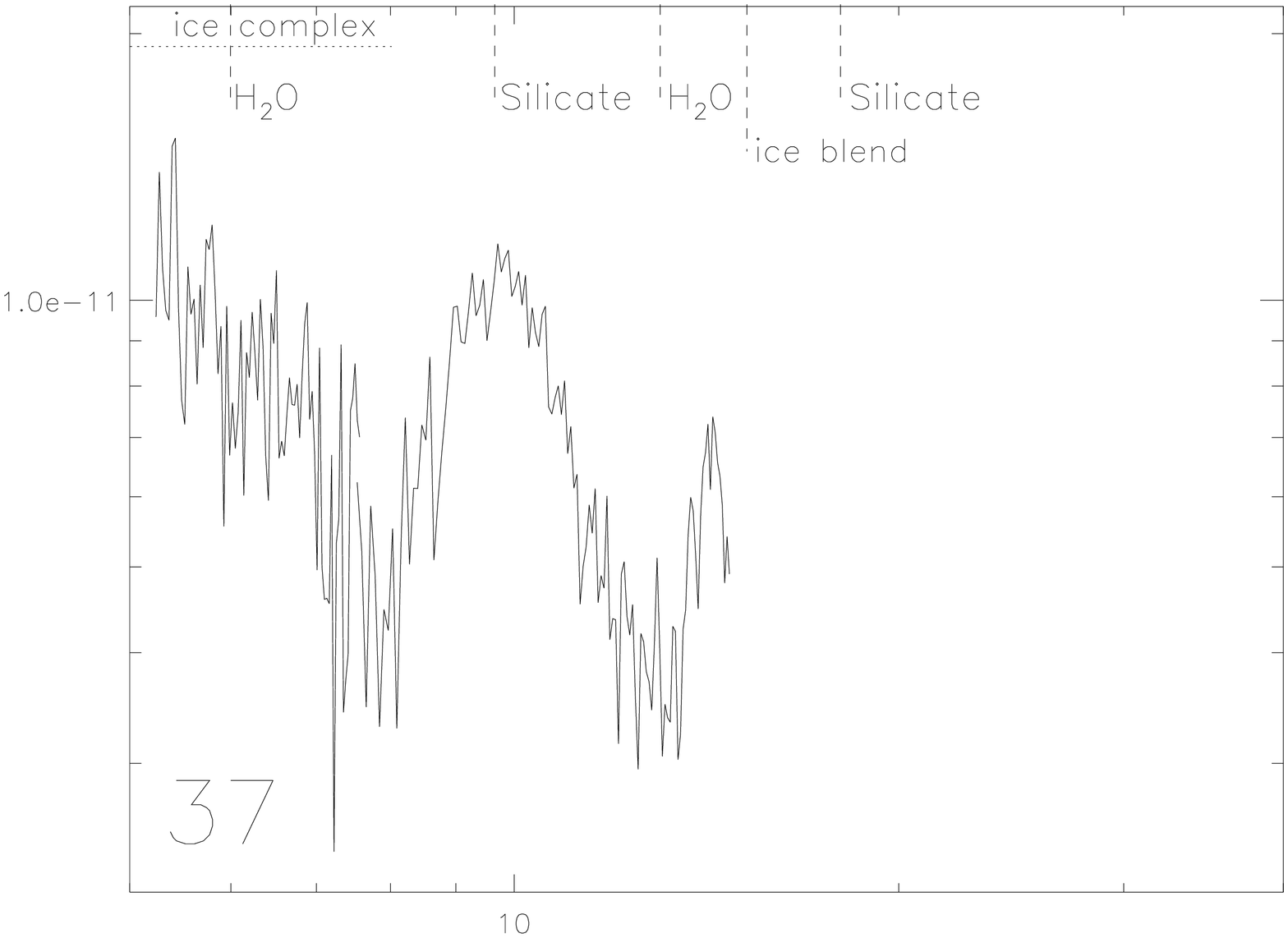}
\includegraphics*[width=1.6in, angle=90, bb = 55 36 577 705]{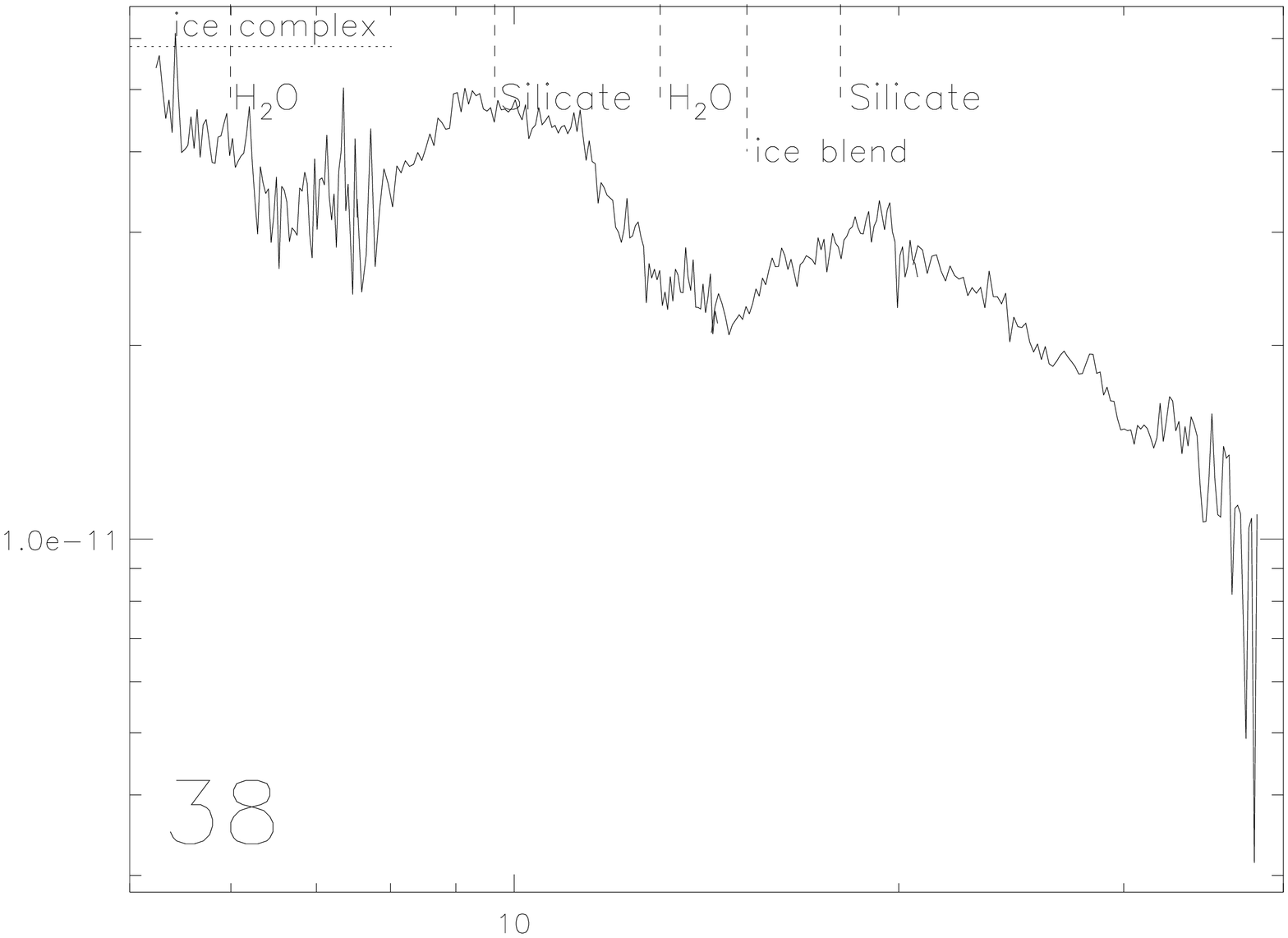}
\includegraphics*[width=1.6in, angle=90, bb = 55 36 577 705]{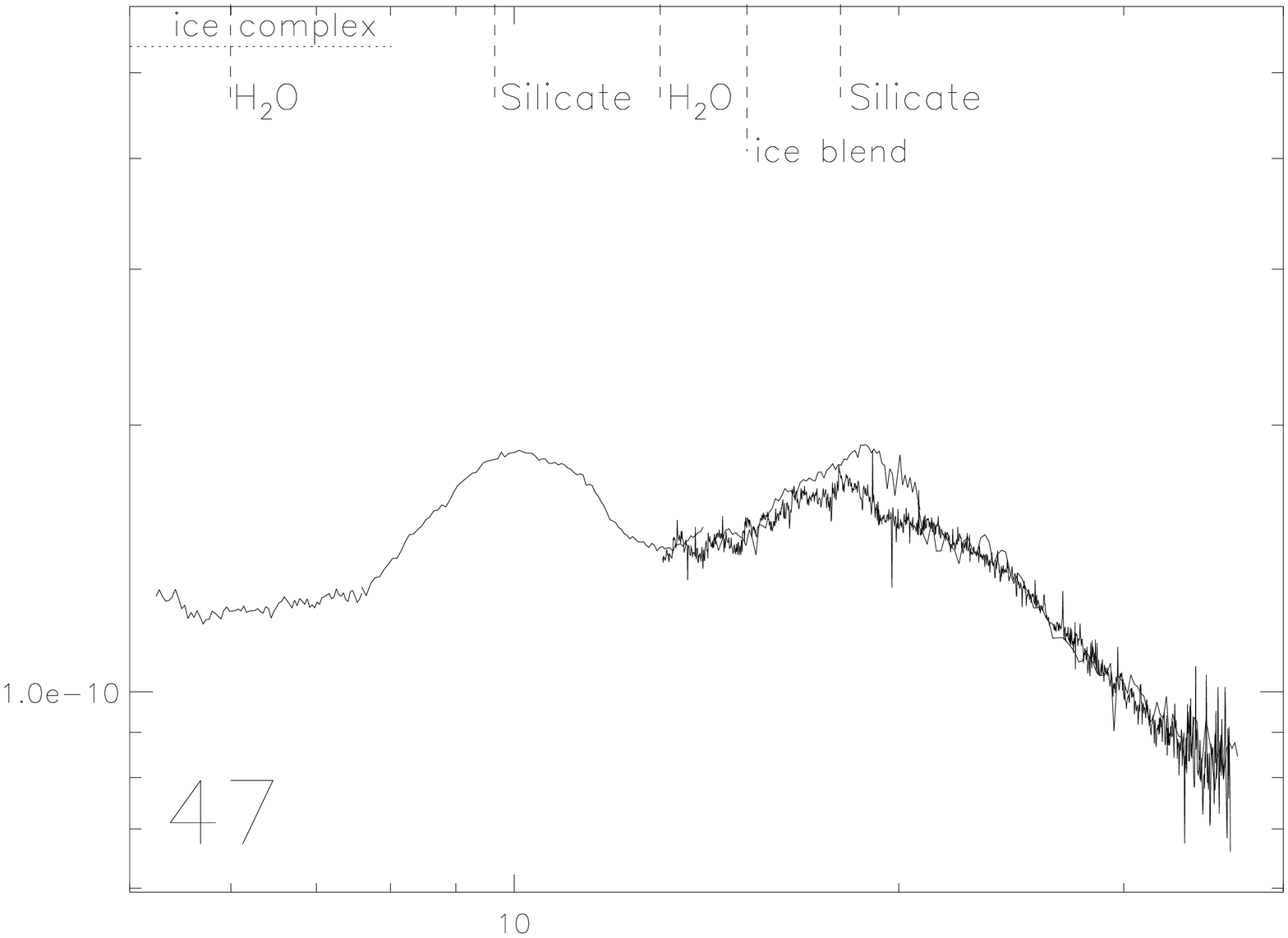}
%{Fig.~\ref{fig_irsplots} --  cntd.}
\caption{IRS spectra of the 12 identified disk sources with low external foreground extinction (stage II sources). \label{fig_irsplots3}}
%\end{center}
\end{figure*}

\begin{figure*}
%\begin{center}
%\includegraphics*[width=1.6in, angle=90, bb = 55 36 577 705]{ic348n2264_08_plotirs.eps}
%\includegraphics*[width=1.6in, angle=90, bb = 55 36 577 705]{ic348n2264_32_plotirs.eps}
%\includegraphics*[width=1.6in, angle=90, bb = 55 36 577 705]{ic348n2264_12_plotirs.eps}
%\includegraphics*[width=1.6in, angle=90, bb = 55 36 577 705]{ic348n2264_16_plotirs.eps}
%\includegraphics*[width=1.6in, angle=90, bb = 55 36 577 705]{ic348n2264_25_plotirs.eps}
%\includegraphics*[width=1.6in, angle=90, bb = 55 36 577 705]{ic348n2264_36_plotirs.eps}
%\includegraphics*[width=1.6in, angle=90, bb = 55 36 577 705]{ic348n2264_06+07_plotirs.eps}
%\includegraphics*[width=1.6in, angle=90, bb = 55 36 577 705]{ic348n2264_41_plotirs.eps}
%\includegraphics*[width=1.6in, angle=90, bb = 55 36 577 705]{ic348n2264_43_plotirs.eps}
\includegraphics*[width=1.6in, angle=90, bb = 55 36 577 705]{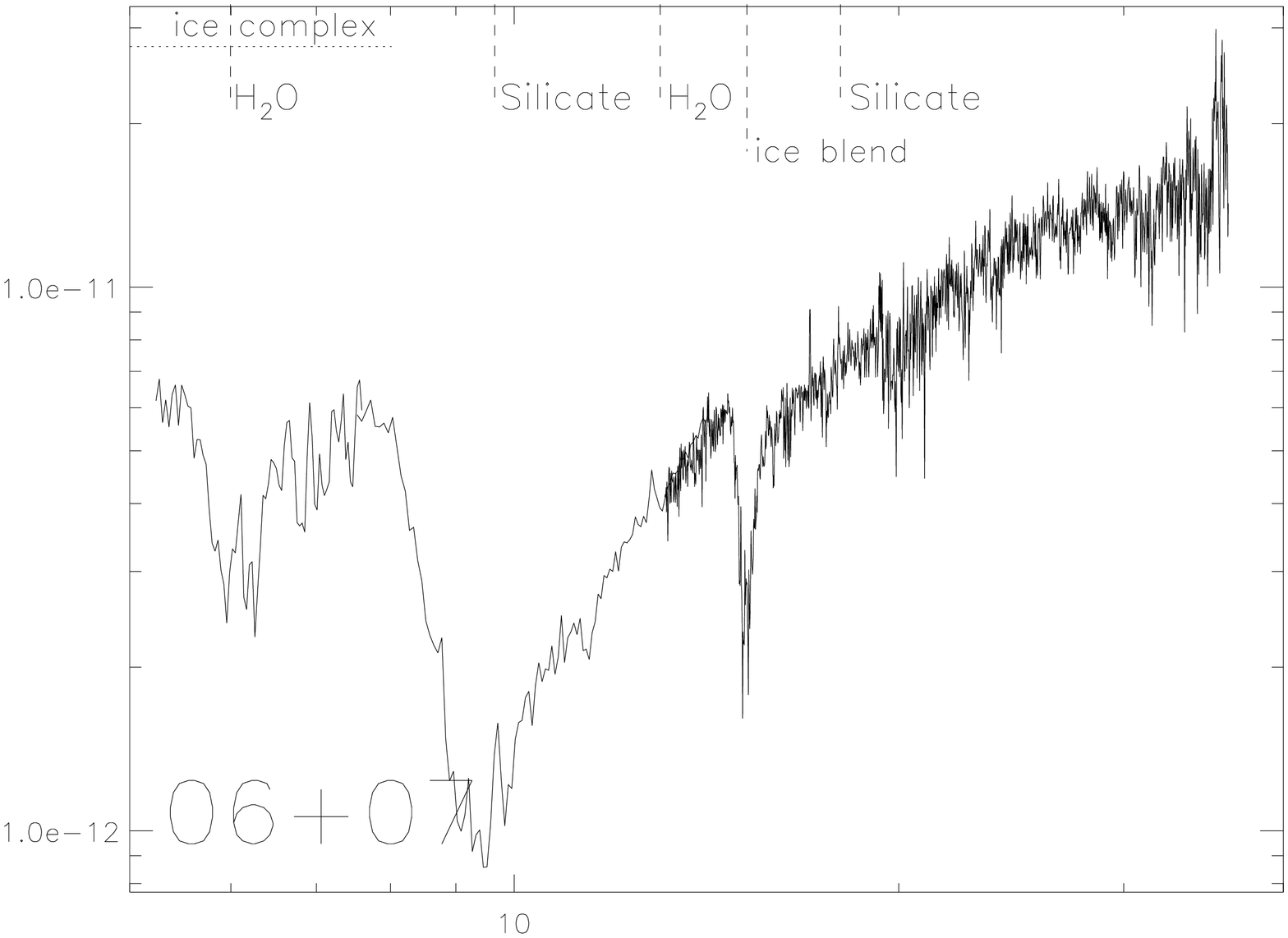}
\includegraphics*[width=1.6in, angle=90, bb = 55 36 577 705]{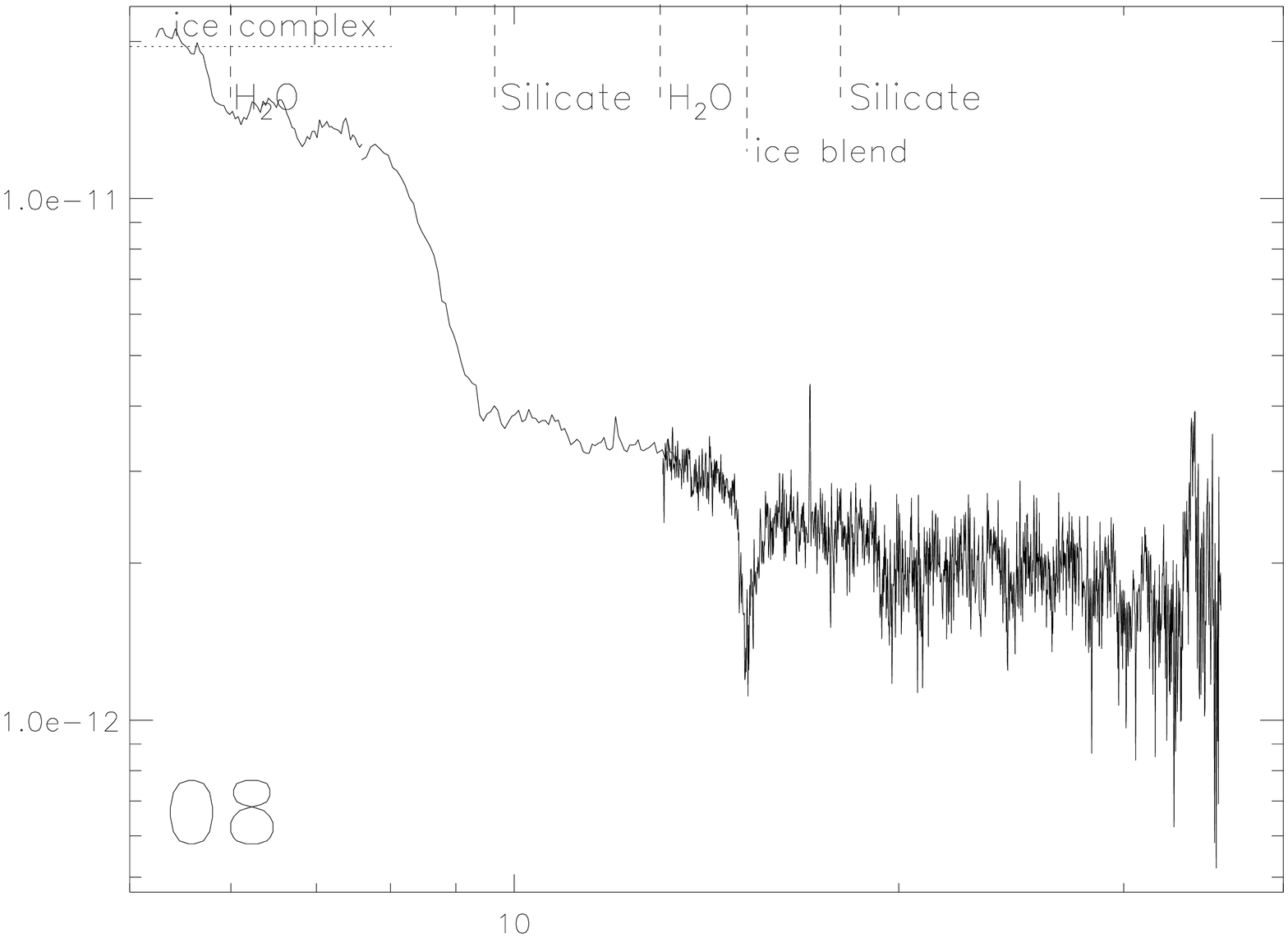}
\includegraphics*[width=1.6in, angle=90, bb = 55 36 577 705]{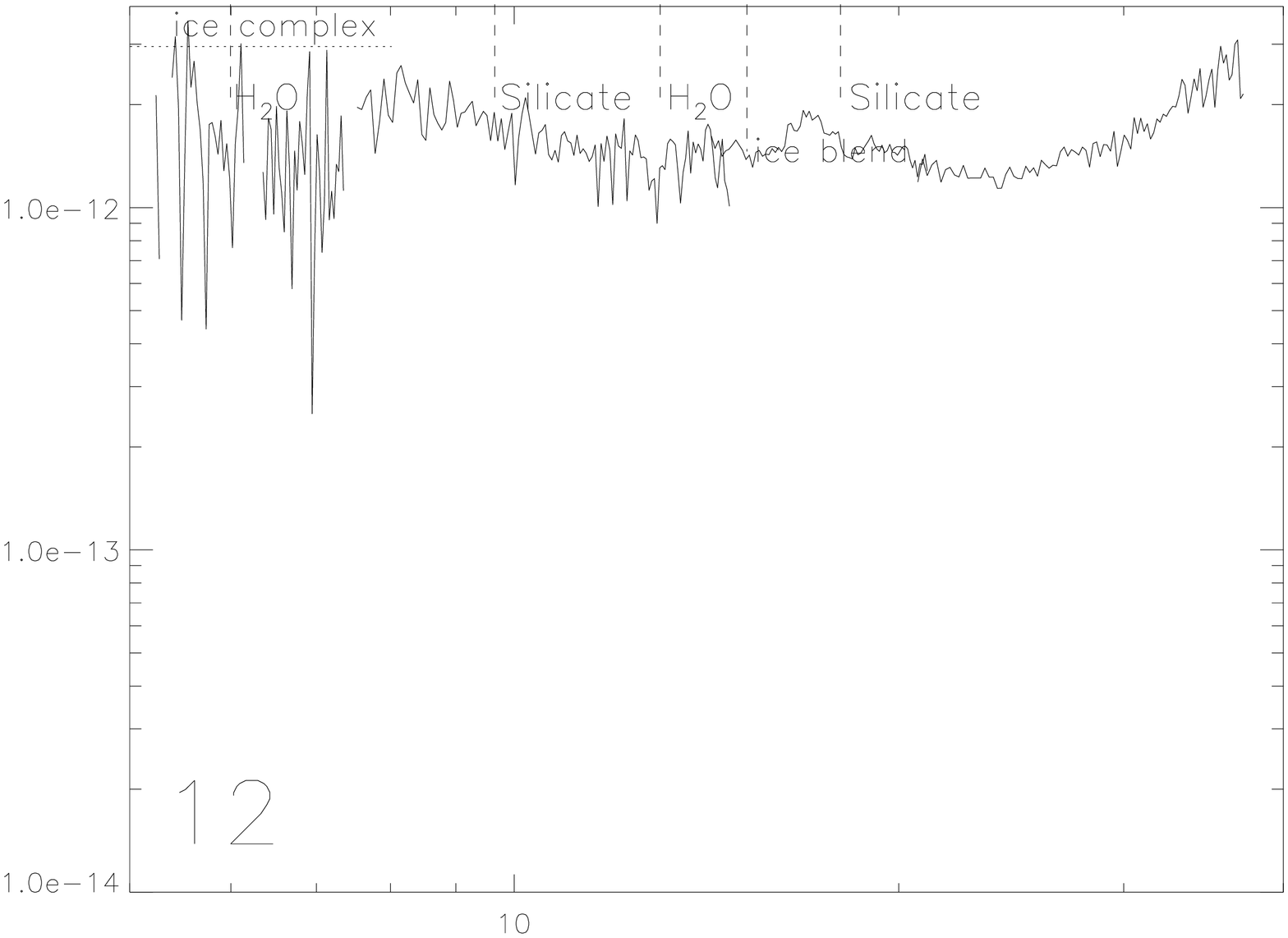}

\includegraphics*[width=1.6in, angle=90, bb = 55 36 577 705]{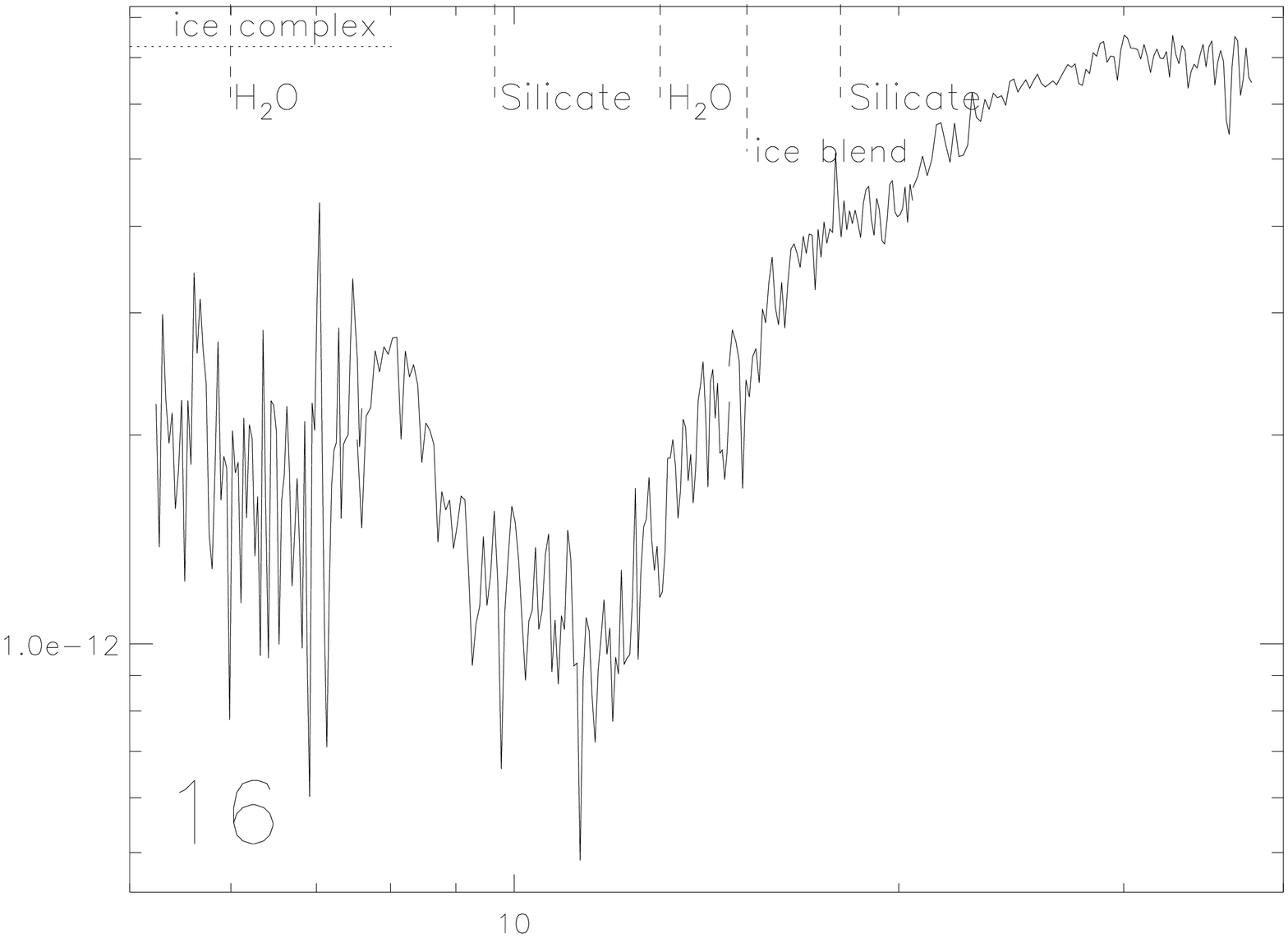}
\includegraphics*[width=1.6in, angle=90, bb = 55 36 577 705]{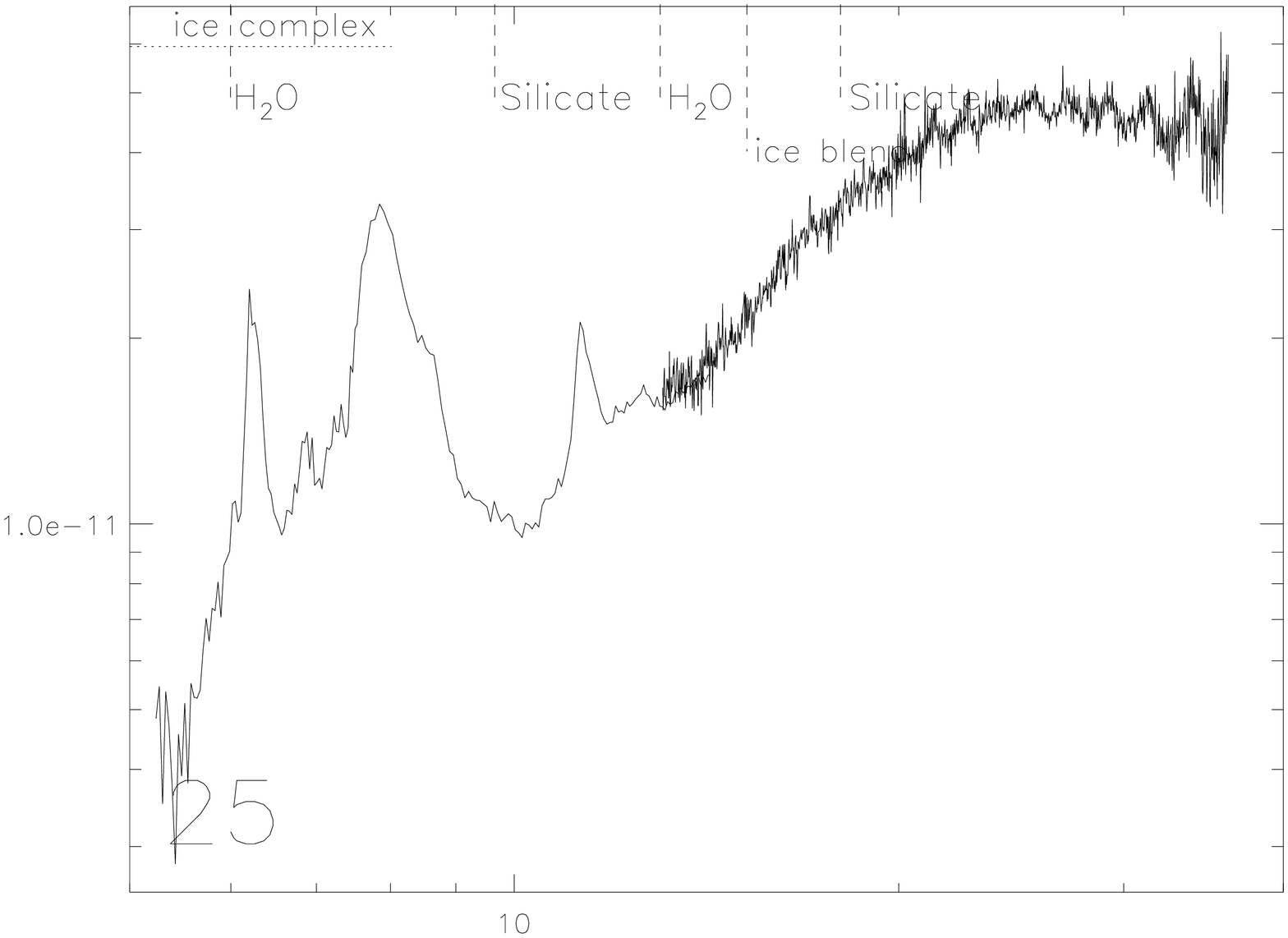}
\includegraphics*[width=1.6in, angle=90, bb = 55 36 577 705]{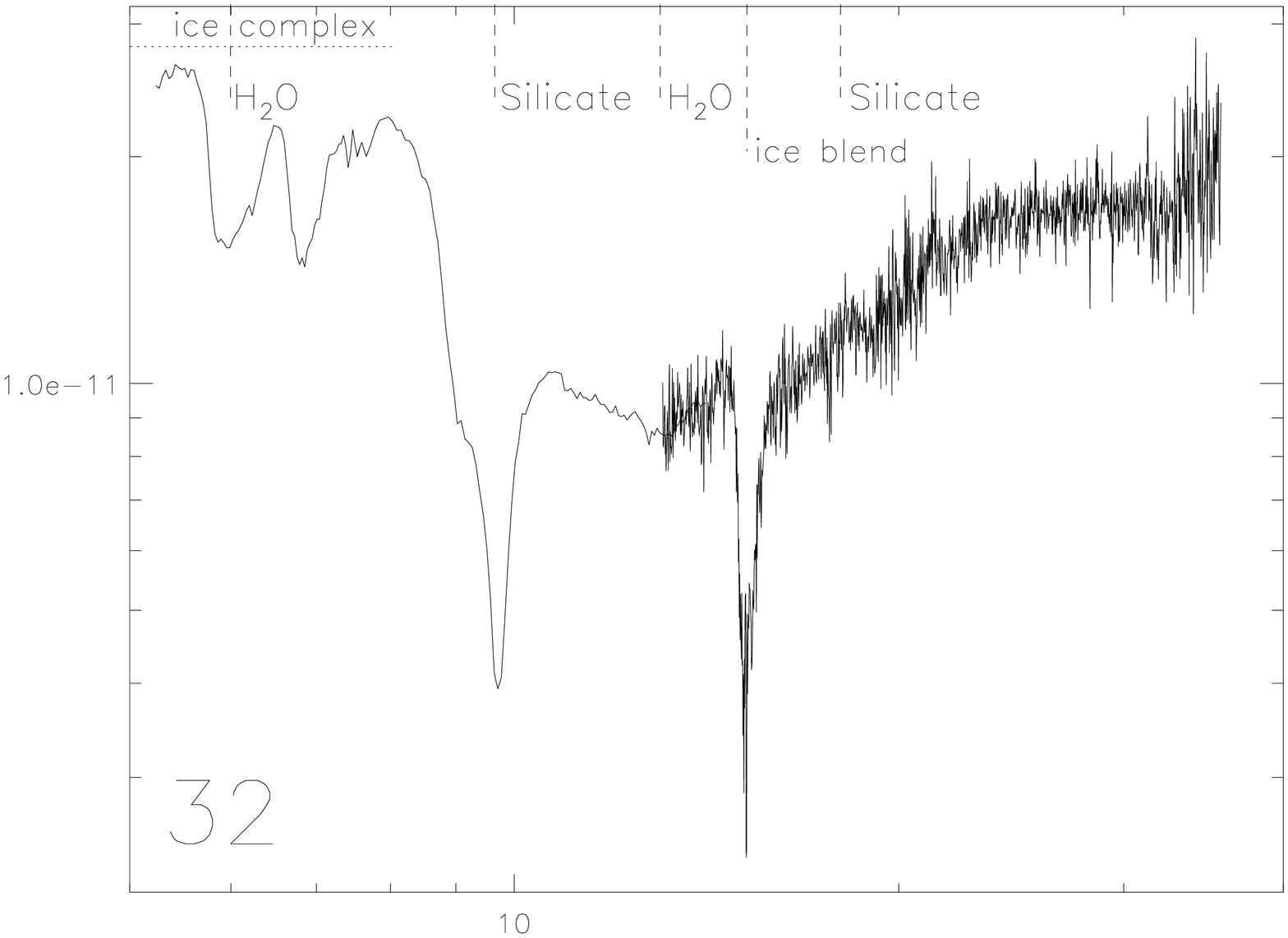}

\includegraphics*[width=1.6in, angle=90, bb = 55 36 577 705]{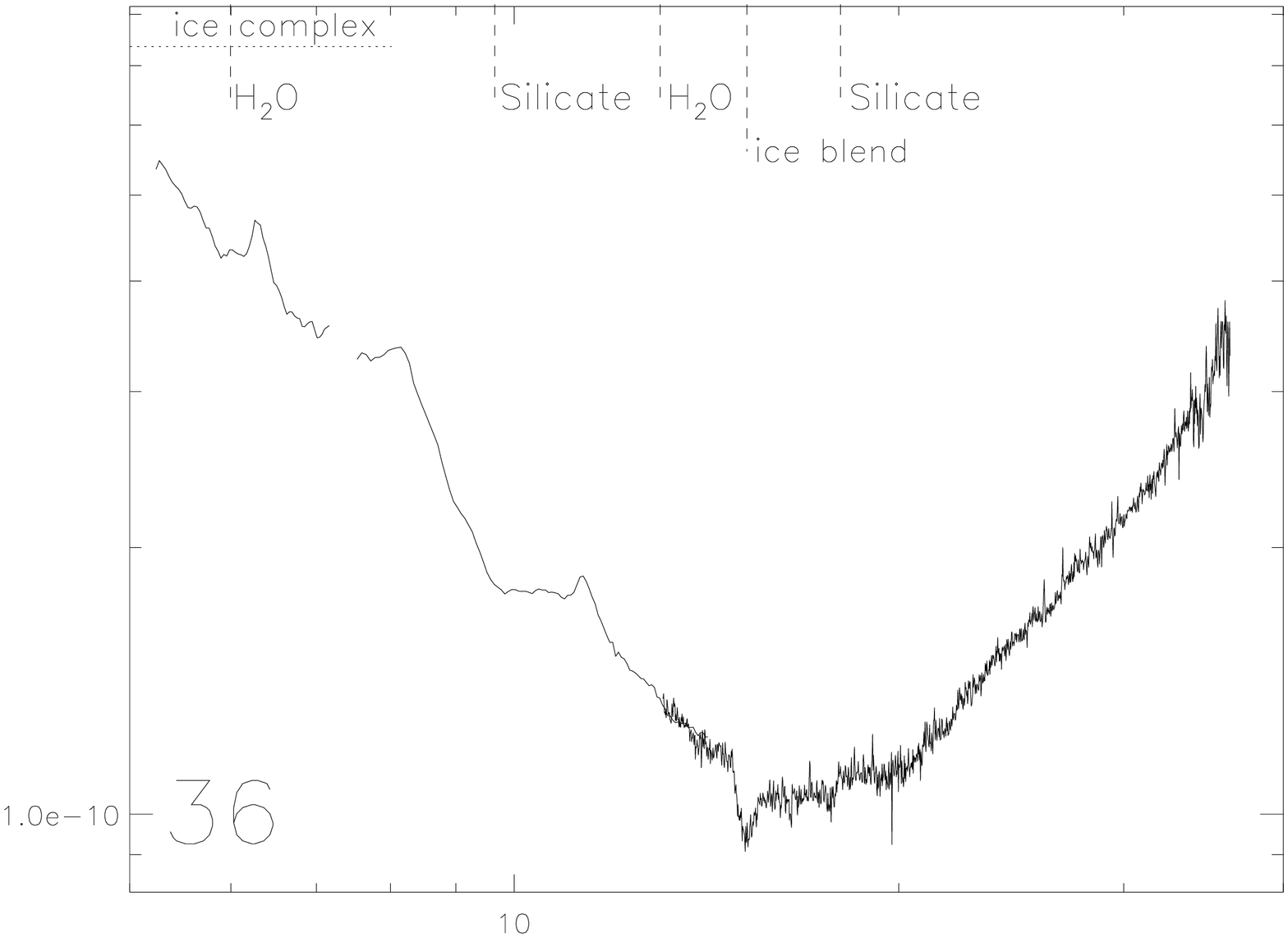}
\includegraphics*[width=1.6in, angle=90, bb = 55 36 577 705]{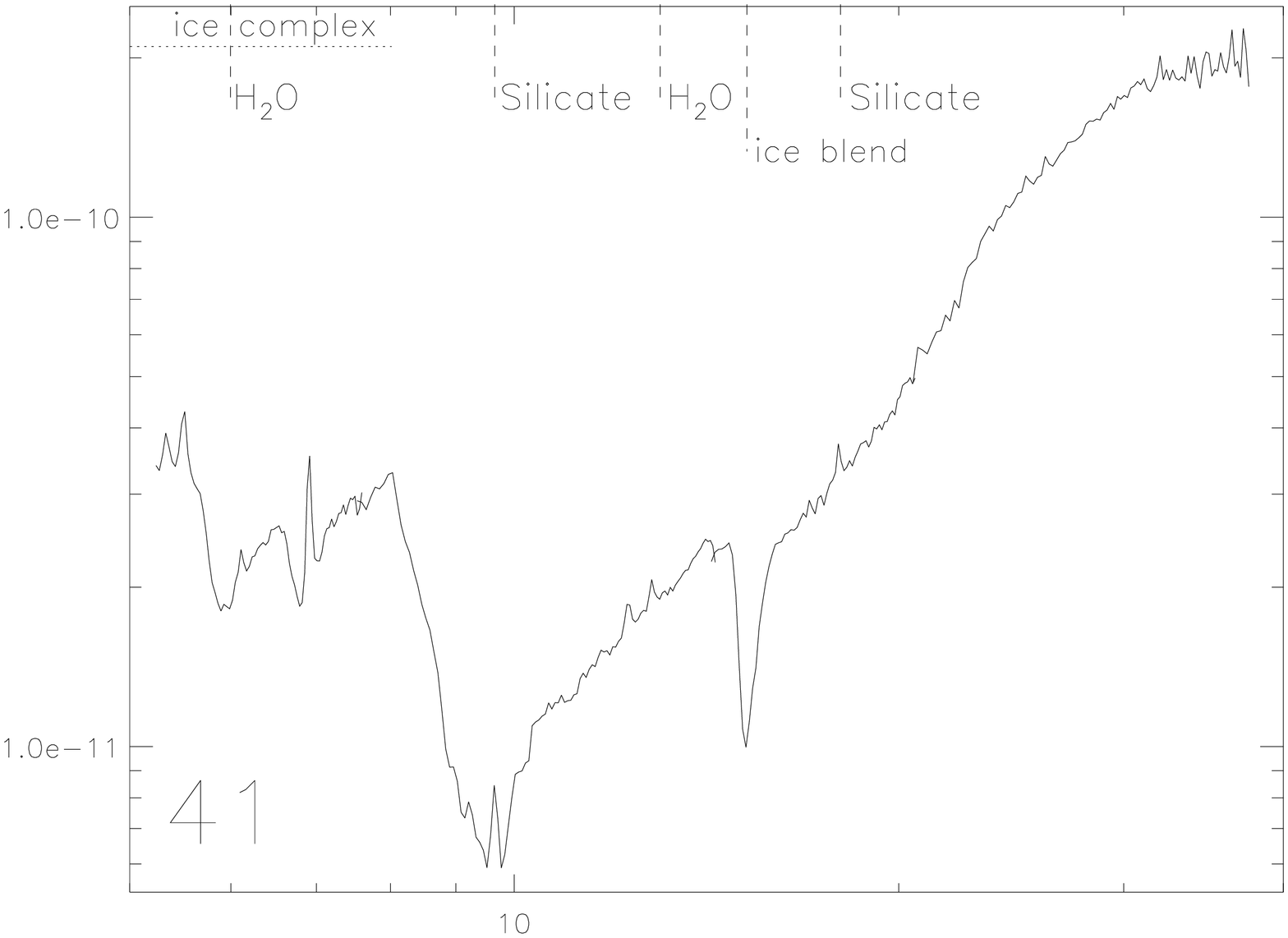}
\includegraphics*[width=1.6in, angle=90, bb = 55 36 577 705]{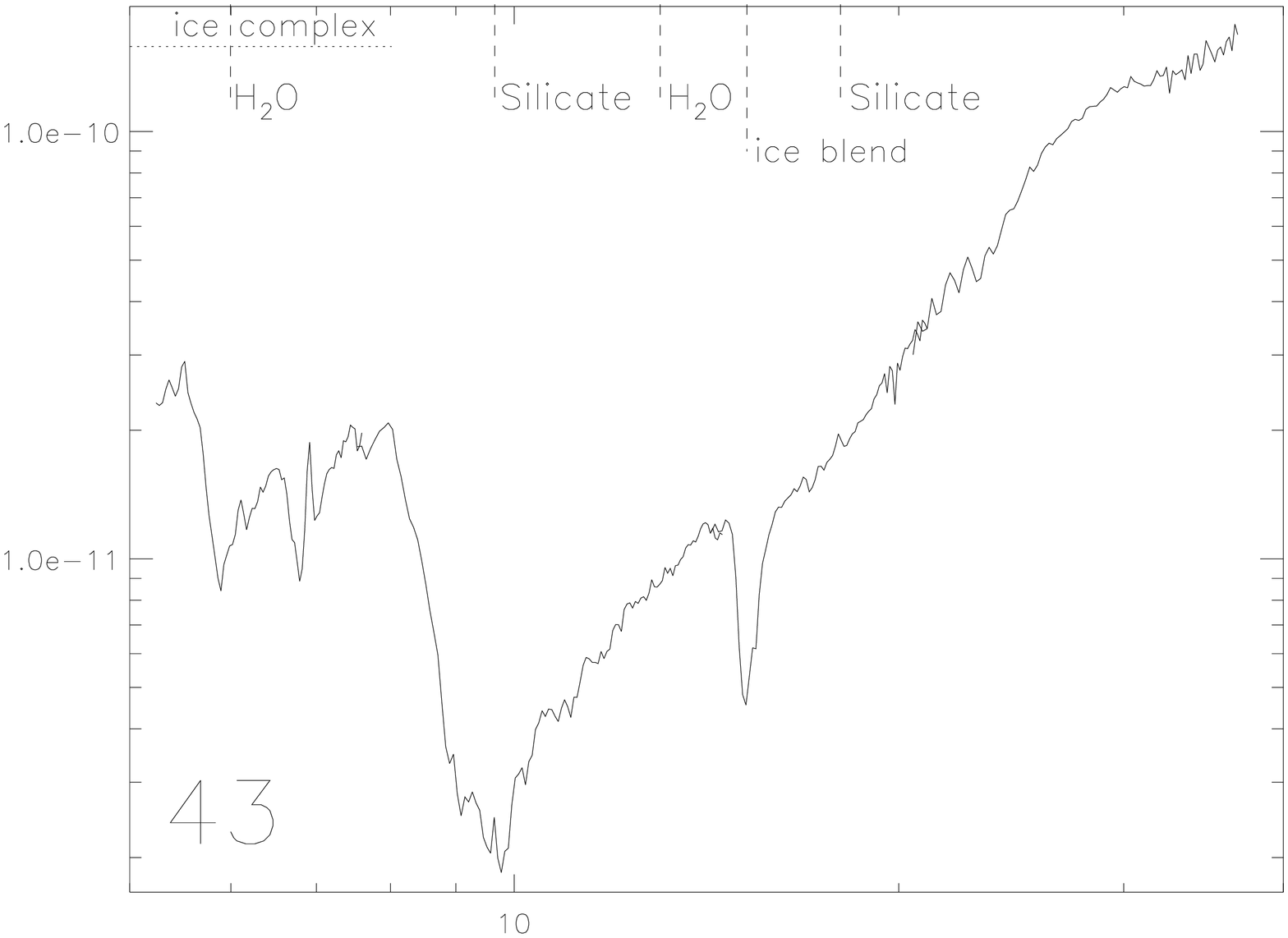}
%{Fig.~\ref{fig_irsplots} --  cntd.}
\caption{IRS spectra of sources where the modeling was found to be unreliable, for example due to blending of several sources (see text).\label{fig_irsplots4}}
%\end{center}
\end{figure*}

\clearpage

\begin{figure*}
\begin{center}
\includegraphics[width=2.2in, angle=-90]{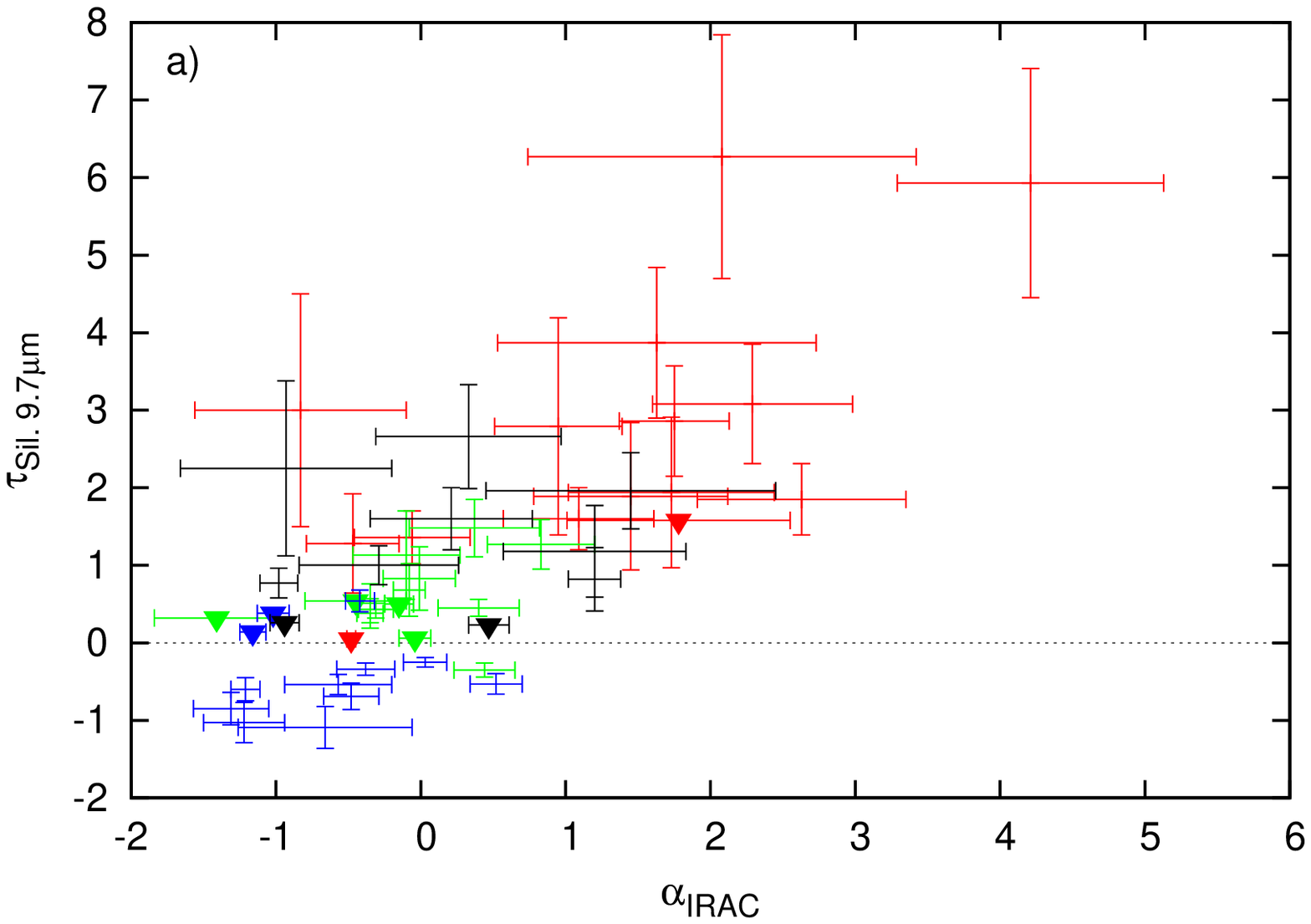}
\includegraphics[width=2.2in, angle=-90]{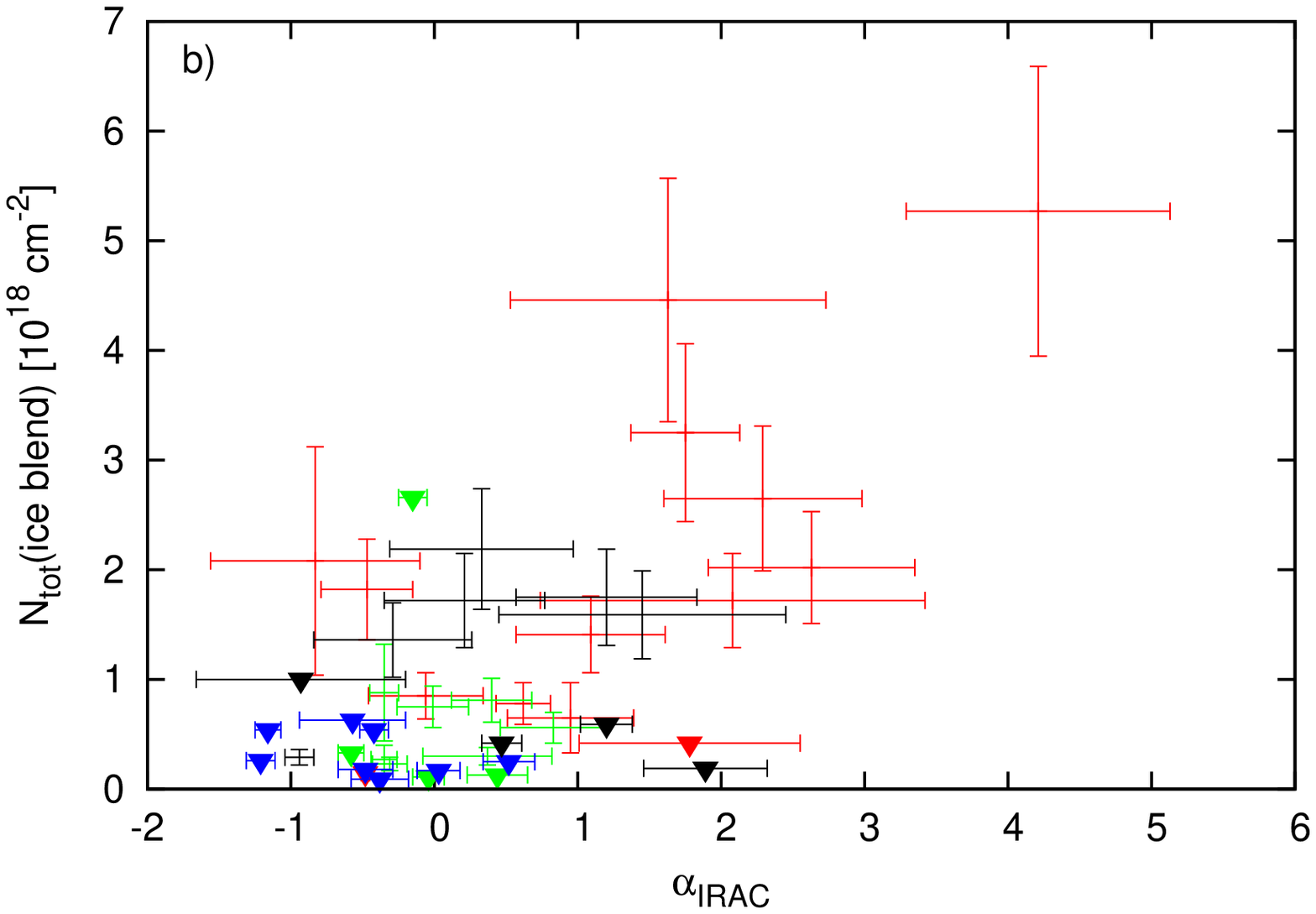}
\includegraphics[width=2.2in, angle=-90]{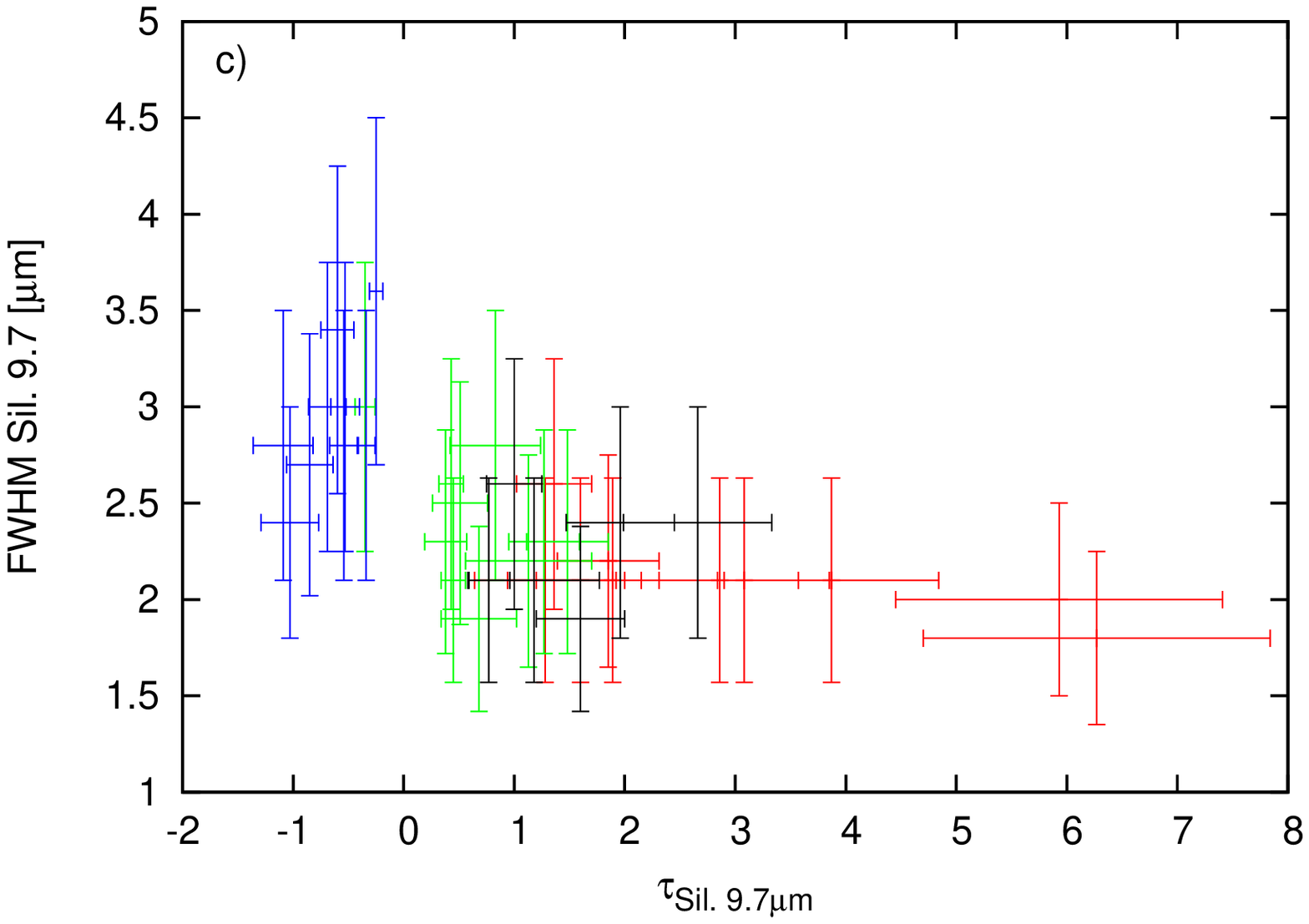}
\includegraphics[width=2.2in, angle=-90]{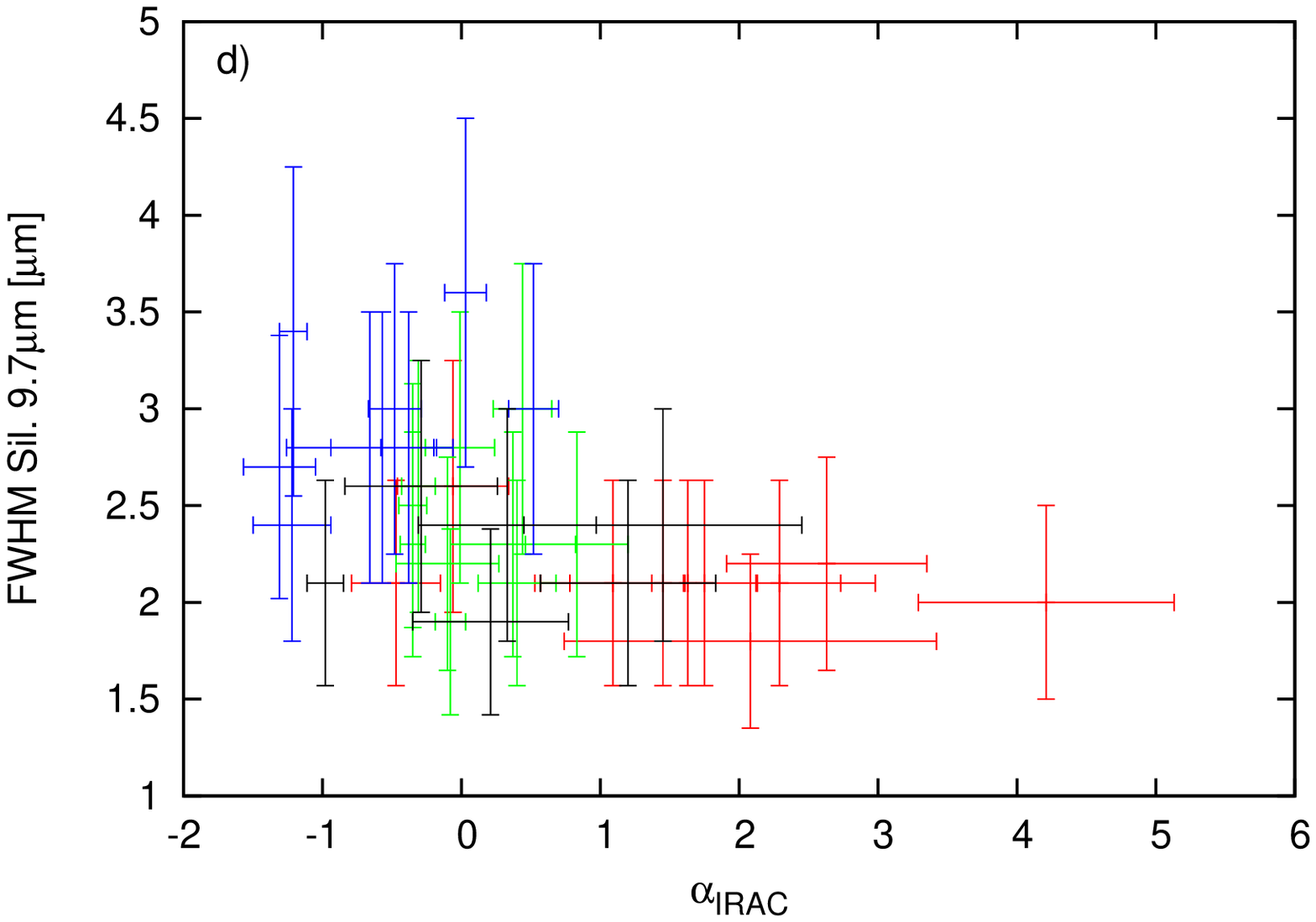}
\includegraphics[width=2.2in, angle=-90]{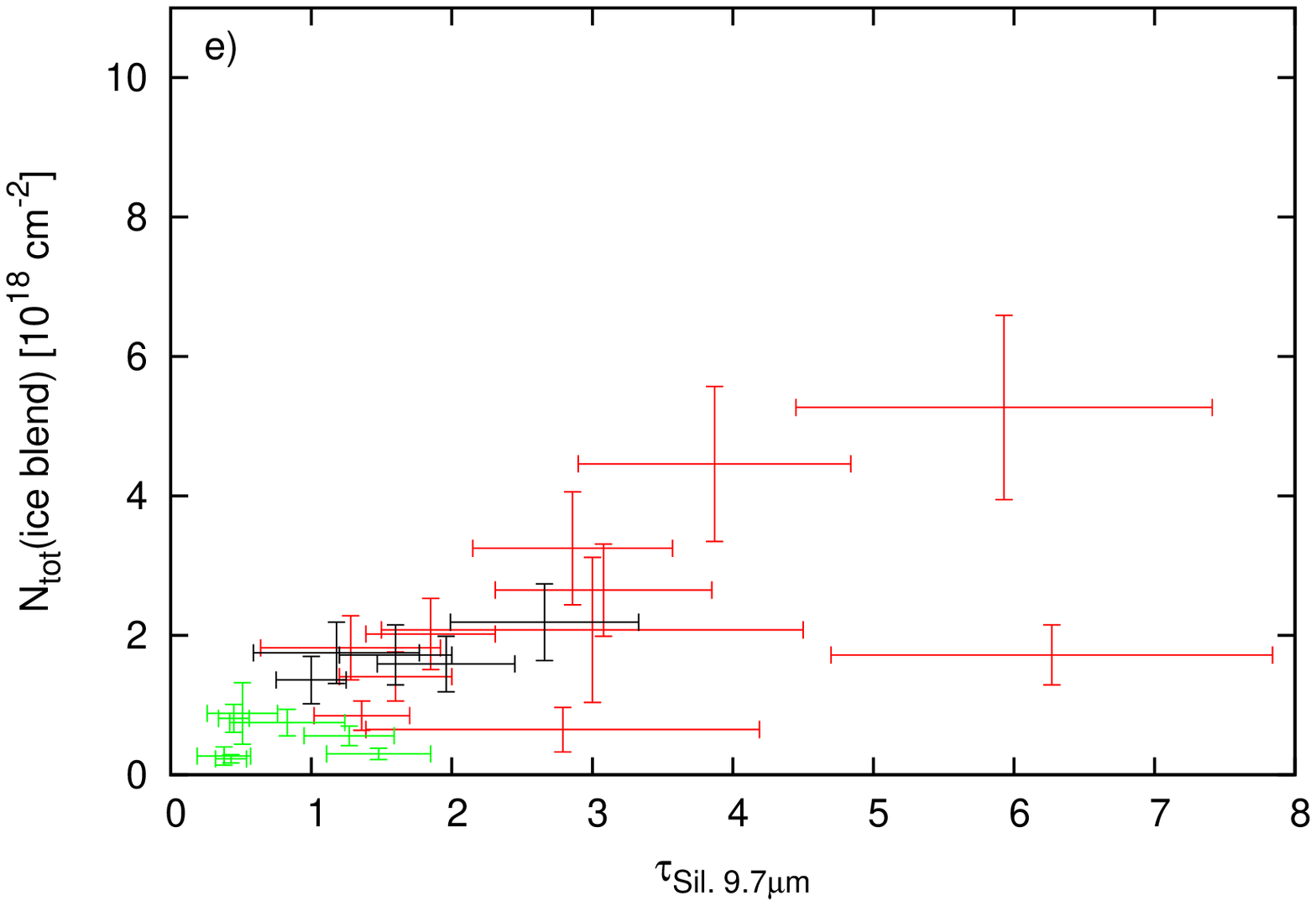}
\includegraphics[width=2.2in, angle=-90]{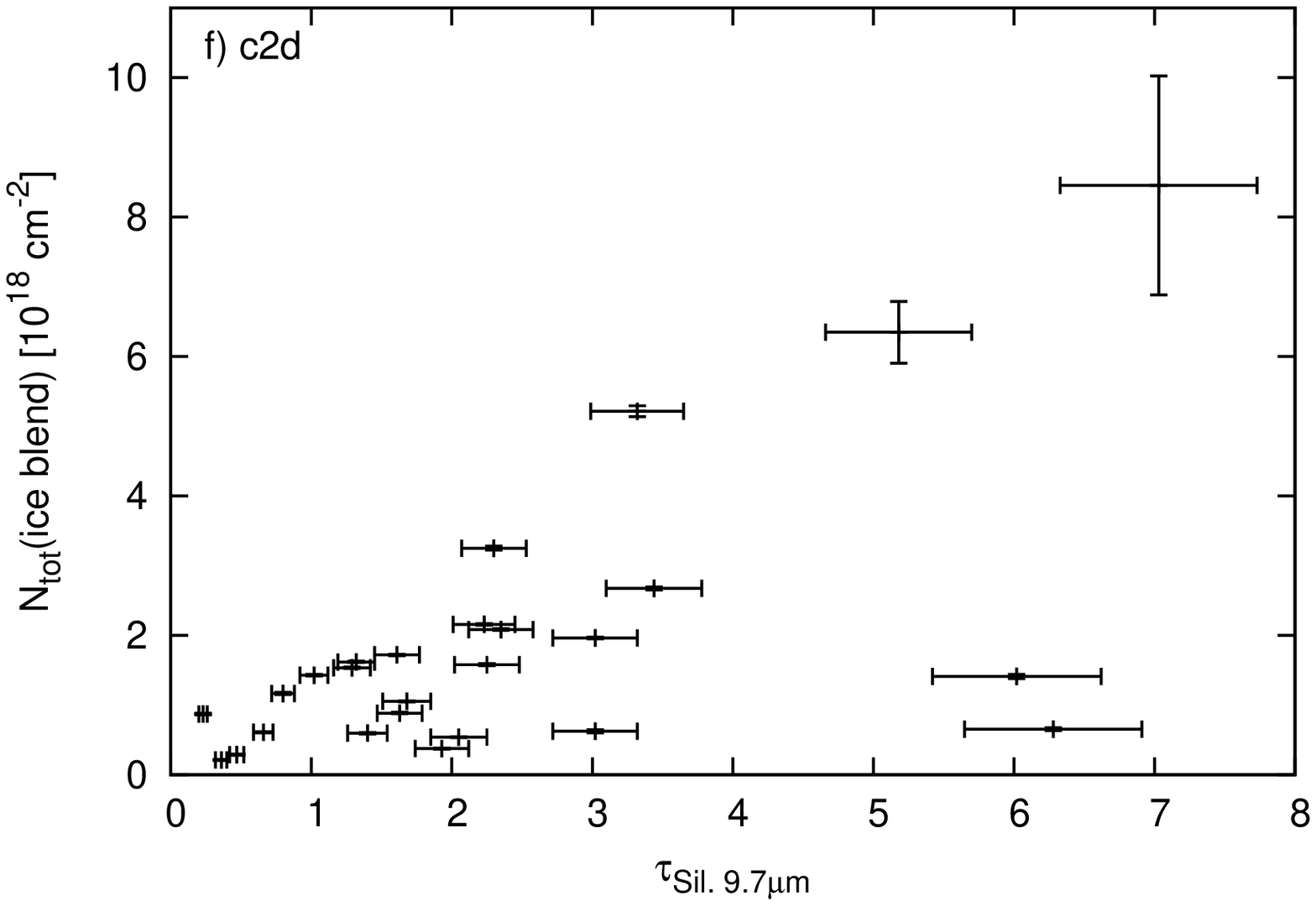}
\caption{a) Correlation of the IRAC spectral index and the optical depth of the 9.7\,$\mu$m silicate feature. Triangles indicate upper limits. b)
 Correlation of the IRAC spectral index and the total ice column density. Triangles indicate upper limits. c) Correlation of the optical depth and the FWHM linewidth of the 9.7\,$\mu$m silicate feature. d) Correlation of the IRAC spectral index and the FWHM linewidth of the 9.7\,$\mu$m silicate feature. e) Correlation of the optical depth of the 9.7\,$\mu$m silicate feature and the total ice column density, without upper limits. f) Correlation of the optical depth of the 9.7\,$\mu$m silicate feature and the total ice column density in the \textsl{Spitzer} c2d survey \citep{pon08,boo08}, without upper limits. In all color plots, red denotes stage I sources, green denotes stage II(ex) sources, blue denotes stage II, and black denotes sources where no evolutionary stage was determined. \label{fig_speccorr}}
\end{center}
\end{figure*}

\begin{figure*}
\begin{center}
\includegraphics[width=\linewidth]{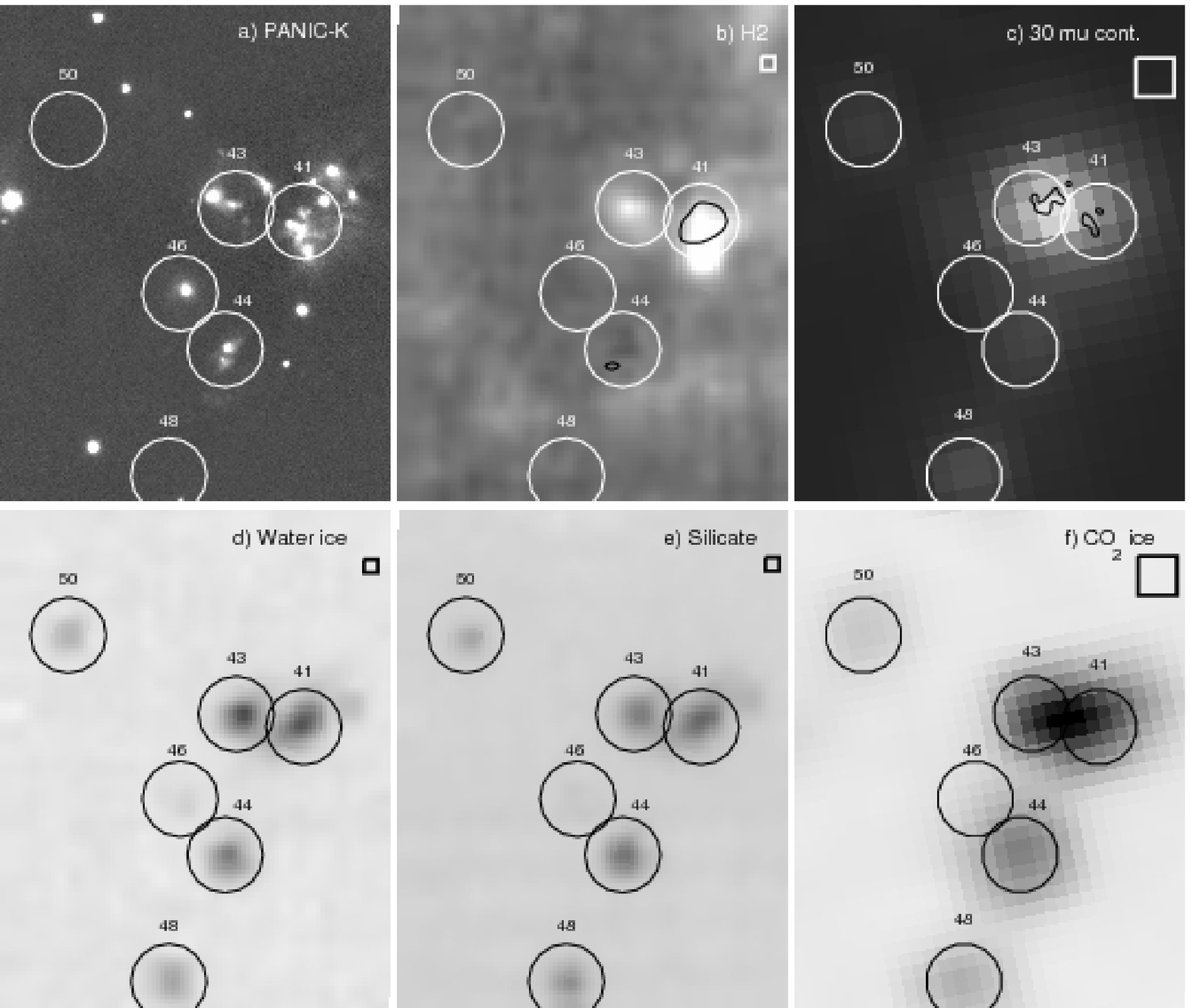}
\caption{Spectral maps for the microcluster region. Wavelength regions: a) Ks 2--2.3\mic\ (includes \htwo\ (1--0)
S(0)-S(2) and \hi\ \bg), b) \htwo\ (0--0) S(7)$+$S(5)$+$S(3)$+$S(2), with \neiif\ 12.8\mic\ contour overlaid (5$\sigma$), c)
30 minus 22\mic\ continuum emission, d) 6.0$+$6.7\mic\ ice absorption (\water\ and \ammpl), e) silicate 9.7\mic\
minus adjacent continuum, f) 15.2\mic\ ice absorption (\cotwo:\co:\water). Boxes denote the original pixel size,
i.e., about the actual angular resolution, for the \textsl{Spitzer} data. In panel c, the submillimeter continuum is marked (5$\sigma$ contour from \citealp{tei06}; note that only sources 41 and 43 were covered in that observation).\label{fig_microclmaps}}
\end{center}
\end{figure*}

\begin{figure*}[htbp]
\begin{center}
\includegraphics[width=6.5in]{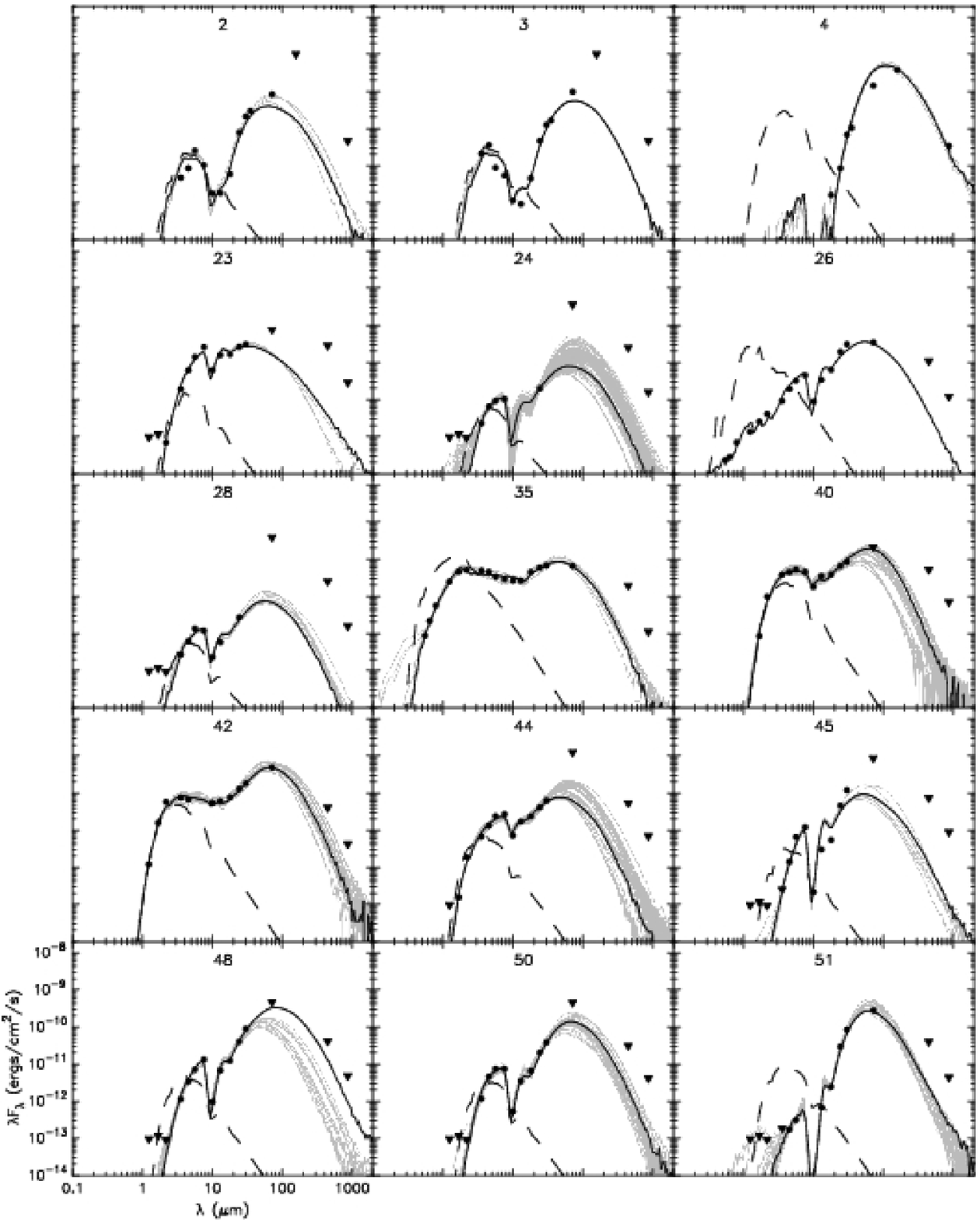}
\caption{Observed SEDs of the 17 genuine protostars (stage I sources), with model SEDs providing a good fit overplotted. The black circles show the observed fluxes, the triangles show upper limits, the black line is the best fit model SED from R06, and the grey lines are other good model SED fits from R06 (where a good fit is defined as one with $\chi^2 - \chi^2_{\rm best} < 3 \times n_{\rm data}$). The dashed line shows the contribution of the stellar photosphere, including the effect of foreground extinction. Note that the data points shown reflect the input for the fitting; upper limits therefore are not necessarily due to non-detections (see text). \label{fig:g1seds}}
\end{center}
\end{figure*}

\addtocounter{figure}{-1}

\begin{figure*}[htbp]
\begin{center}
\includegraphics[width=4.3in]{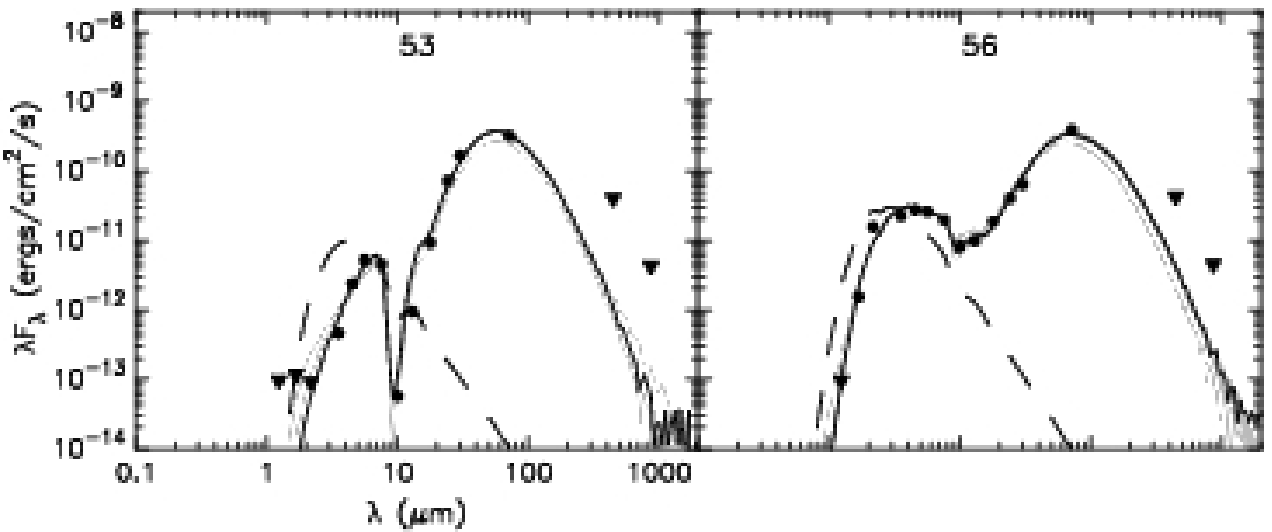}
\caption{continued}
\end{center}
\end{figure*}

\begin{figure*}[htbp]
\begin{center}
\includegraphics[width=6.5in]{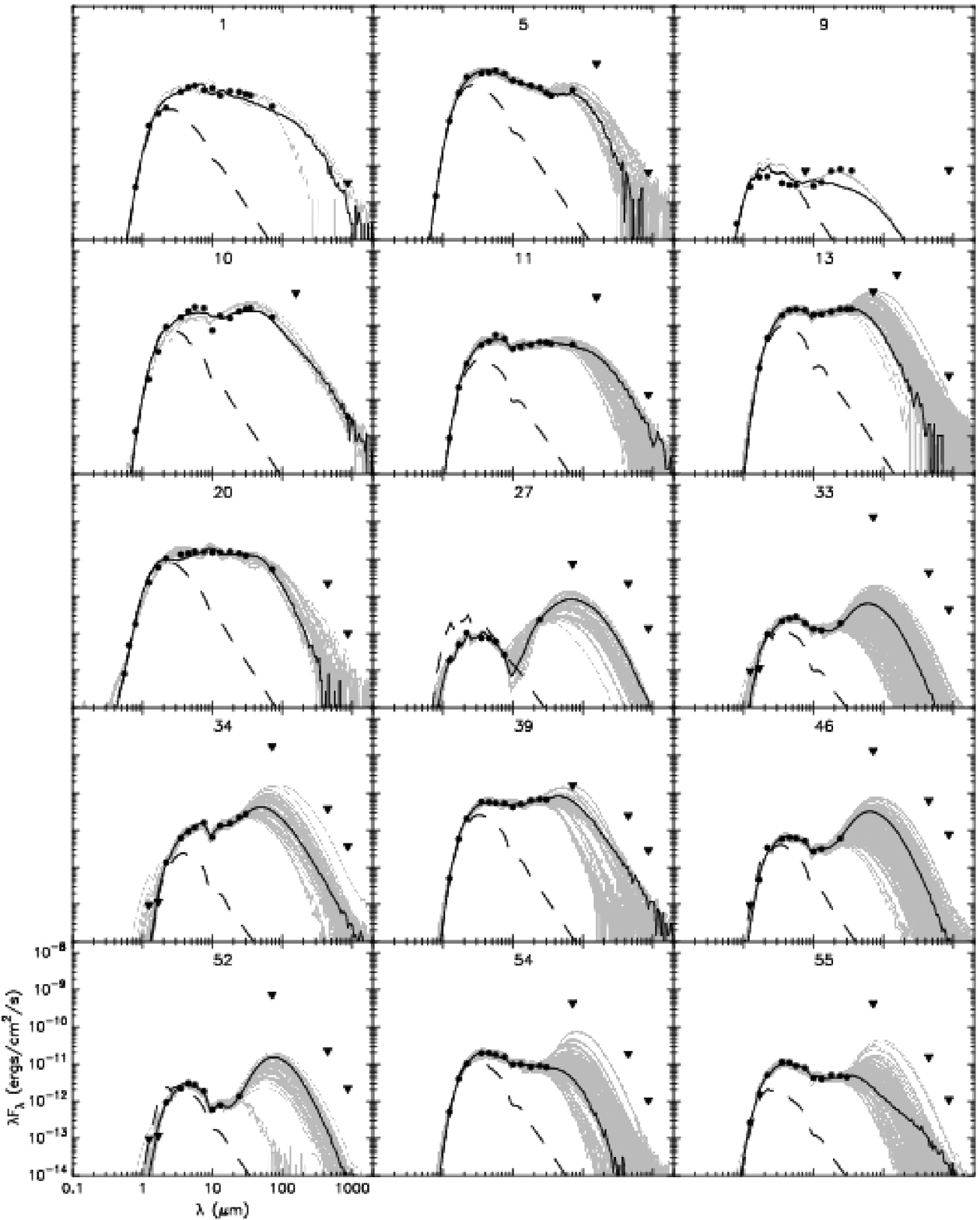}
\caption{Observed SEDs of the 15 disk sources with high external foreground extinction (stage II(ex) sources), with model SEDs providing a good fit overplotted. The symbols and lines are as in Figure~\ref{fig:g1seds}.\label{fig:g2seds}}
\end{center}
\end{figure*}

\begin{figure*}[htbp]
\begin{center}
\includegraphics[width=6.5in]{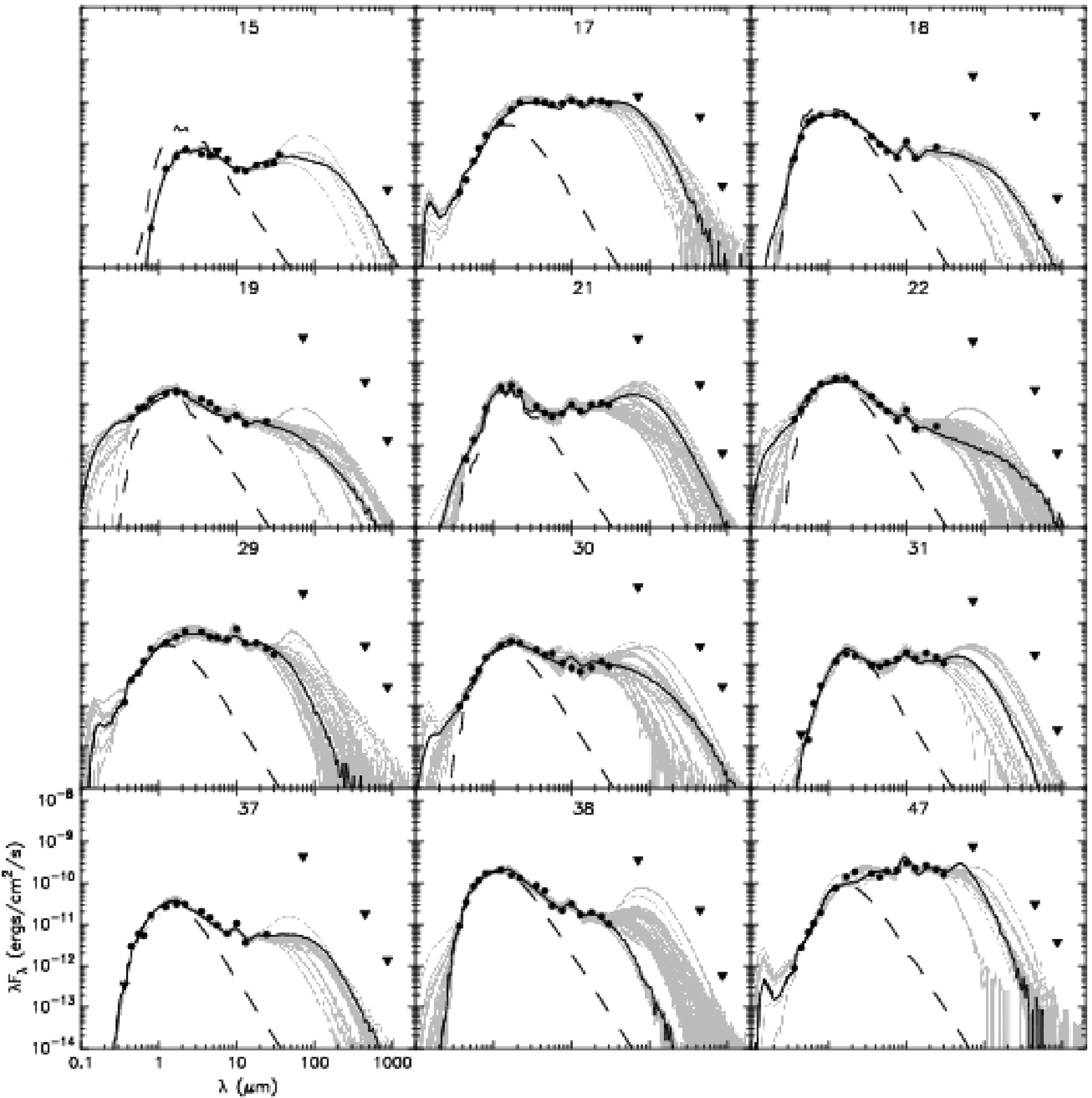}
\caption{Observed SEDs of the 12 disk sources with low external foreground extinction (stage II sources), with model SEDs providing a good fit overplotted. The symbols and lines are as in Figure~\ref{fig:g1seds}.\label{fig:g3seds}}
\end{center}
\end{figure*}

%\addtocounter{figure}{-1}

\begin{figure*}[htbp]
\begin{center}
\includegraphics[width=6.5in]{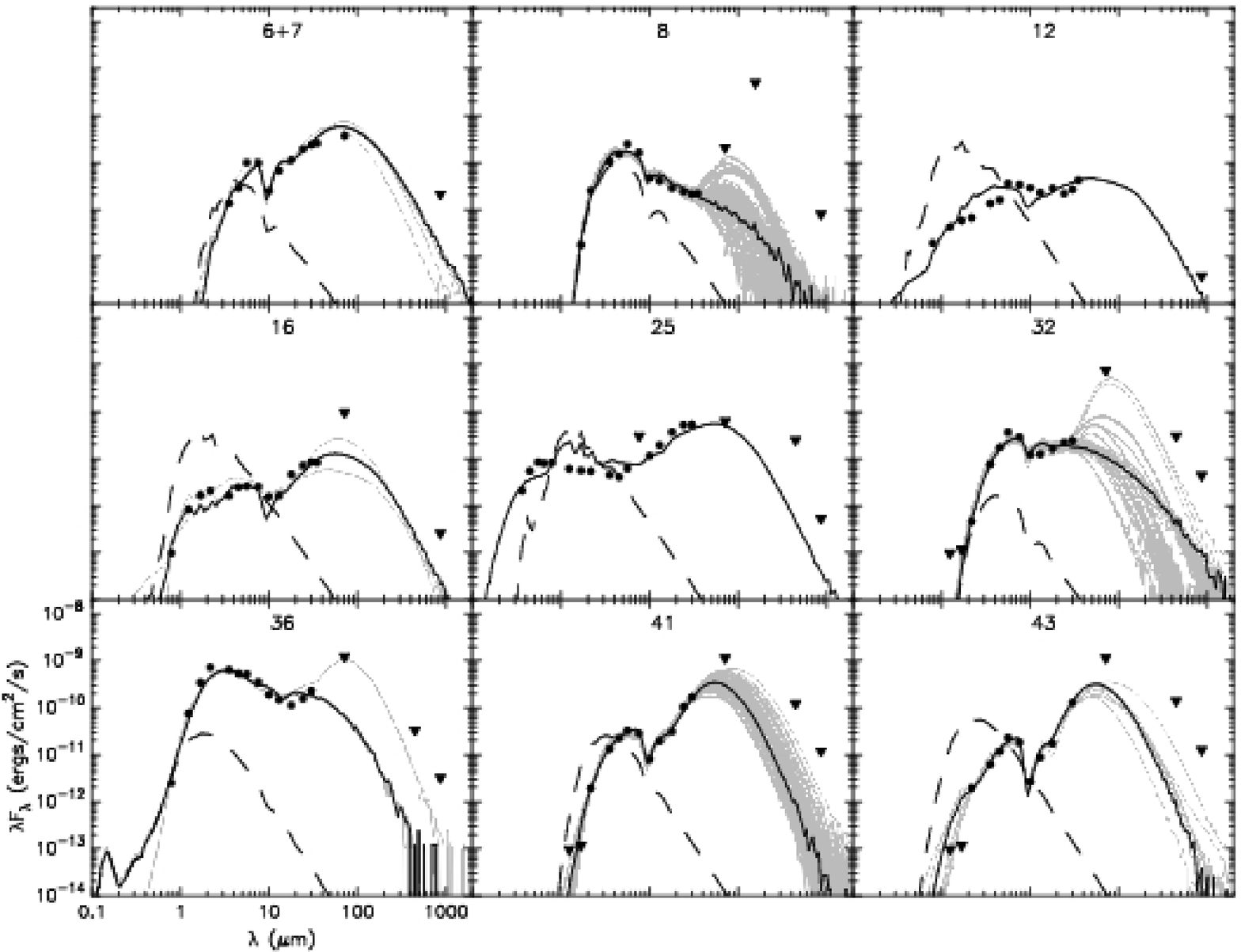}
\caption{Observed SEDs of sources where the modeling results cannot be trusted, as described in the text. Best fit results are only shown for guidance; note, for example, that good fits were obtained for the unresolved microcluster sources 41 and 43. The symbols and lines are as in Figure~\ref{fig:g1seds}.\label{fig:g4seds}}
\end{center}
\end{figure*}

\begin{figure*}[t]
\begin{center}
\includegraphics[height=3.5in]{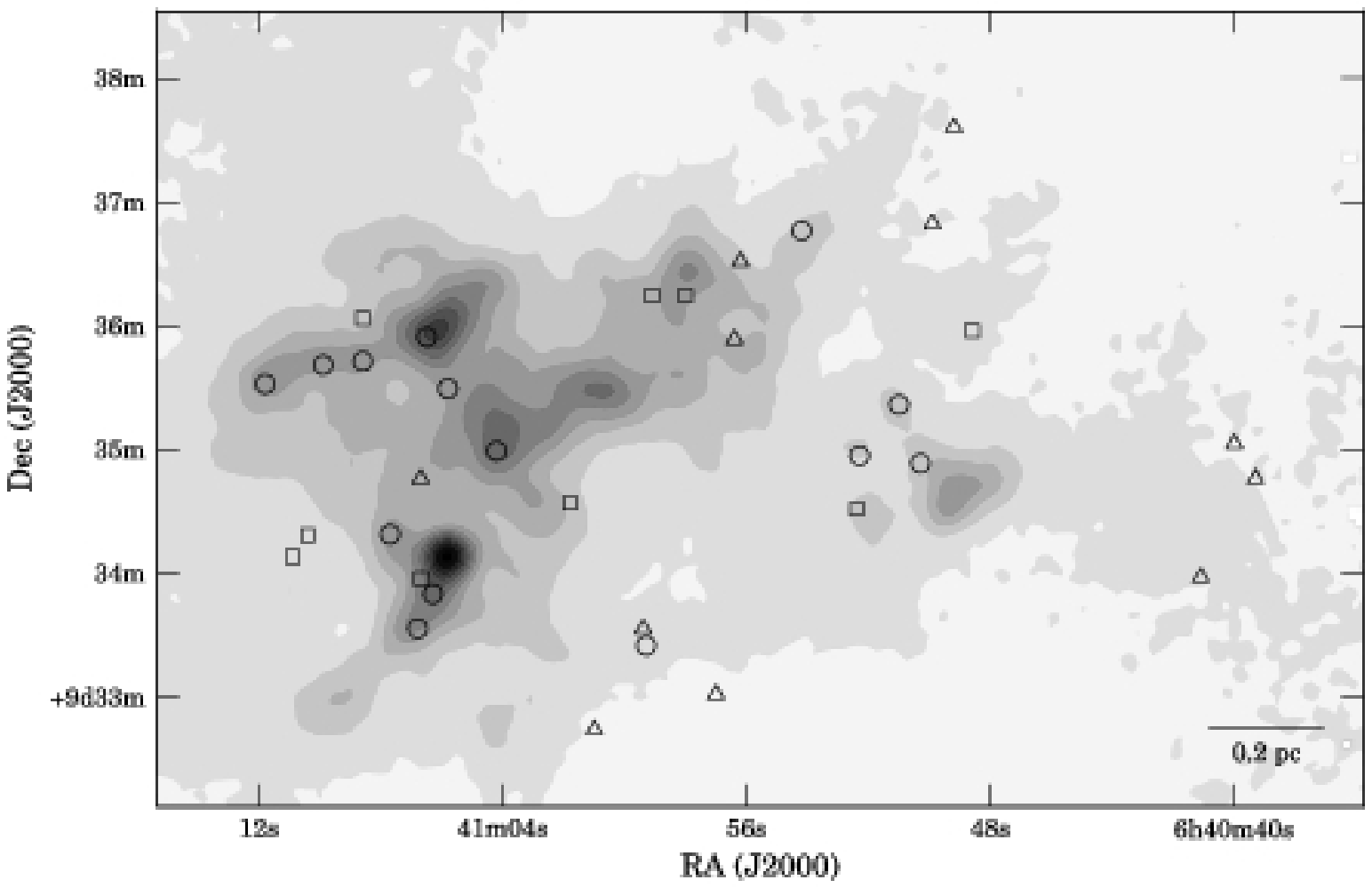} \\
\includegraphics[height=3.5in]{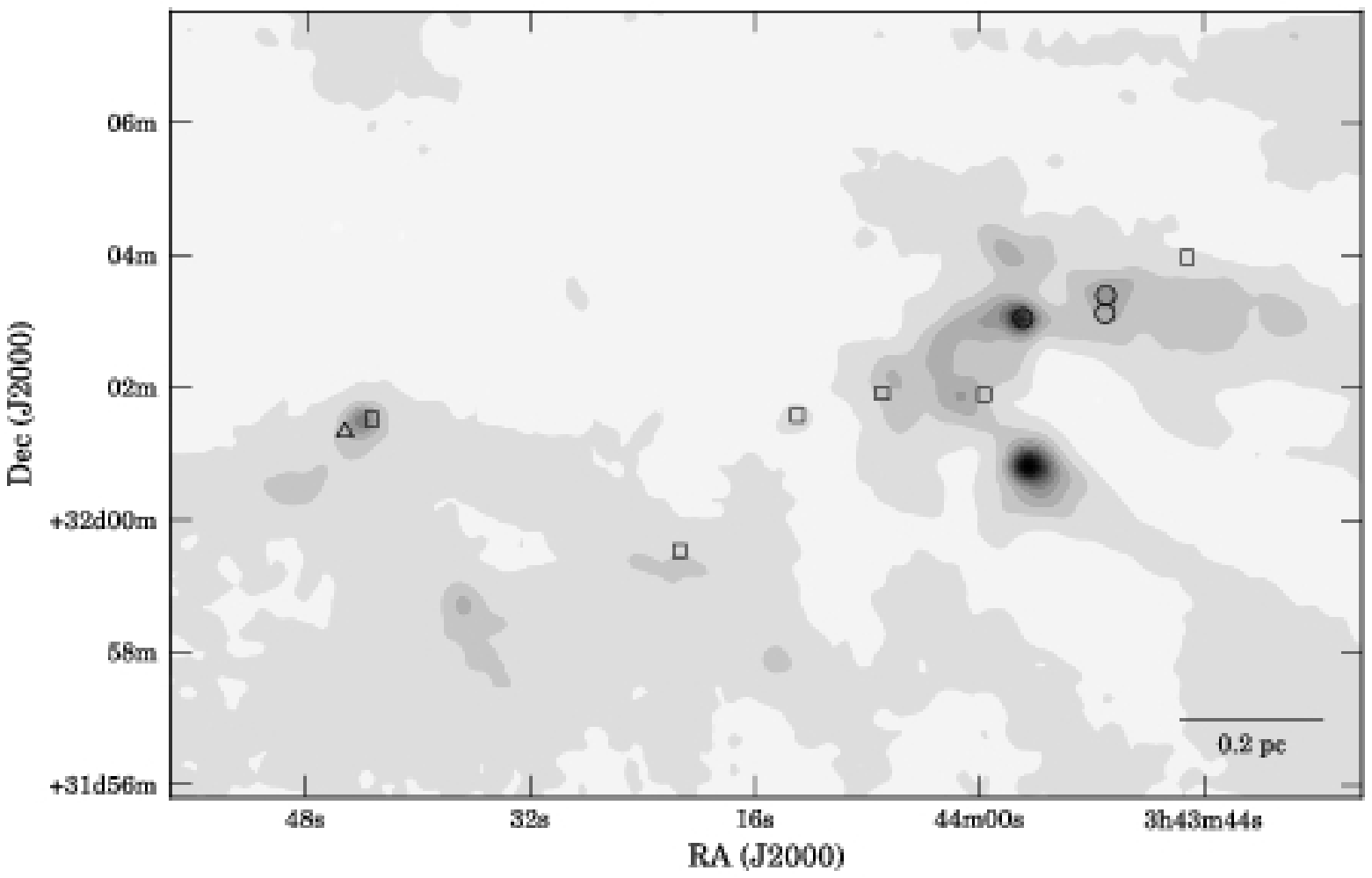}
\caption{Position of the sources overplotted on 850\microns SCUBA maps of NGC\,2264 (upper panel) and IC\,348 (lower panel). The symbols denote stage I (circles), stage II(ex) (squares), and stage II (triangles) sources respectively. Grayscale steps are in steps of 0.2~Jy\,beam$^{-1}$ up to 2~Jy\,beam$^{-1}$.\label{fig:scuba_map}}
\end{center}
\end{figure*}

\end{document}